\documentclass[twocolumn]{aastex701}

\usepackage{amsmath}
\usepackage{array, makecell}

\begin{document}

\title{Anisotropy-driven constraints on the transition from Galactic to extragalactic cosmic-ray sources
}

\author[0000-0003-4005-0857]{Teresa Bister}
\email{teresa.bister@ru.nl}
\affiliation{Nationaal Instituut voor Subatomaire Fysica (NIKHEF), Science Park, Amsterdam, The Netherlands}
\affiliation{Institute for Mathematics, Astrophysics and Particle Physics, Radboud University Nijmegen, Nijmegen, The Netherlands}
\author[0000-0002-0525-3758]{Foteini Oikonomou}
\email{foteini.oikonomou@ntnu.no}
\affiliation{%
 Department of Physics, Norwegian University of Science and Technology (NTNU),\\
 Høgskoleringen 5, Trondheim 7491, Norway
}
\author[0000-0003-4927-9850]{Damiano F.\ G.\ Fiorillo}
\email{damianofg@gmail.com}
\affiliation{Gran Sasso Science Institute, Viale F. Crispi 7, L’Aquila, 67100, Italy}
\affiliation{Istituto Nazionale di Fisica Nucleare (INFN), Sezione di Napoli,
Complesso Universitario di Monte Sant’Angelo, Via Cintia, 80126 Napoli, Italy}

\begin{abstract}
The spatial distribution of the cosmic-ray (CR) flux above $8\,\mathrm{EeV}$ exhibits a significant dipole in agreement with expectations from extragalactic sources following the large-scale structure. At the ankle, the direction changes suddenly and is compatible with the Galactic center for energies $\lesssim4\,\mathrm{EeV}$. 
We leverage the direction and strength of the dipole anisotropy as diagnostics of the possible sources of the intermediate-mass nuclei below the ankle that cannot be associated with the continuation of the Peters cycle of lower-energy Galactic CRs. Using 3D-simulations of CR propagation in up-to-date Galactic magnetic field models, we show that the dipole anisotropy of both continuous and transient individual Galactic sources and source distributions exceeds the measurements by $>3\sigma$. An extragalactic origin of the intermediate nuclei below the ankle instead agrees with all data, where the observed change in dipole direction is guided by the fading heavy Galactic component dominant at the second knee.
\end{abstract}


\section{Introduction} 
The spectrum of diffuse cosmic rays (CRs) observed at Earth is approximately described by a broken power law in energy from $10^{10}$~eV to above $10^{20}$~eV with a few notable features, namely, the \emph{knee} at energy $E = (3\text{--}5) \times 10^{15}$~eV, the second knee at $E = (1\text{--}4) \times 10^{17}$~eV, and the \emph{ankle} at $E \simeq  4 \times 10^{18}$~eV. At low energies, the observed CRs are commonly associated with Galactic accelerators. At energies above $E = 8 \times 10^{18}$ eV, a highly significant dipole anisotropy which points away from the Galactic Center strongly supports an extragalactic origin of the highest-energy CRs~\citep{PierreAuger:2017ScienceDipole}.  

In general, the knee~\citep{Kulikov:1958zh,KASCADE:2005ynk,TIBETIII:2008qon,IceCube:2019hmk,LHAASO:2024knt} is interpreted as the cutoff of the light (protons, helium) Galactic component. A rigidity-dependent scaling of the maximum energy~\citep{Peters:1961mxb} means that the Galactic iron cutoff is at $E \simeq  10^{17}$~eV, roughly at the energy where the second knee is observed~\citep{KASCADEGrande:2011kpw}, well below the ankle~\citep{HiRes:2007lra,PierreAuger:2020Spectrum}.
In this picture, the energy range between the second knee and the ankle remains unaccounted for. Initially, \citet{Hillas:2006ms} suggested that it could be filled by the so-called `Component B'', which he proposed to be Galactic. The idea was later developed and featured in the popular three-population model of~\cite{Gaisser:2013bla}.

The question of the Galactic-extragalactic transition energy and the origin of Component B remains unsettled. On the one hand, 
there is an argument that the second knee, a steepening of the spectrum, is an unlikely feature to mark an emerging new source population~\citep{Parizot:2014ixa} (but see~\citealp{Hillas:2006ms} Fig.\ 8 for an alternate view). According to this argument, a more natural transition feature is marked by the ankle. Several recent studies subscribe to this view and propose the continuation of a Galactic population of sources as the origin of the flux between the second knee and the ankle~\citep{Thoudam:2016syr,Bhadra_2024}. 

On the other hand,~\cite{Giacinti:2011Transition} showed that the expected dipole amplitude from a population of Galactic sources at $E \sim 10^{18}$ eV
exceeds the dipole bounds measured by the Pierre Auger Collaboration (hereafter Auger), if the primary composition is dominated by light (p, He) or intermediate (CNO) nuclei. Therefore, a transition at the ankle or above would require
a heavy, iron-dominated, composition, or a rather extreme random Galactic magnetic field with strength $\sim 10~\mu$G (see also~\citealp{PierreAuger:2013AnisotropyApJL}). Together with Auger composition measurements indicating that the iron fraction in this energy range is small and decreasing~\citep{PierreAuger:2026MassComposition}, this favors a Galactic-extragalactic transition at the second knee~\citep{Kachelriess:2019oqu}. 

Another possibility is that Component B is dominated not by a quasi-continuous source population, but by a single or a handful of Galactic sources~\citep{Farrar:ICRC2021}. In that case, the dipole anisotropy does not have to point to the Galactic center. Moreover, the dipole induced by these few sources would add vectorially to any extragalactic dipole, as well as to any contribution from a quasi-continuous Galactic source population. If this scenario is correct, the anisotropy constraints derived for a quasi-continuous Galactic source population in~\cite{Giacinti:2011Transition,PierreAuger:2013AnisotropyApJL} would therefore be affected.

Although the argument of~\cite{Giacinti:2011Transition} makes a strong case against a predominantly Galactic origin up to the ankle, the conclusion depends on the assumed Galactic magnetic field model and on the details of the CR composition in the relevant energy range. The most recent composition results from Auger~\citep{PierreAuger:2026MassComposition} and TALE~\citep{TelescopeArray:2026rdu} show that in this energy range the composition is predominantly protons and CNO-type nuclei, with a possible helium contribution, while only a small and decreasing iron component is allowed at the level of $\lesssim 20\%$. A common interpretation of the sub-ankle protons is that they arise from photodisintegration of extragalactic UHECR nuclei, which themselves contribute substantially to the CR flux above the ankle~\citep{unger:2015_ufa}. Even if the sub-ankle protons are assigned an extragalactic origin, the origin of the sub-ankle CNO component remains open: are these nuclei produced by an additional extragalactic population, or by an additional Galactic one? This question is particularly relevant since the CNO component does not appear to be part of the Galactic Peters cycle between the knee and second knee, nor the extragalactic one above the ankle.

Conclusions about the origin of Component B are sensitive to the assumed Galactic Magnetic Field (GMF). In the relevant energy range, $\sim10^{17}-10^{18}$~eV, propagation lies in an intermediate rigidity regime: particles are still significantly affected by Galactic magnetic deflections, but their rigidity is high enough that the diffusion approximation becomes increasingly unreliable. \citet{Giacinti:2011Transition} adopted a custom-made turbulent GMF, while the Auger Collaboration analysis of~\citet{PierreAuger:2013AnisotropyApJL} used the coherent GMF model of~\citet{Pshirkov:2011} together with a simplified turbulent component whose RMS strength was taken to be three times that of the coherent field. 
Subsequent GMF models have become substantially more data-constrained. In particular,~\citet{Jansson:2012pc,Jansson:2012b_Random} introduced a GMF model (\texttt{JF12}) whose random and coherent components were fit to all-sky rotation measure and synchrotron emission data. More recently,~\citet{Unger:2024uf23} introduced a suite of coherent-GMF models (\texttt{UF23}) fitted to full-sky extragalactic rotation measures and synchrotron maps, designed to quantify current systematic uncertainties in the coherent GMF and in UHECR deflections. The GMF prescriptions used in earlier studies of Component-B anisotropy differ from one another and from more recent data-constrained GMF models, both in terms of the large-scale geometry, and crucially in terms of the total strength of the turbulent GMF by up to a factor of a few, motivating a reassessment of the origin of Component B using updated GMF models. 

The observational constraints on the large-scale anisotropy have also improved substantially, with roughly a fivefold increase in exposure since previous works~\citep{Giacinti:2011Transition, PierreAuger:2013AnisotropyApJL}. The latest Auger analysis with 19 years of data reports dipole amplitudes and phases down to $E = 3\times10^{16}$~eV~\citep{PierreAuger:2024Anisotropy19Y}. Above 8\,EeV, the dipole is measured with more than $5\sigma$ significance. The direction of the dipole at the highest energies points away from the Galactic center, confirming an extragalactic origin at these energies. The direction can be well described by sources following the extragalactic matter distribution when deflections in the GMF are considered, and the observed rise in dipole amplitude can be well-explained by the decreasing energy loss length of extragalactic CRs~\citep{PierreAuger:2017ScienceDipole, Ding:2021, Allard:2022, Bister:2024Anisotropies, Bister:2024GMF, PierreAuger:2024Anisotropy19Y}.
At $\sim4\,$EeV, the dipole phase changes rapidly. At and below this energy down to $\sim0.1\,$EeV, it agrees with the right ascension of the Galactic center, no longer matching the direction predicted by sources following the extragalactic source distribution. In this energy range, the dipole is not significantly measured, so upper limits $\lesssim1\%$ apply. Note however that the dipole phase is steady with the energy, which makes a statistical fluctuation an unlikely origin (see discussion and Fig.~8 in~\citet{Deligny:2019}). Instead, the dipole phase pointing towards the Galactic center $<4\,$EeV raises the question of a possible Galactic contribution in this energy range.

A renewed interest in determining the Galactic-extragalactic transition energy comes from the recent detection of tens of Galactic PeVatron sources with the Large High Altitude Air Shower Observatory (LHAASO)~\citep{LHAASO:2023rpg}. Among these sources, a small number of microquasars has been shown to accelerate particles to PeV energies: their $\gamma$-ray spectra extend beyond $E_\gamma \sim 100~\mathrm{TeV}$, implying parent protons or electrons with energies $\gtrsim \mathrm{PeV}$~\citep{LHAASO:2024psv}.
The most striking among them is the microquasar Cygnus X-3, from which LHAASO has reported photons with energy $E \lesssim 4 $~PeV~\citep{LHAASO:2025ysm}. This observation, together with $3.2\sigma$ evidence for day-scale orbital modulation, suggests acceleration of particles up to $E_{p}^{\rm max}\sim 40$~PeV in a very compact region. Only protons and nuclei can be efficiently accelerated in such a compact environment without severe energy losses. This makes Cygnus~X-3 the clearest Galactic candidate for hadronic particle acceleration beyond $10$~PeV to date. By a simple rigidity-scaling argument, Cygnus X-3 could plausibly accelerate CNO nuclei into the lower Component~B energy range
$E_{\rm CNO}^{\max}~\sim Z E_p^{\max}~\simeq~Z \times 40~{\rm PeV}\simeq~2.4\text{--}3.2 \times 10^{17}~{\rm eV}$. 

In this work, we revisit the possibility that the CNO component observed in the Component-B energy range has a Galactic origin. Motivated by the recent observations of Galactic accelerators reaching tens of PeV energies, we consider Galactic source scenarios capable of accelerating CNO nuclei into the subankle regime. For the first time, we consider also the interplay of the anisotropy of the three components -- the end of the heavy Galactic sources dominant at the second knee, the onset of the extragalactic component around the ankle, and the intermediate-mass Component B below the ankle.

We study scenarios in which an entire source population contributes to this flux component, but also individual powerful sources. We consider the nearest PeVatron microquasar, Cygnus X-1 ($d \approx 2.22$~kpc~\citealp{Miller-Jones:2021plh}) as well as Cygnus X-3 which is among the most distant known $\gamma$-ray sources~\citep{Reid:2023ksq}, to bracket the effect of source distance on the predicted dipole. 
We also study the Galactic Center itself, whose very-high-energy $\gamma$-ray emission has long hinted at hadronic acceleration to $\gtrsim$~PeV energy, for example due to past activity of Sgr A$^{*
}$~\citep{HESS:2016pst,Albert:2024aaa}. Additionally, we consider young stellar clusters, which are among the most promising PeVatron source classes. High-energy $\gamma$-ray emission has been detected from several of them, whereas LHAASO has mapped a giant ultra-high-energy $\gamma$-ray bubble containing both massive star associations (the Cygnus Cocoon) and Cygnus X-1~\citep{LHAASO:2024_Cygnus_Cocoon, LHAASO:2024psv}. In addition, the Milky Way's most massive young stellar cluster, Westerlund 1, while outside LHAASO's field of view, shows emission up to 100 TeV~\citep{HES:2022_Westerlund1}.

We propagate UHECR nuclei from these sources/source populations through updated models of the GMF and compare the resulting amplitudes and phases with the latest Auger measurements. Our aim is to determine whether a Galactic CNO-dominated Component B can remain compatible with present anisotropy constraints.

\section{A Three-Component Model for the Sub-Ankle Dipole} \label{sec:fit}
We build a combined model consisting of three parts, an extragalactic component (``XGal"), a Galactic component (``Component A"), and an additional component that can be either Galactic or extragalactic (``Component B"). Determining what dipole amplitude and direction Component B should have to reproduce the measured dipole amplitudes and the observed shift in dipole direction is the goal of this analysis.

\subsection{Three-Component Model}
\paragraph{Assumed mass composition and energy range}
The relative contributions of the three components as functions of energy are based on the observed mass composition fractions by Auger~\citep{PierreAuger:2026MassComposition}, as only these are available for the two up-to-date hadronic interaction models compatible with measurements - EPOS-LHC-R~\citep{pierog:2025_eposlhcr} and Sibyll 2.3e~\citep{Riehn:2020_Sibyll}. The mass composition fractions by Auger as well as the energy-dependent contributions of the three components in our model are visualized in Fig.~\ref{fig:comp} in the Appendix, and the exact assumed relative flux fractions as a function of energy are summarized in Table~\ref{tab:fit}. We take into account the limited field of view of the Pierre Auger Observatory when applying the relative weights of the three components.

Mass fractions using these hadronic interaction models are only reported $>0.5$ EeV, which sets the lower limit of this analysis. Above 4 EeV, the contribution of Component B is not necessary to describe the dipole amplitude and direction, as both are well described by the extragalactic model alone. Thus, as observables to determine the dipole amplitude and direction of Component B, we use the equatorial dipole amplitudes $\delta_\bot$ and phases measured by Auger~\citep{PierreAuger:2024Anisotropy19Y} in the three energy bins $(0.5-1)\,$EeV, $(1-2)\,$EeV, and $(2-4)\,$EeV. The equatorial dipole amplitude $\delta_\bot$ is related to the total amplitude $\delta$ via a multiplication with the cosine of the declination angle
\begin{equation}  \label{eq:dipole_eq}
    \delta_\bot = \delta \cos(\mathrm{dec}).
\end{equation}
The declination of the dipole, as well as higher multipole moments like the quadrupole, are not reported below 4\,EeV and hence cannot be used to constrain the scenario.

\paragraph{Galactic Component A}
We assume a Peters cycle of (at least one) Galactic population where iron is dominant at the second knee, at $E \approx 10^{17}$~eV, consistent with composition measurements in this energy range~\citep{Apel:2013uni,IceCube:2019hmk,TelescopeArray:2026rdu}. We test two different cases, one where a subdominant iron tail of Component A still contributes above $0.5\,$ EeV, and one where it is negligible. These two cases are represented by the median values of the mass composition fractions using the two hadronic interaction models~\citep{PierreAuger:2026MassComposition}.
For EPOS-LHC-R, the relative contribution of iron is roughly $15\%$ between 0.5\,EeV and 1\,EeV, $10\%$ between 1\,EeV and 2\,EeV, and $5\%$ between 2\,EeV and 4\,EeV. For Sibyll 2.3e on the other hand, the relative contribution is compatible with zero above 0.5\,EeV. Note however that both models agree within systematic uncertainties. The rigidity of Component A (i.e.~of iron) in the three energy bins is $R\simeq2\times 10^{16}$\,V, $R\simeq5\times10^{16}$\,V. and $R\simeq9\times10^{16}$\,V.

We use a generic model based on the distribution of matter along the spiral arms in the Galactic disk to model Component A. Concretely, we use the distribution of pulsars (\texttt{SourcePulsarDistribution} as provided by CRPropa3). As our aim is not the characterization of Component A, this choice should be understood as a tracer of the Galactic disk and not a physical identification of Component A with pulsars. Detailed 3D-simulations using the \texttt{UF23} GMF model suite~\citep{Unger:2024uf23} in combination with the \texttt{JF12} Planck-tuned random field model~\citep{Jansson:2012b_Random, Planck:2016GMF} predict an average dipole amplitude of $\sim5\%$ pointing to the Galactic center direction for iron from the pulsar distribution at $1\,\mathrm{EeV}$, as described in more detail below in Sec.~\ref{sec:crpropa}. We use that average dipole amplitude for Component A\footnote{Concretely, we use $\delta_\bot = $5\%, 6\%, 7\% for the three energy bins (0.5-1)\,EeV, (1-2)\,EeV, (2-4)\,EeV, which is the dipole amplitude of nitrogen from the pulsar distribution averaged over all 8 \texttt{UF23} GMF models and 2 random variations of the \texttt{JF12-Pl} random field, see Fig.~\ref{fig:amp_ra}.}, and keep its direction fixed towards the Galactic center.

\paragraph{Extragalactic Component}
For the extragalactic component, we adopt the model of~\citet{Bister:2024Anisotropies, Bister:2024GMF} which is based on sources following the large-scale structure (LSS), and deflections modeled using again the \texttt{UF23} GMF model suite with the \texttt{JF12} Planck-tuned random field model. The model was fit to the energy spectrum, mass composition, and dipole measurements above 8\,EeV, and we extend it here to lower energies $E>0.5\,$EeV.
For the model of the LSS, we use both the constrained simulations based on CosmicFlows2~\citep{Hoffman:2018_CF2} and the updated CosmicFlows4~\citep{Valade:2026_CF4}.  
The sources emit a hard spectrum and mixed composition following a Peters cycle (see \textit{baseline model} in~\citet{Bister:2024GMF} Table 2, in agreement with the fit parameters found by Auger in~\citet{PierreAuger:2022AcrossAnkle}). Due to that, the extragalactic component around the ankle is predominantly light. We assume that the protons observed below the ankle~\citep{PierreAuger:2026MassComposition, Apel:2013ura} are produced by the same extragalactic source population through in-source interactions~\citep{unger:2015_ufa}\footnote{Note that \textit{Galactic} proton sources in the EeV energy range would induce an anisotropy that is significantly larger than observed, see Sec.~\ref{sec:crpropa}. An upper limit of 1.3\% has been explicitly placed on the contribution of Galactic EeV protons in~\citet{Abbasi:2017_Galprotons}.}.
For both hadronic interaction models, the relative contribution of the extragalactic population is roughly $40\%$ between 0.5\,EeV and 1\,EeV and then rises up to $70\%$ between 2\,EeV and 4\,EeV for EPOS-LHC-R and $85\%$ for Sibyll2.3e (see Fig.~\ref{fig:comp} in the Appendix).

Below 1\,EeV, the energy losses of extragalactic protons are dominated by adiabatic losses due to the Hubble expansion, and the energy-loss length is of the order of the Hubble scale ($\sim4\,$Gpc)~\citep{Berezinsky:2002nc}. The contribution of individual extragalactic sources becomes negligible, and the expected dipole amplitude becomes very small. However, a non-negligible dipole amplitude is still expected due to the movement of the Galaxy in relation to the Cosmic Microwave Background -- the Compton-Getting effect~\citep{Compton_Getting:1935, Kachelriess:2006}. As the amplitude of the Compton-Getting dipole of 0.6\% (before GMF deflections) becomes comparable or larger than the one predicted by the LSS model below $\sim2$\,EeV, this effect must be taken into account in our model. Note however that the dipole amplitude is significantly suppressed by the GMF due to demagnification~\citep{Bister:2024GMF} and deflections. In the (0.5-1)\,EeV energy bin, the amplitude of the Compton-Getting dipole is only $\sim0.03\%$ on average after propagation through the \texttt{UF23} GMF models, which is significantly below the observed amplitude of $0.4^{+0.3}_{-0.2}\%$~\citep{PierreAuger:2024Anisotropy19Y}. A similar suppression of the Compton-Getting dipole was also reported for the \texttt{JF12} GMF model in~\citet{Mollerach:2022}.

The dipole direction predicted by the extragalactic model including the Compton-Getting effect and GMF deflections is quite stable as a function of energy, as shown in Fig.~\ref{fig:XGal} in the Appendix. We also include the \texttt{JF12}~\citep{Jansson:2012pc} and \texttt{KST24}~\citep{Korochkin:2025kst} GMF models in the figure, which predict similar dipole directions and amplitudes to the ones based on the \texttt{UF23} model suite.

\paragraph{Component B}
The relative contribution of Component B can be deduced from the nitrogen (and helium) fractions~\citep{PierreAuger:2026MassComposition}. For both hadronic interaction models, it is around 50\% between 0.5\,EeV and 1\,EeV and decreases to around 20\% between 2\,EeV and 4\,EeV. The rigidity of Component B (i.e.~of nitrogen) in the three energy bins is $R\simeq1\times 10^{17}$\,V, $R\simeq2\times10^{17}$\,V. and $R\simeq3.5\times10^{17}$\,V.

For the anisotropy of Component B, we test two scenarios. In the first scenario, we assume that the flux is \textit{isotropic}. Negligible anisotropy would be expected in case Component B is extragalactic. The extragalactic proton component is already nearly isotropic at these energies (as we show later, see Fig.~\ref{fig:fit_sibyll}), and nitrogen nuclei have a larger energy-loss length at these energies and undergo stronger magnetic deflections.

As an alternative, we allow Component B to be \textit{anisotropic}. In that case, the dipole amplitude and direction of Component B is determined by a fit to the measured equatorial dipole amplitude and right ascension in the three energy bins. We take into account the limited field of view of the Pierre Auger Observatory for the calculation of the dipole. The amplitudes of Component B in the three energy bins are individual fit parameters. We only allow for a constant or increasing dipole amplitude with the energy (as expected as transport becomes more ballistic with increasing rigidity). Because in our fit range below 4\,EeV only the equatorial dipole amplitude is measured and its declination is not known, a large total dipole amplitude at directions close to the equatorial poles would be compatible with the limits on the equatorial dipole amplitude (eq.~\ref{eq:dipole_eq}). Thus, a point source close to the equatorial South pole (the North pole is not in the Auger field of view) is in principle not in tension with the upper limits on the equatorial dipole, as long as it does not contribute above 4\,EeV where both components of the dipole are measured. Note however that such a source is not visible in flux maps from 2012~\citep{Auger:2012_anisotropy} (Fig. 7), which are the last ones released by the Pierre Auger Collaboration below 4~EeV. To circumvent such unlikely point sources close to the equatorial poles, we set another constraint, which is that the total dipole amplitude below 4~EeV must not exceed the one above 4~\,EeV where both dipole components are measured.

The direction of the Component B anisotropy is kept fixed over the three energy bins. This allows for a good fit without introducing more parameters than observables. Below, in Sec.~\ref{sec:crpropa}, we will verify our assumptions through 3D-simulations, i.e.~that the dipole amplitude of specific simulated Galactic sources always increases with energy, and that the dipole direction is rather stable with energy.

The two scenarios for Component B that we are testing are similar to the ones used in the combined fit across the ankle by Auger~\citep{PierreAuger:2022AcrossAnkle}. In that work, they fit two scenarios to the spectrum and mass composition: one where the nitrogen component below the ankle is Galactic, and one where it is extragalactic. No assumption about the origin of the protons below the ankle was made in~\citet{PierreAuger:2022AcrossAnkle}, only that they are extragalactic. We however follow~\citet{unger:2015_ufa} and assume that the protons are produced from in-source interactions in the same extragalactic source population present above the ankle that traces the LSS, in order to be able to predict their anisotropy. Also, the possible contribution of sub-ankle iron was neglected in the model of~\citet{PierreAuger:2022AcrossAnkle}. Based on the goodness-of-fit to spectrum and composition, no scenario was preferred over the other in~\citet{PierreAuger:2022AcrossAnkle}. In this work, however, we will leverage the anisotropy to differentiate the two scenarios. 

Like~\citet{Auger:magnetic}, we do not consider a suppression of the low-energy flux of Component B due to diffusion in the extragalactic magnetic field. This is because Component B, if extragalactic, is assumed to be from an abundant class of sources with a large density. The flux for such a source population would not be significantly influenced by a magnetic horizon in the energy range $\gtrsim0.5\,$EeV used in this analysis.

\subsection{Fit results}
The goodness of fit as well as the fit parameters of Component B and their uncertainties in case Component B is anisotropic are summarized in Table~\ref{tab:fit}. The reported values are based on the extragalactic model using CosmicFlows4~\citep{Valade:2026_CF4} for the source distribution and \texttt{UF23-twistX}~\citep{Unger:2024uf23} for the GMF deflections. Neither the fit parameter values nor the goodness of fit depend on that choice, and similar values are reached for CosmicFlows2 and any other GMF model of the \texttt{UF23} model suite, as well as the \texttt{KST24} model. This is mainly because all GMF models predict a stable dipole direction of the extragalactic component as a function of energy, which never points towards the Galactic center (see Fig.~\ref{fig:XGal}).

\begin{table*}[ht]
\centering
\begin{tabular}{l | l l | l l}
Scenario & \multicolumn{2}{c|}{\textbf{No Iron} (Sibyll 2.3e)} & \multicolumn{2}{c}{\textbf{Iron-tail} (EPOS-LHC-R)} \\
Component B & Isotropic & Anisotropic & Isotropic  & Anisotropic  \\
\hline
$f_\mathrm{XGal}$  (LSS+\texttt{UF23-twistX})                & \multicolumn{2}{c|}{[0.4, 0.7, 0.8]}  & \multicolumn{2}{c}{[0.4, 0.5, 0.7]} \\
$f_\mathrm{compA}$   (Gal. pulsar distribution, Fe)              & \multicolumn{2}{c|}{[0, 0, 0]}  & \multicolumn{2}{c}{[0.15, 0.1, 0.05]} \\
$f_\mathrm{compB}$                  & \multicolumn{2}{c|}{[0.6, 0.3, 0.2]}  & \multicolumn{2}{c}{[0.45, 0.4, 0.25]} \\
\hline
$\delta_\mathrm{B}^\mathrm{0.5-1 EeV} / \%$  & 0                & $0.8^{+0.5}_{-0.3}$& 0               &      $0.04^{+0.14}_{-0.02}$         \\
$\delta_\mathrm{B}^\mathrm{1-2 EeV} / \%$    & 0                & $1.7^{+1.4}_{-0.7}$ & 0               &     $0.3^{+0.8}_{-0.2}$            \\
$\delta_\mathrm{B}^\mathrm{2-4 EeV} / \%$    & 0                & $6.7^{+3.5}_{-2.1}$ & 0               &    $2.8^{+2.5}_{-1.2}$           \\
RA$_\mathrm{B} / ^\circ$                    & /                & $-60^{+17}_{-16}$    & /               &      $-22^{+26}_{-33}$         \\
dec$_\mathrm{B} / ^\circ$                   & /                & $-1^{+52}_{-47}$      & /               &      $-13^{+60}_{-50}$         \\
\hline
99\% C.L. upper limit on $\delta_\mathrm{B,\bot}^\mathrm{0.5-1 EeV} / \%$  & 0                & 1.4 & 0               &      0.6        \\
99\% C.L. upper limit on $\delta_\mathrm{B,\bot}^\mathrm{1-2 EeV} / \%$    & 0                & 3.5 & 0               &     1.8            \\
99\% C.L. upper limit on $\delta_\mathrm{B,\bot}^\mathrm{2-4 EeV} / \%$    & 0                & 9.0 & 0               &    4.5           \\
\hline
$\chi^2$ / ndf  (MAP)                     & 59.8 / 6         & 1.44 / 1             & 11.1 / 6        &    1.8 / 1           \\
\hline
Conclusion                     & Not a good fit         & Comp. B required             & Comp A sufficient        &    No gain           \\
\end{tabular}
\caption{Overview of fits to the dipole amplitude and phase measurements by Auger in the three energy bins 0.5-1 EeV, 1-2 EeV, and 2-4 EeV. The model consist of a Galactic Component A (iron, and based on the distribution of Galactic pulsars), extragalactic component (based on the extragalactic matter distribution, mixed composition above the ankle), and Component B (mostly nitrogen below the ankle), which is either fully isotropic (i.e.~extragalactic), or whose dipole amplitudes $\delta_\mathrm{B}$ in the three energy bins and its direction are fit parameters. The relative flux contributions $f$ of the three components in the three energy bins are based on the mass fractions measured by Auger~\citep{PierreAuger:2026MassComposition} and summarized in the first three rows of the table. The posterior mean and $1\sigma$ uncertainties of the model parameters, the resulting 99\% C.L. upper limits on the equatorial dipole amplitudes $\delta_\mathrm{B,\bot}$ in the three energy bins, the goodness of fit $\chi^2$/ndf of the maximum a posteriori (MAP), and a conclusion about the scenario, are provided in the other rows. }
\label{tab:fit}
\end{table*}

\begin{figure*}[ht]
\includegraphics[width=0.49\textwidth]{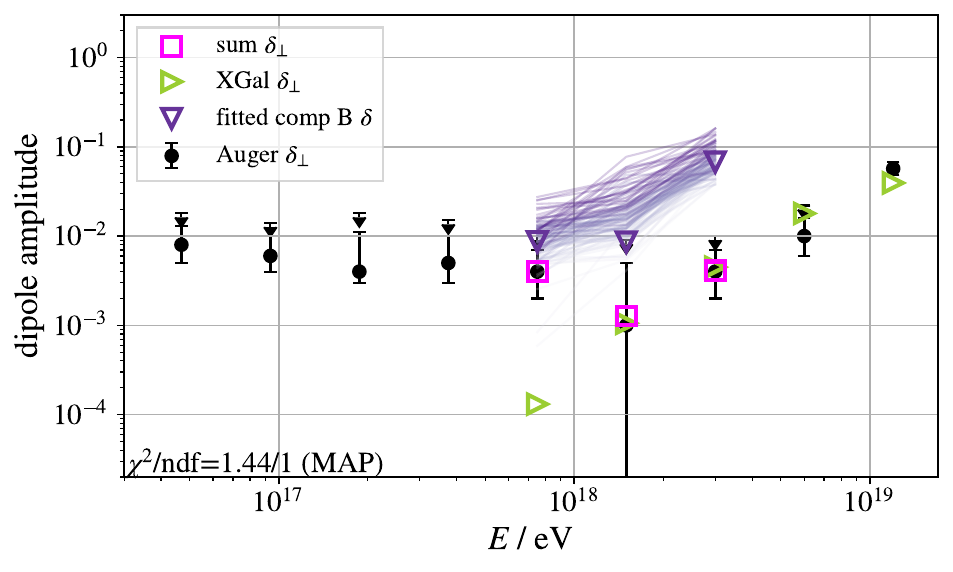}
\includegraphics[width=0.49\textwidth]{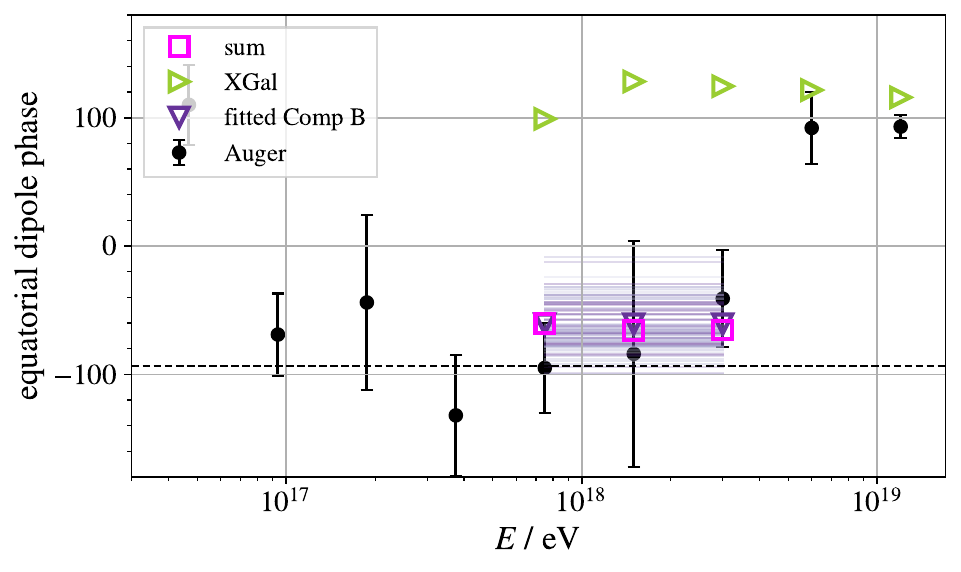}\\
\includegraphics[width=0.49\textwidth]{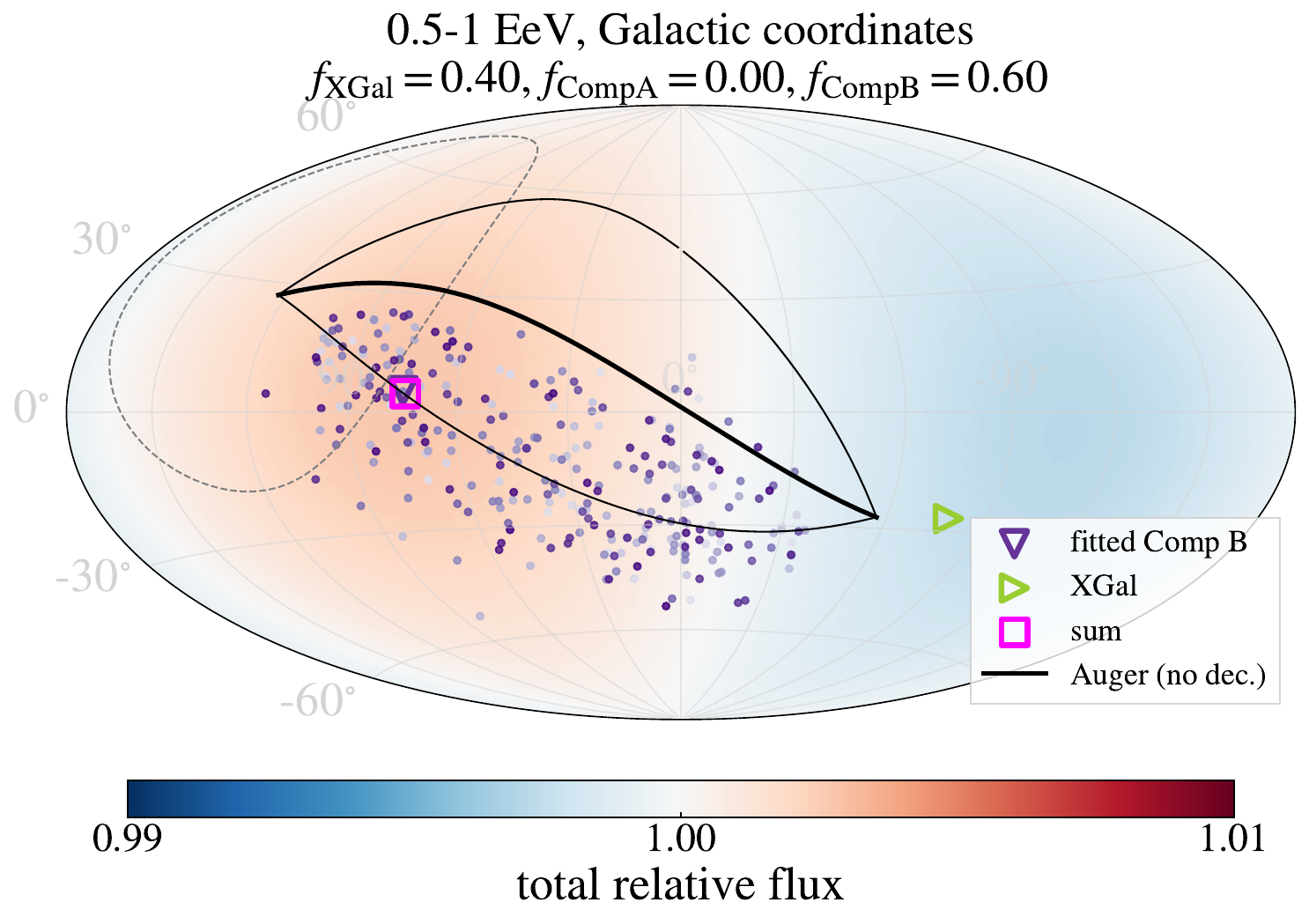}
\includegraphics[width=0.49\textwidth]{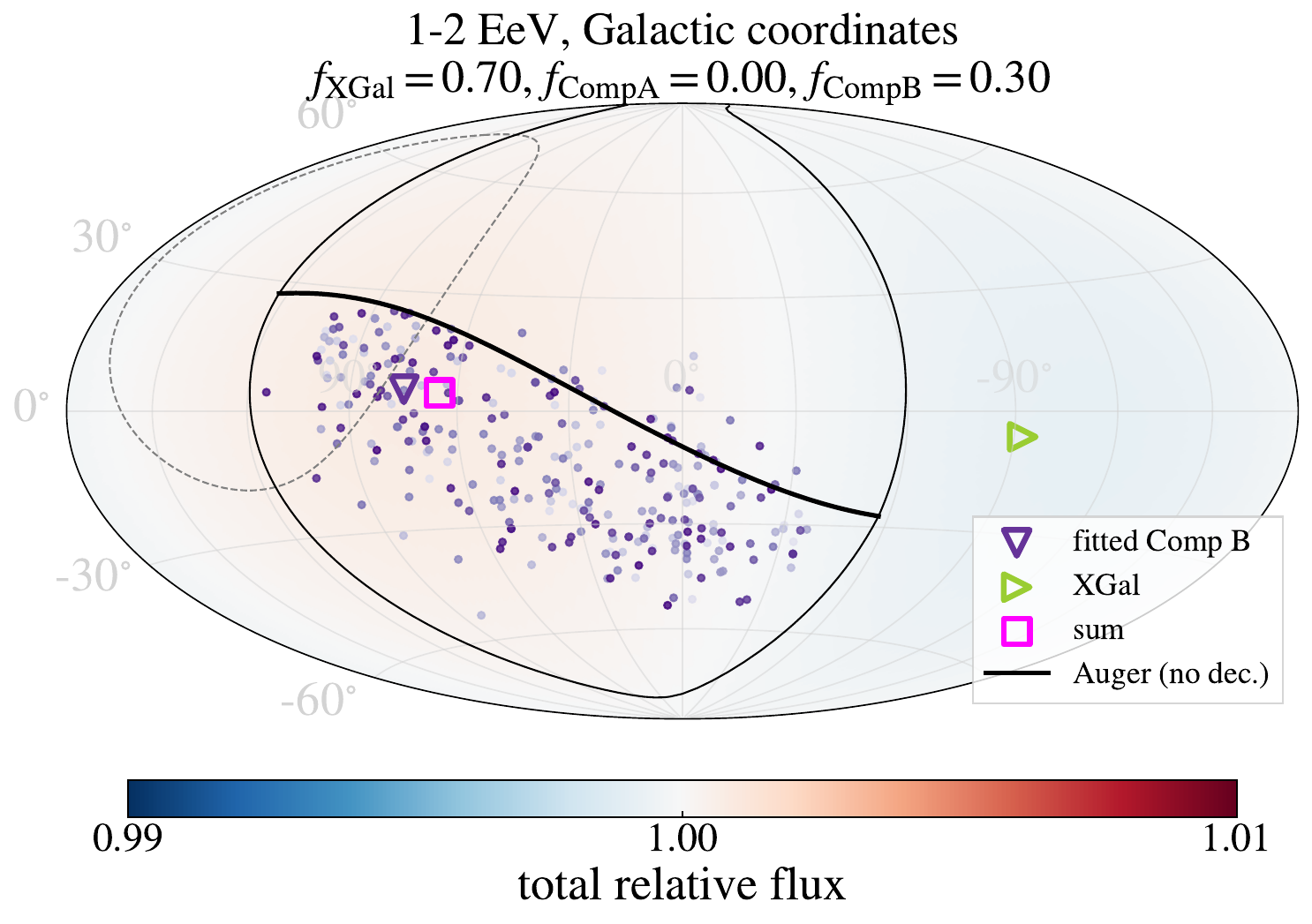}
\includegraphics[width=0.49\textwidth]{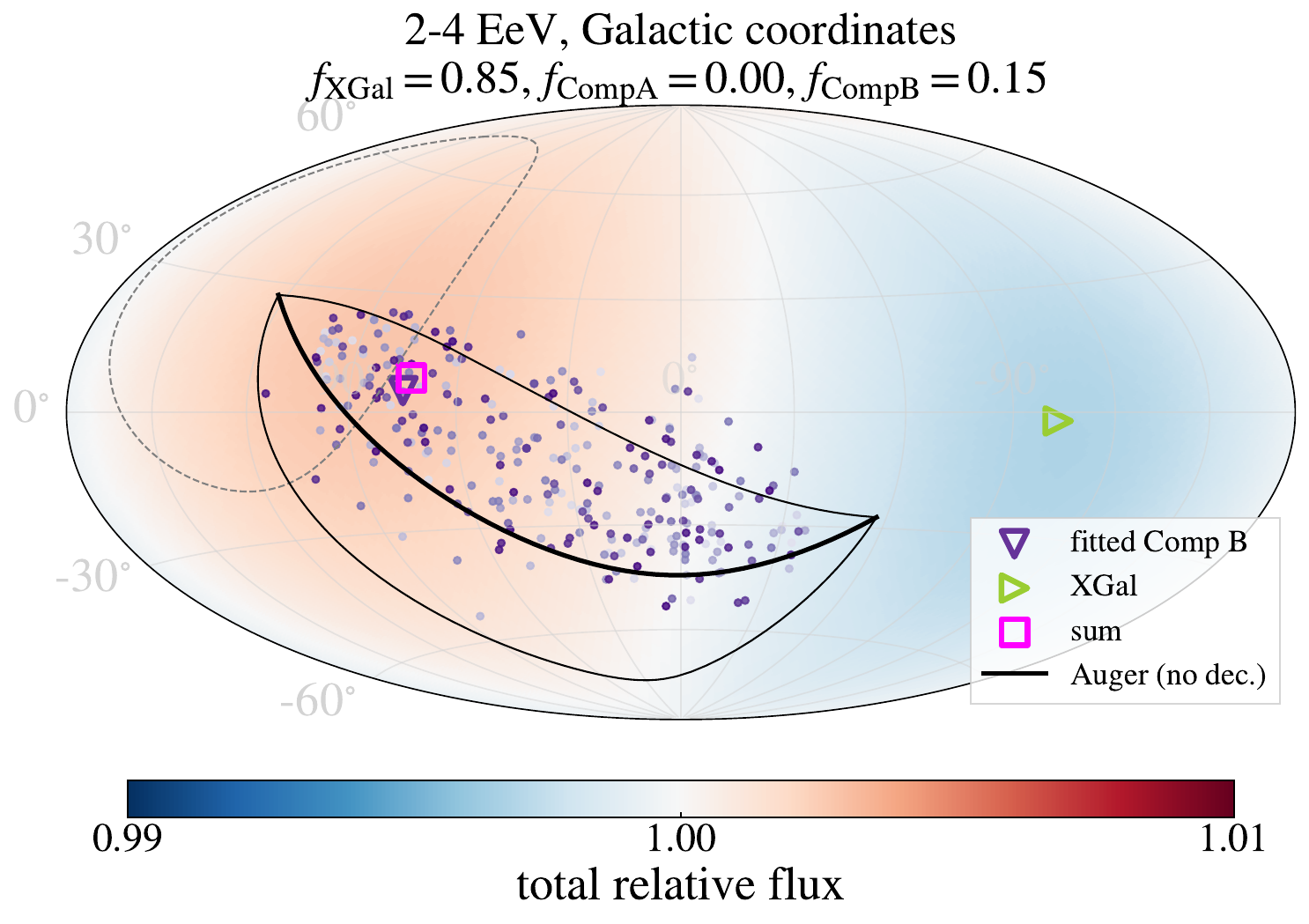}
\includegraphics[width=0.49\textwidth]{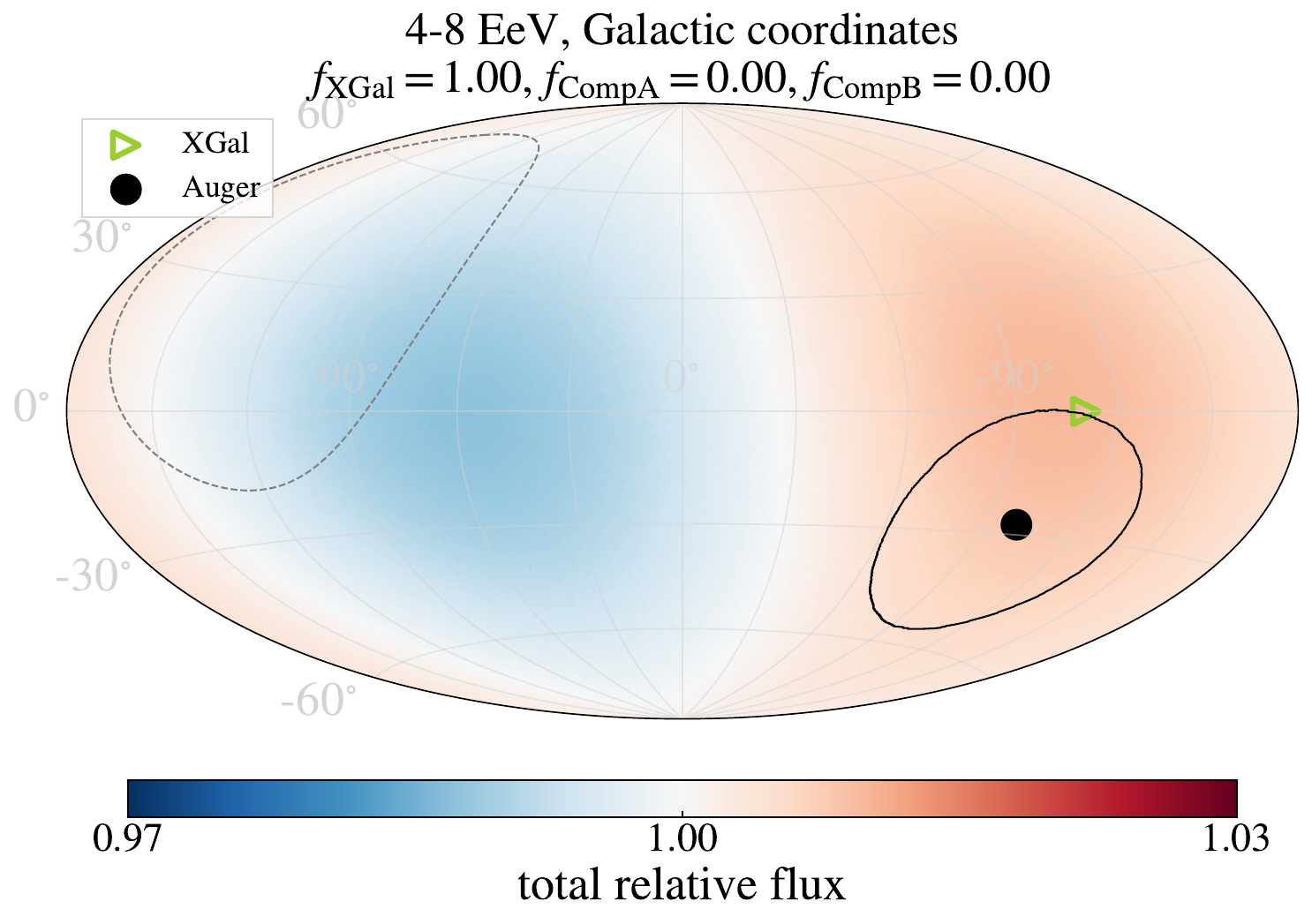}
\caption{Visualization of the combined model in the \textbf{no-iron scenario} (no Galactic Component A above $0.5\,$EeV), and Component B is anisotropic.
In the upper row, the dipole amplitude and equatorial dipole phase (i.e.~right ascension) are shown, and in the middle and lower rows the resulting flux in Galactic coordinates in four energy bins, smoothed with a $45^\circ$ Gaussian smearing. 
The green / purple / pink unfilled markers represent the extragalactic component / Component B best-fit / summed dipole direction, taking into account the flux weights $f$ as stated in the skymap titles and Table~\ref{tab:fit}. The purple dots and lines show draws from an MCMC sampler to visualize the uncertainty on the dipole of Component B. The color visualizes the mean dipole amplitude over the three energy bins (darker purple = larger amplitude). The dashed line in the upper right plot is the right ascension of the Galactic center.
The black points in the upper row show the Auger data from~\citet{PierreAuger:2024Anisotropy19Y}, where the error bars represent $1\sigma$ uncertainties. For energy bins where the dipole is not yet established at discovery level, 95\% C.L. upper limits are shown. In the skymap for (4-8)\,EeV, the Auger measurements are shown as a central marker and $1\sigma$ contour. Below 4\,EeV where the declination is not measured, the measured right ascension is represented as a thick line, and thinner lines visualize the uncertainty. The edge of the Auger exposure is shown as a dashed circle.}
\label{fig:fit_sibyll}
\end{figure*}

\paragraph{No-iron scenario}
In the no-iron scenario (as in the case for the median Sibyll 2.3e composition fractions, i.e.\ the Galactic Component A iron has ended well below our fitting range), an isotropic Component B cannot describe the dipole observations ($\chi^2/\mathrm{ndf}\simeq60/6\simeq10$, corresponding to $>5\sigma$ discrepancy). This is as expected for two independent reasons: the change of dipole direction towards the Galactic center cannot be reproduced by sources following the extragalactic matter distribution (see Fig.~\ref{fig:XGal}), and the measured dipole amplitude in the (0.5-1)\,EeV bin of $0.4\%$~\citep{PierreAuger:2024Anisotropy19Y} is too large to be reproduced by extragalactic sources at this energy even when the Compton-Getting effect is taken into account, as described above. 
Thus, under the assumptions of our model, at least a subdominant part (Component A and/or Component B) of the CR flux between 0.5\,EeV and 4\,EeV must be Galactic to reproduce the change in dipole phase at $\sim4\,$EeV and the observed amplitudes on percent level $<1\,$EeV.  

Allowing Component B to be anisotropic significantly improves the fit quality ($\chi^2$/ndf=1.44/1). A visualization of this scenario is shown in Fig.~\ref{fig:fit_sibyll}. In this case Component B must be Galactic: the preferred dipole amplitude of Component B is around 1\% below 2\,EeV, and increases to $\sim10$\% in the 2-4\,EeV bin. Such large amplitudes cannot be produced by extragalactic sources at these energies\footnote{Note that the preferred dipole amplitude can be larger than the naively expected maximum value of measured total amplitude divided by the relative flux of Component B ($0.004 / 0.2 = 2\%$ at 2-4\,EeV). This is due to the extragalactic component which points in a different direction, partly canceling the dipole of Component B. That effect leads to a total dipole amplitude that is smaller than the amplitudes of the two components separately above 2\,EeV, as visible in Fig.~\ref{fig:fit_sibyll} (\textit{upper left}).}

The preferred right ascension of Component B is $-60^{+17}_{-16}$ degrees. The declination is not well constrained which is due to the missing measurement of that component of the total dipole below 4\,EeV. This leads to a large spread of possible dipole directions including Galactic longitudes $-45^\circ\lesssim l\lesssim 110^\circ$ and Galactic latitudes close to the Galactic plane with $-50^\circ\lesssim b\lesssim 25^\circ$.

\begin{figure*}[ht]
\includegraphics[width=0.49\textwidth]{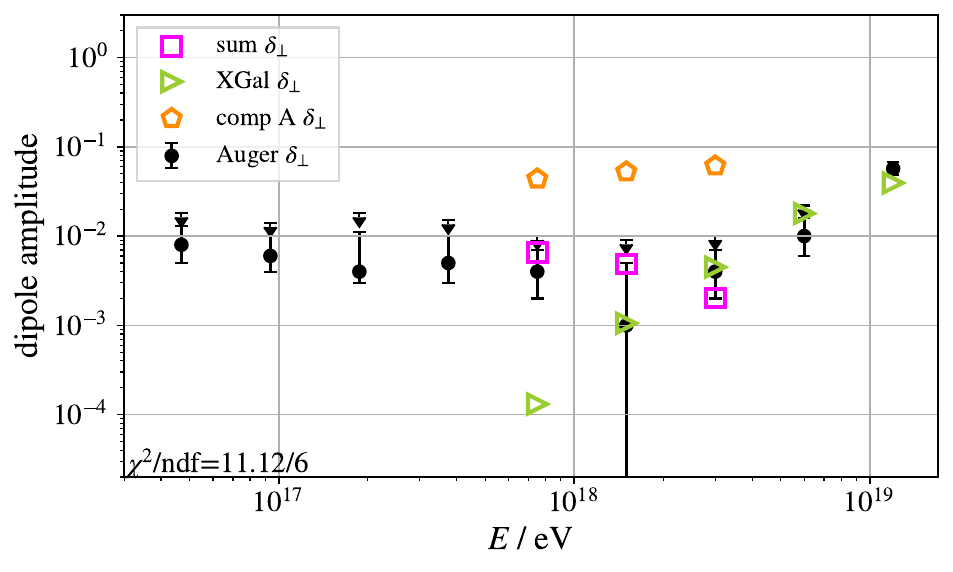}
\includegraphics[width=0.49\textwidth]{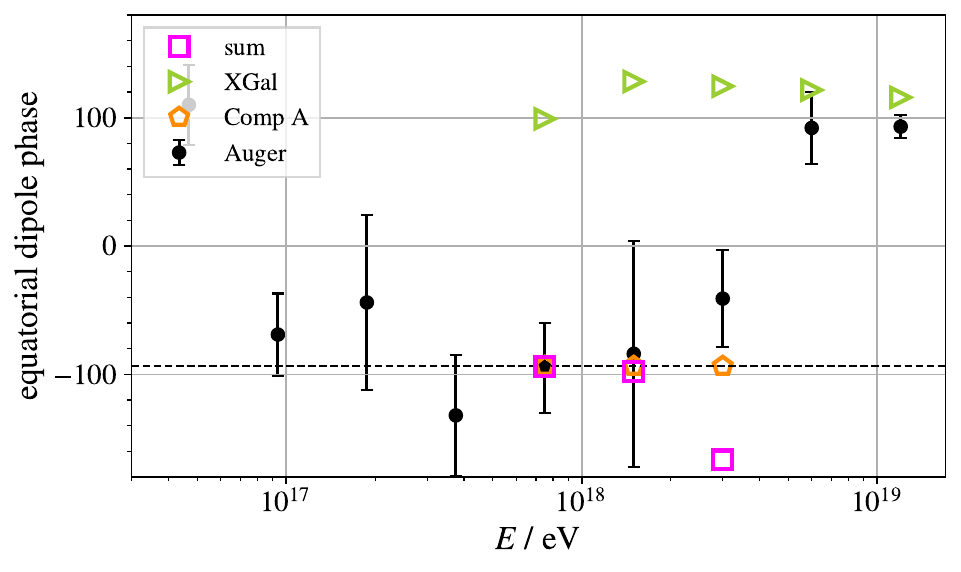}\\
\includegraphics[width=0.24\textwidth]{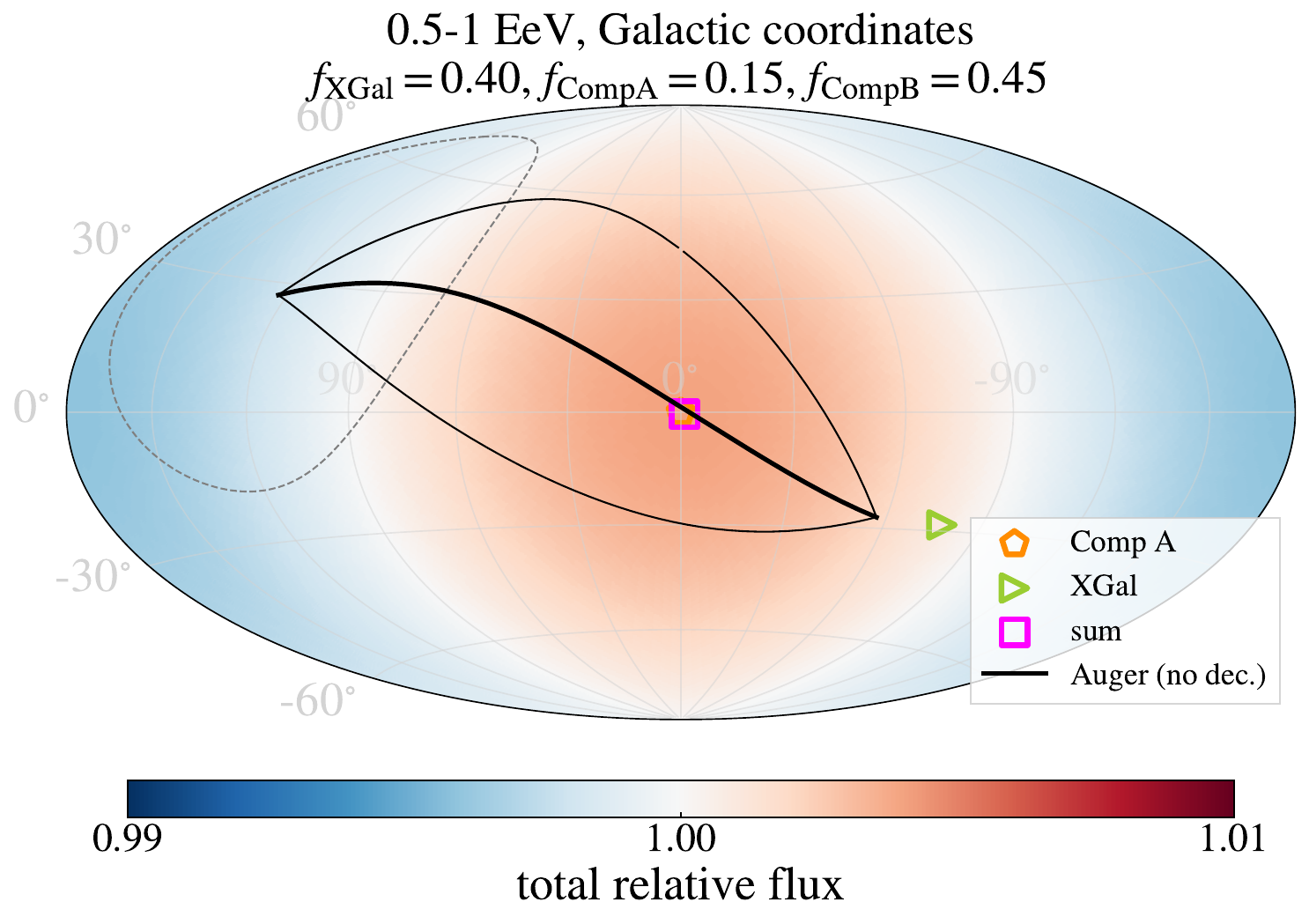}
\includegraphics[width=0.24\textwidth]{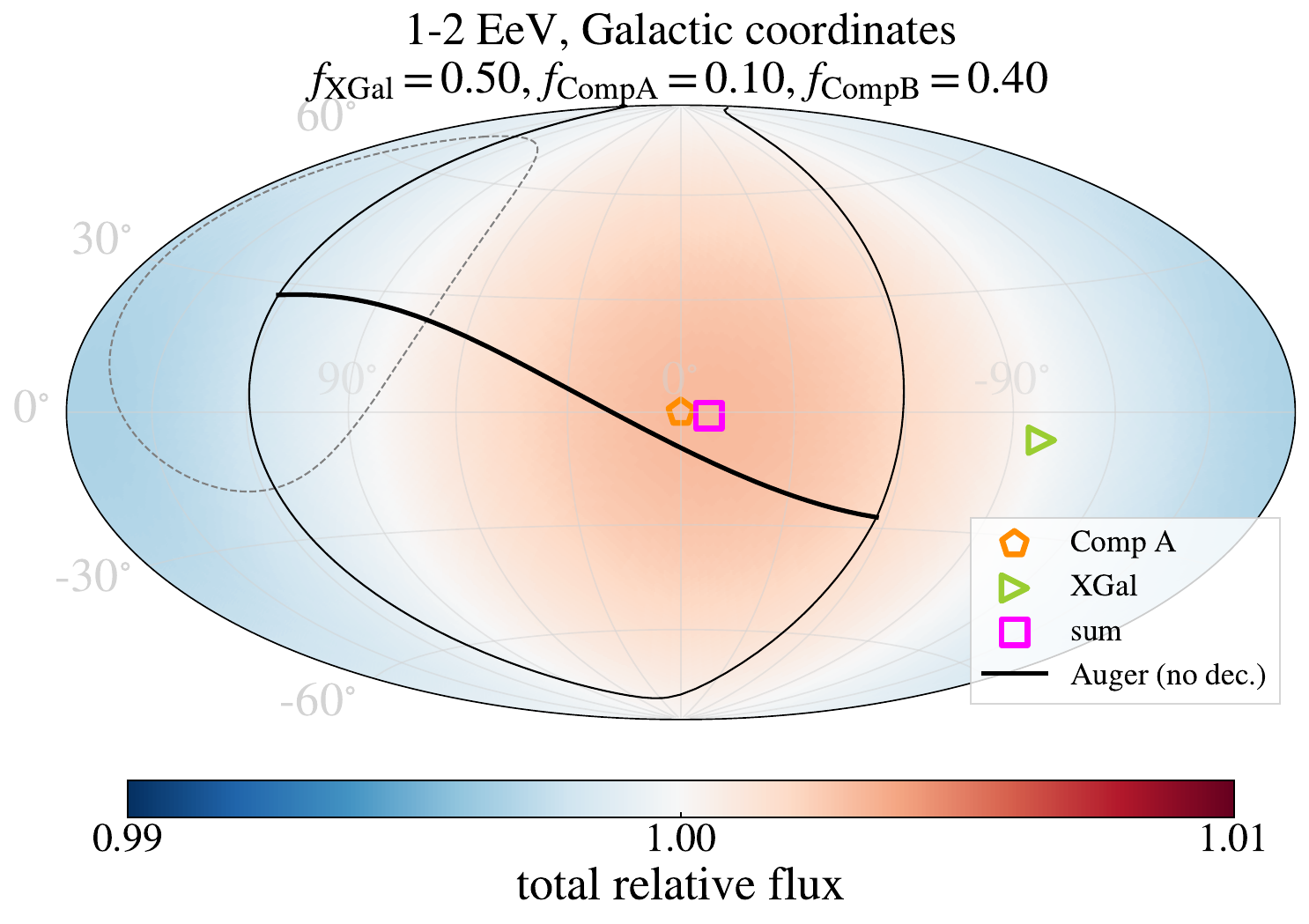}
\includegraphics[width=0.24\textwidth]{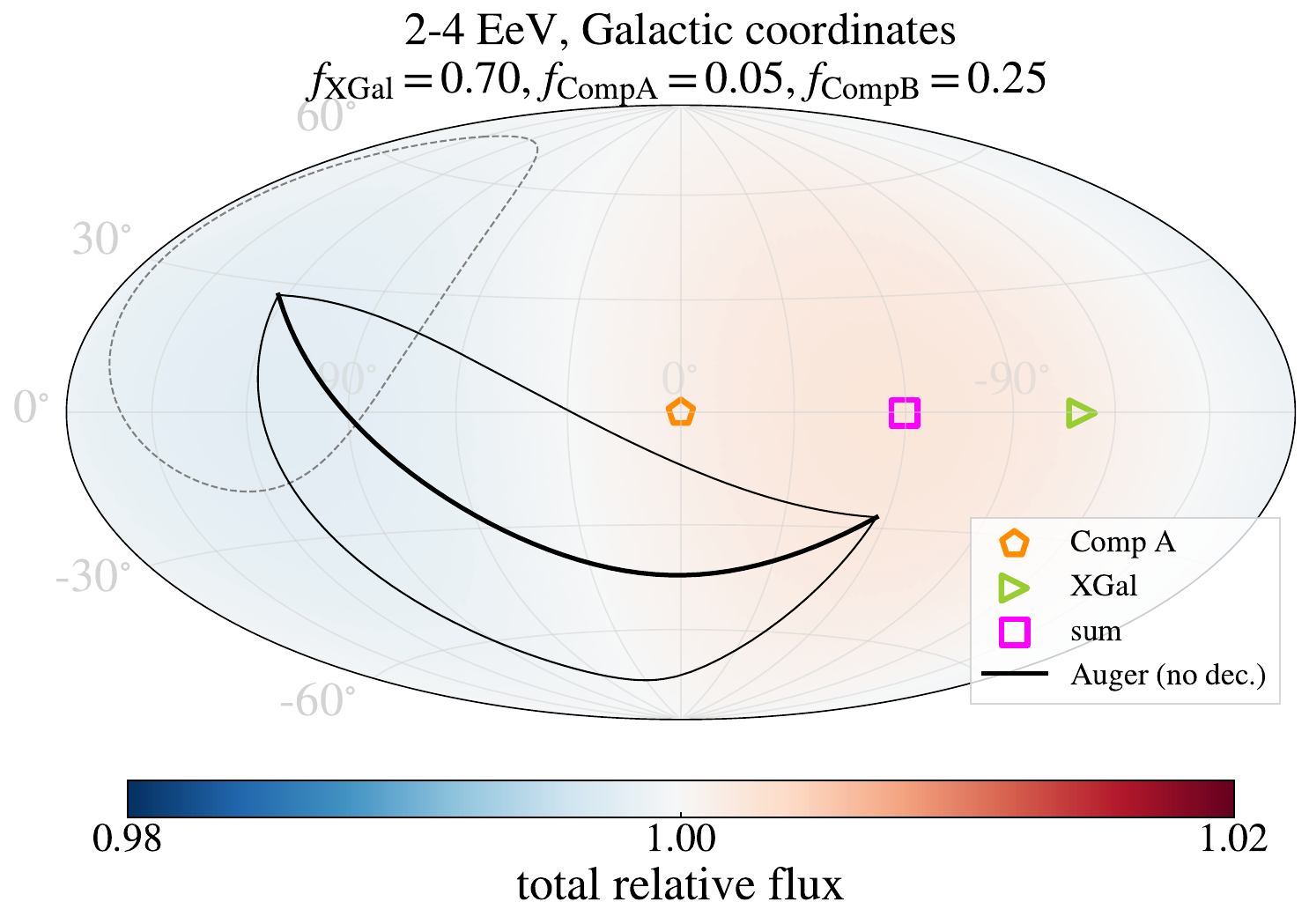}
\includegraphics[width=0.24\textwidth]{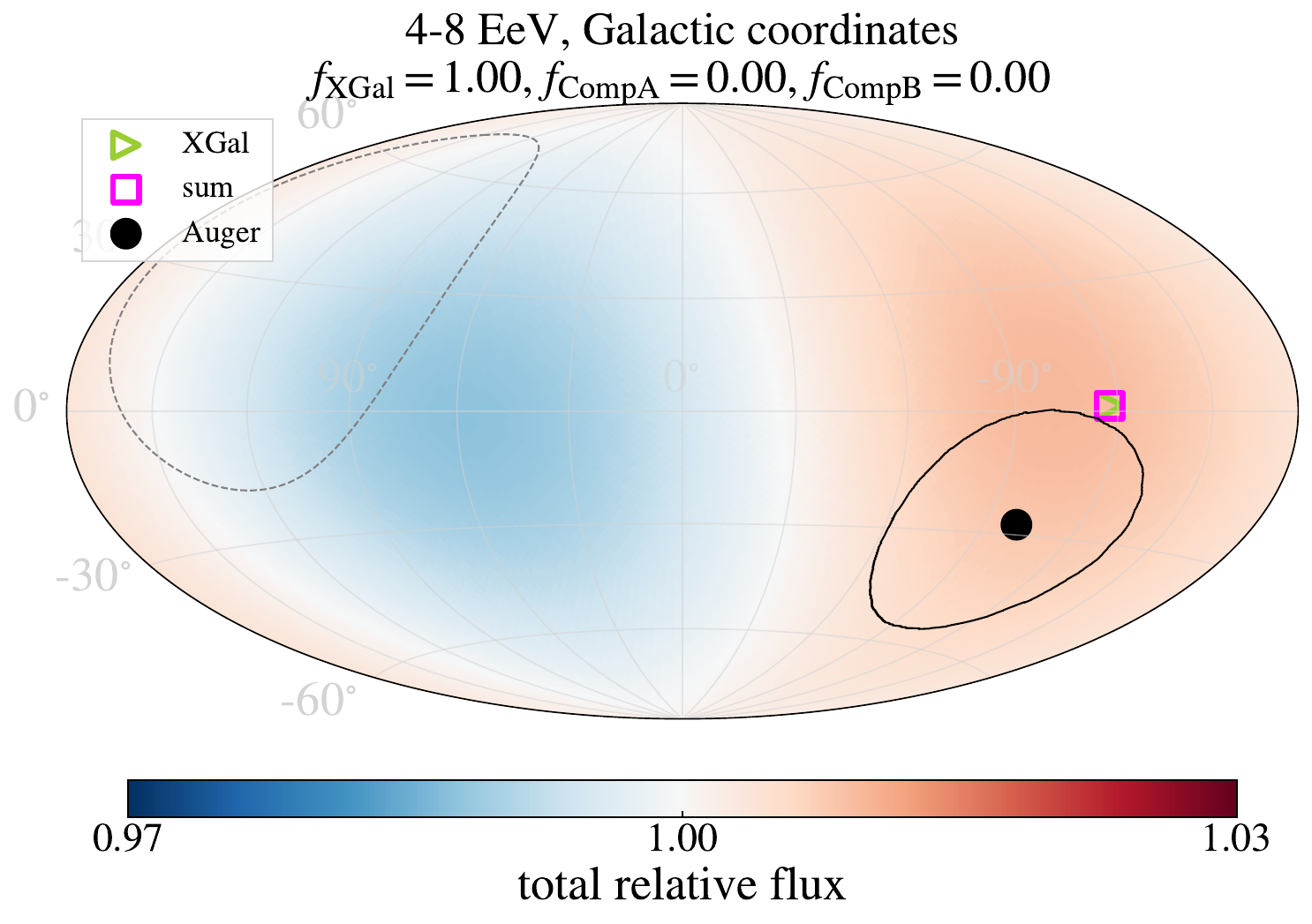}
\caption{Same as Fig.~\ref{fig:fit_sibyll}, but in the \textbf{iron-tail scenario} (subdominant Galactic Component A present between 0.5\,EeV and 4\,EeV pointing to the Galactic center), while Component B is isotropic.}
\label{fig:fit_epos}
\end{figure*}

\paragraph{Iron-tail scenario}
If there is still a subdominant contribution of iron from Galactic Component A sources in the energy range $>0.5\,$EeV (as is the case for the median mass fractions with EPOS-LHC-R, or the upper limits with Sibyll 2.3e), the data can be described well even if Component B is isotropic ($\chi^2$/ndf=$11.1/6\approx2$). In that case, the change of right ascension towards the Galactic center is driven by the subdominant tail of Component A, as is visible in Fig.~\ref{fig:fit_epos}. Note that no fine-tuning was performed - the equatorial dipole amplitude and phase data are well described taking the median mass composition fractions from EPOS-LHC-R, and the median dipole amplitude from the pulsar distribution using the \texttt{UF23} suite of GMF models, in combination with the \texttt{JF12+Planck} random field (see sec.~\ref{sec:crpropa}). 

Allowing Component B to be anisotropic does not significantly improve the fit quality for the case where a subdominant Galactic Component A is present up to 4 \,EeV ($\chi^2$/ndf=$1.8/1$). The preferred dipole amplitudes of Component B in that case are smaller, but still compatible with the ones found if Component A is not present.

\bigskip
In summary, a model where Component B is isotropic and extragalactic can describe both the measured dipole amplitude and its change in direction, provided that the subdominant Galactic iron component is not too small.
If there is no Fe-tail of Component A above 0.5\,EeV, Component B must be anisotropic to describe the change in right ascension dipole amplitude on percent level below 1\,EeV. In that case, Component B must be Galactic, with maximum amplitudes summarized in Table~\ref{tab:fit}. 

The preferred zone of the Component B dipole direction (Galactic longitudes $-45^\circ\lesssim l\lesssim 110^\circ$) includes several powerful Galactic $\gamma$-ray sources, like the microquasars Cygnus X-1 or Cygnus X-3, the Cygnus Cocoon, and Westerlund 1. The Galactic center direction is also compatible.
In the following section, we will calculate the dipole amplitudes and directions of different Galactic sources and source distributions using simulations. This will allow us to determine if there are any suitable source candidates that are compatible with the dipole directions and amplitudes required for Component B.

\section{Dipole from Galactic sources \label{sec:crpropa}}
In the following, we will evaluate the dipole anisotropy expected from Galactic CR sources using simulations, for a population of sources in sec.~\ref{sec:aniso_distr}, for a individual continuously emitting source candidates in sec.~\ref{sec:aniso_single}, and for transient sources in sec.~\ref{sec:aniso_transient}. Afterwards, we will validate the numerical results by comparing them to expectations from diffusion theory in sec.~\ref{sec:aniso_theo}.

To evaluate the anisotropy induced by Galactic sources, we perform forward simulations using CRPropa3.2~\citep{CRPropa}. While computationally expensive, forward propagation has the advantage that the number of particles reaching Earth can be recorded -- allowing for conclusions on the necessary source power -- and it allows to study the temporal variation of the flux.
First, we test two continuous distributions of sources, either following a homogeneous cylinder with height 200~pc and galactocentric radius 20~kpc, or the distribution of pulsars as provided by CRPropa and based on~\citet{Blasi:2012JCAP, FaucherGiguere:2006ApJ}. The former will allow us to compare to previous works based on simple homogeneous source distributions in the disk~\citep{PierreAuger:2013AnisotropyApJL, Giacinti:2011Transition, Kaapa:2021icrc, Kaapa:2022gmf}, and the latter is the one used for Component A as described above.
Additionally, we propagate CRs from individual source candidates, including the Galactic center (GC), Cygnus X-1 (distance 2.2\,kpc, $l=71.33^\circ$, $b=3.0^\circ$, and roughly compatible also with the location of the Cygnus Cocoon\footnote{Although we explicitly simulate only Cygnus X-1, our results on this source are a useful benchmark for any source in the Cygnus Cocoon region, whose centroid is $\sim 8^{\circ}$ away at a comparable, somewhat smaller distance of $d\sim1.4–1.8\,$kpc.}), Cygnus X-3 (distance 9.4\,kpc, $l=79.85^\circ$, $b=0.68^\circ$~\citep{Reid:2023ksq}), and Westerlund 1 (distance 4.03\,kpc, $l=339.^\circ$, $b=-0.4^\circ$~\citep{Rocha:2022}). We place several observers at Earth's position, inside of each other (with radii 50~pc, 100~pc, 200~pc, and 500~pc), to be able to ensure that the observer size has no spurious effect on the resulting anisotropy, and to be able to maximize statistics where it is necessary. An additional observer is placed at the Galactic boundary of galactocentric radius $r=20\,$kpc, which records when and where particles are leaving the Galaxy.

\begin{figure}[ht]
\includegraphics[width=0.49\textwidth]{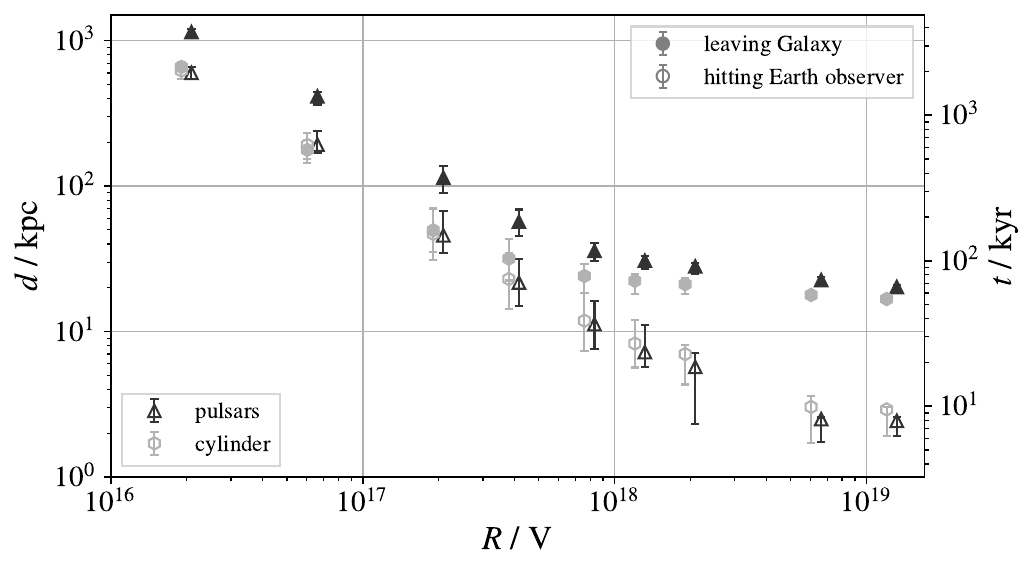}
\includegraphics[width=0.49\textwidth]{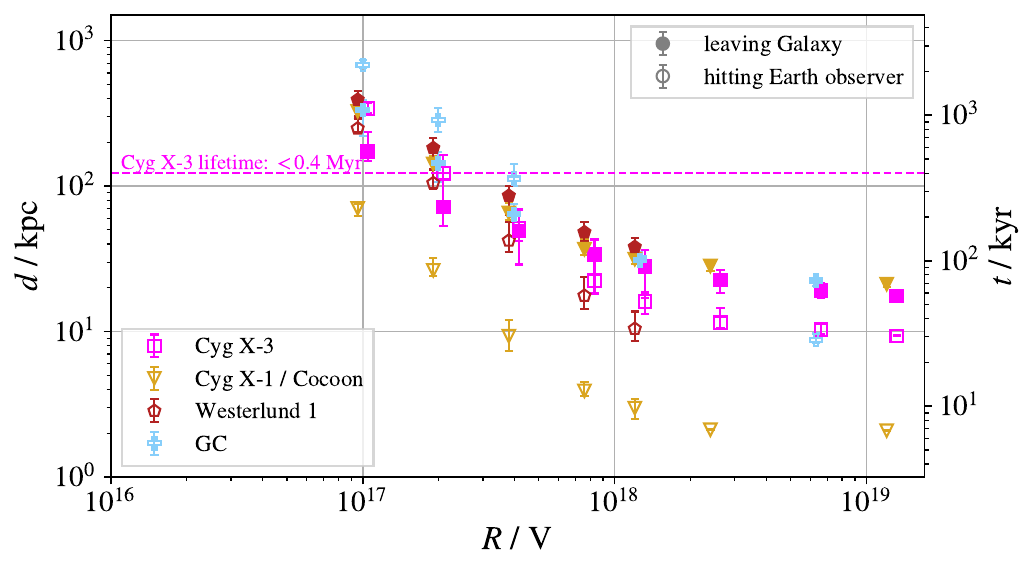}
\includegraphics[width=0.49\textwidth]{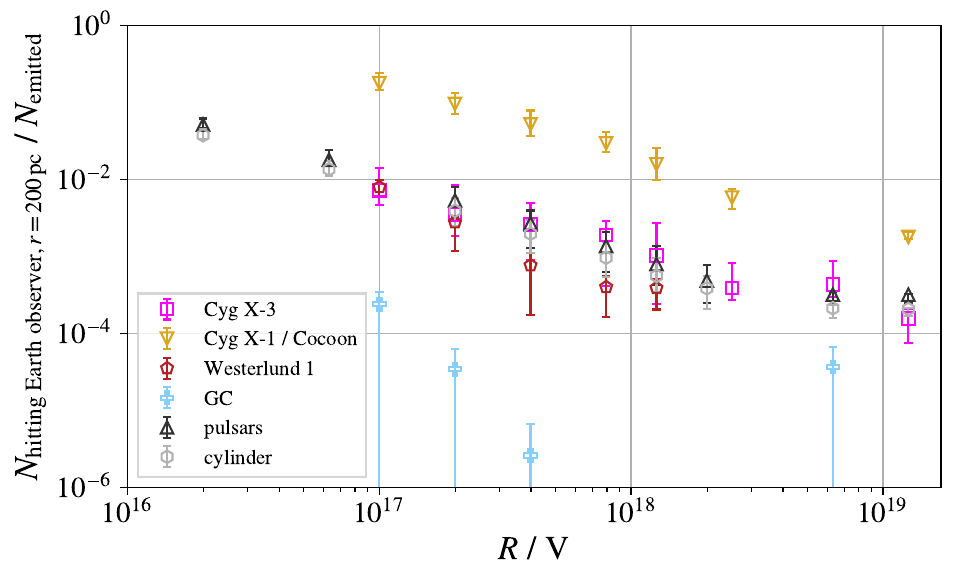}
\caption{\textit{Top and middle:} median distance $d$ and corresponding time $t$ that it takes CRs from different sources to either reach Earth (unfilled markers), or leave the Galaxy (filled markers). Markers are slightly offset on the x-axis for better visibility.
\textit{Bottom:} number of events hitting the observer at Earth with radius 200 pc divided by the emitted number for different sources. In both figures, the uncertainties represent the minimum and maximum over the 8 \texttt{UF23} GMF models and up to 3 random seeds for the random GMF model. Note that the rigidity of Component B (i.e.~of nitrogen) in the three energy bins between 0.5\,EeV and 4\,EeV used in this analysis is $R\simeq1\times 10^{17}$\,V, $R\simeq2\times10^{17}$\,V. and $R\simeq3.5\times10^{17}$\,V.}
\label{fig:time_N}
\end{figure}

We vary the rigidity of the propagated protons between $10^{16.8}\,\mathrm{eV}\simeq20\,\mathrm{PeV}$ and $10^{19.1}\,\mathrm{eV}\simeq12\,\mathrm{EeV}$, neglecting possible interactions. Results for other nuclei can then be calculated by scaling with the corresponding charge numbers. Neglecting interaction is justified because for component B, we consider nitrogen particles which traverse less than a few hundred kpc in the Milky Way before escaping (see Fig.~\ref{fig:time_N}).
The expected number of interactions per nucleus can be calculated via $N_{\rm int} = X/\lambda_{\rm int}$, with the traversed grammage $X$ and the interaction length $\lambda_\mathrm{int}$, which we calculated by rescaling the measured proton cross section $\sigma_{p\rm Air}=505$~mb at $\sqrt{s}=57$~TeV~\citep{PierreAuger:2012egl} to nitrogen via $\sigma_{\rm Np}=\sigma_{p\rm Air}(A_{\rm N}/A_{\rm Air})^{0.691}\approx490$~mb~\citep{Gaisser:2016uoy}. This gives a number of interactions of
$N_{\rm int} \approx 0.046 \cdot \left(n_{\rm ISM}/1~\mathrm{cm^{-3}}\right) \cdot \left(d/30~\rm kpc\right),$
confirming a negligible interaction probability over the relevant scales.

For the GMF, we use the suite of eight models from~\cite{Unger:2024uf23} (\texttt{UF23}), in combination with the Planck-tuned \texttt{JF12} random field~\cite{Jansson:2012b_Random, Planck:2016GMF} and with a coherence length of 56~pc. The models have a maximum field strength in the disk of $\sim6\mu$G, so that the simulated rigidities result in gyroradii $r_g \simeq 20 \,\mathrm{pc} \, (R / 0.1 \, \mathrm{EV}) \, (6\,\mu \mathrm{G} / B)$ between approximately 4\,pc and 2.5\,kpc, i.e.~covering the regime from diffusive to ballistic propagation in the Galaxy. In Fig.~\ref{fig:topview} in the appendix, examples of how CRs are propagating in the Galaxy are shown for different rigidities between the diffusive and ballistic regime.

The distance / time that CRs spend in the Galaxy decreases strongly with increasing rigidity as shown in Fig.~\ref{fig:time_N} (\textit{top}). It resembles a powerlaw below $R\lesssim3\,$EV. Above $\gtrsim3\,$EV, the residence time becomes flat, indicating fully ballistic propagation so that CRs are not confined by the Galaxy anymore.
There is a mild dependency on the source location - CRs from sources in the inner part of the disk like the GC or Westerlund 1 typically take longer to leave than from sources at larger galactocentric radii like Cygnus X-3.

The time it takes for CRs to reach Earth has a stronger dependency on the source position than the residence time, both because of the varying source distances to Earth as well as the structure of the coherent field. For CRs from the nearby Cygnus X-1 (which is also connected to Earth through the local magnetic field arm, see Fig.~\ref{fig:topview}), it takes around 200\,kyrs ($\sim$60\,kpc) at $R=0.1\,$EV to reach Earth, while for Cygnus X-3 it takes roughly three times longer.

For CRs from the GC it takes 2000 kyrs ($\sim$600\,kpc) to reach Earth at $R=0.1\,$EV, and no CRs with travel times smaller than $\sim800\,$kyrs are observed (note that that sets the relevant time where the CRs we see at Earth today must have been accelerated if they originate in the GC).
The reason for the large travel times for CRs from the Galactic Center is the coherent magnetic field structure. Instead of crossing multiple arms vertically (see Fig.~\ref{fig:topview}), CRs are much more likely to be transported up or down directly by the poloidal halo field (the X-field), and leave the Galaxy at the North / South pole -- as is recorded by the observer spanning the entire Galaxy with $r=20\,$kpc (see also~\citet{Cerri:2017} where this effect is discussed at lower rigidities). This effect also leads to an incredibly small fraction of events detected at the Earth observer from the GC, as shown in Fig.~\ref{fig:time_N} (\textit{bottom}).
When propagation becomes fully ballistic above $\gtrsim3\,$EV, however, CRs are able to cross the GMF between the GC and Earth and the number of particles detectable from the GC recovers~(see also, Fig.~\ref{fig:topview} in the Appendix). 

For all other tested sources, no such behavior is visible. Instead, a powerlaw-like decrease of CR flux reaching Earth is observed with increasing rigidity. This is expected because lower-rigidity particles travel larger distances in the plane due to diffusion (see above), and are hence more likely to hit Earth, see also discussion in~\citet{Tinyakov:2016}. In general, this effect can soften the spectra of Galactic sources, and increase the detection rate of heavy particles compared to light ones at the same energy, see also~\citet{Kaapa:2021icrc, Kaapa:2022gmf}.

\subsection{Anisotropy from a Galactic source distribution}  \label{sec:aniso_distr}
\begin{figure}[ht]
\includegraphics[width=0.49\textwidth]{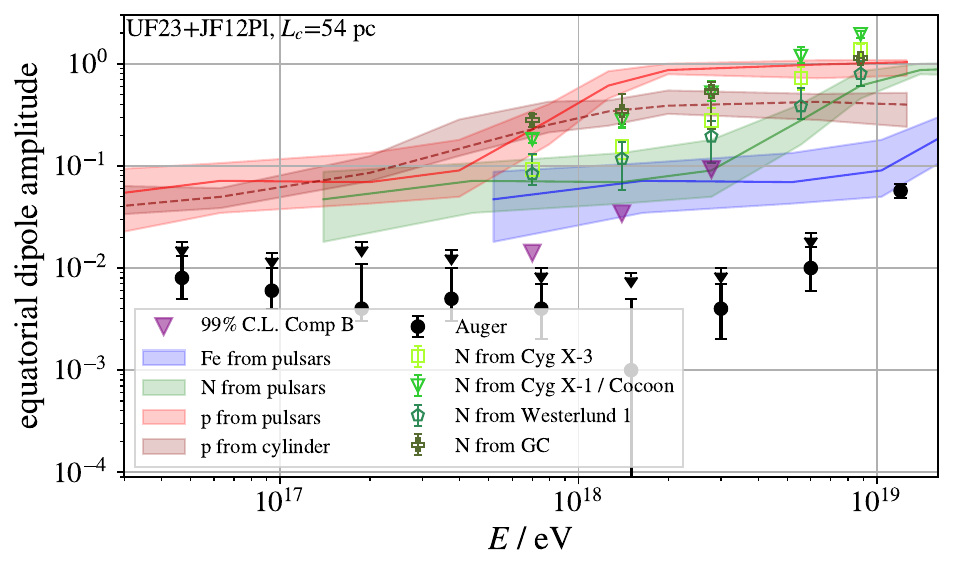}
\includegraphics[width=0.49\textwidth]{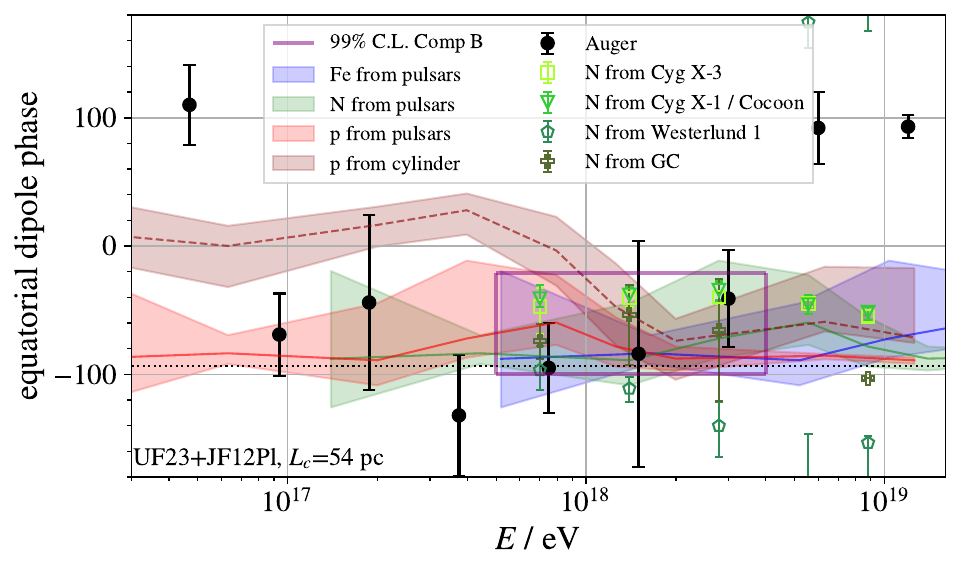}
\caption{Equatorial dipole amplitude (\textit{top}) and right ascension (\textit{bottom}) as a function of energy for multiple source candidates and mass composition assumptions. The shaded areas / error bars show the median and inner 90\% over the 8 \texttt{UF23} GMF models and up to 3 random seeds for the random GMF model. Black error bars show the Auger data from~\cite{PierreAuger:2024Anisotropy19Y}. The purple triangles (\textit{top}) and purple box (\textit{lower}) mark the 99\% C.L. limits on the dipole amplitude and right ascension of Component B (Table~\ref{tab:fit}) in the energy range used for the fit in Sec.~\ref{sec:fit}.}
\label{fig:amp_ra}
\end{figure}

The equatorial dipole moment and phase of the different sources are reported in Fig.~\ref{fig:amp_ra} as a function of energy. For the cylinder distribution, a good agreement of the dipole amplitude with the previous work by~\citet{Giacinti:2011Transition} (based on the older coherent GMF model \texttt{PT11} of~\citet{Pshirkov:2011}, and a simplified random field model) is observed: the dipole amplitude grows as a function of energy and reaches a maximum value of around $\sim40\%$. It never reaches dipole amplitudes larger than one even with ballistic propagation due to the homogeneous distribution of sources around Earth. Compared to \citet{PierreAuger:2013AnisotropyApJL}, the dipole amplitude is overall larger, but that is because \citet{PierreAuger:2013AnisotropyApJL} used a random magnetic field of three times the strength of the coherent field - which is not supported by the newer measurements by the Planck satellite~\citep{Planck:2016GMF}. The phase of the dipole from the homogeneous cylinder distribution shows an unexpected behavior: it does not point to the GC direction, especially at lower energies. This is due to the coherent magnetic field structure and connected to the effect described above -- CRs from close to the GC are deflected up-/ downwards and do not significantly contribute to the flux observed at Earth, unless propagation is ballistic. That way, sources in the outer Galaxy contribute more CRs than sources in the inner Galaxy. Due to the dipole direction of the cylinder distribution not pointing in the observed direction close to the GC, we are not using it to model Component A and instead rely on the pulsar distribution.

For the pulsar distribution, the dipole direction always points towards the Galactic center. This is due to the source density increasing with decreasing radius. Because of that, the pulsar distribution also leads to a larger dipole amplitude at higher rigidities than the homogeneous cylinder distribution. Overall, the dipole amplitudes of both the pulsar distribution and the cylinder distribution are not in agreement with the measured ones for any mass composition. Instead, we find that a homogeneous distribution of Galactic sources should not contribute more than 50\% of the flux above 0.5\,EeV to be compatible with the observed dipole amplitude at 99\% C.L. for the most favorable GMF model, even for an iron composition. Taking instead the median over all tested GMF models, only $\gtrsim10\%-15\%$ of the flux above 0.5\,EeV can be from Galactic iron sources at 99\% C.L.. That is in agreement with the measured fraction of iron~\citep{PierreAuger:2026MassComposition} as discussed above. This result improves on and supersedes previous works based on old composition and anisotropy data and GMF models~\citep{PierreAuger:2013AnisotropyApJL}, where an iron fraction of 100\% was still allowed above 1\,EeV.

Additionally, nitrogen from a continuous source distribution also overshoots the 99\% C.L. upper limit on the equatorial dipole amplitude of Component B in the (0.5-1)\,EeV energy bin and the (1-2)\,EeV bin as visible in the upper panel of Fig.~\ref{fig:amp_ra}. At 99\% C.L., less than 40\% of the Component B flux $\lesssim1\,$EeV can be from Galactic sources emitting intermediate elements in the most conservative case (to bring the lowest dipole amplitude from the pulsar distribution in agreement with the 99\% C.L. upper limit on the amplitude of Component B)\footnote{I.e., less than $f_\mathrm{Comp B}^{1-2\,\mathrm{EeV}} \times 40\% = 60\% \times 40\%\simeq25\%$ of the total CR flux $\lesssim1\,$EeV can be from any Galactic source or source population emitting intermediate elements or lighter at 99\% C.L. in order to not overshoot the measured dipole amplitude.}. Thus, a continuous Galactic source distribution cannot be the source of Component B.

\subsection{Anisotropy from a steady Galactic source}   \label{sec:aniso_single}
In addition to the source distributions, we tested different individual Galactic sources to evaluate if they could be responsible for Component B. For that, we selected a few promising PeVatron source candidates that lie in the part of the sky preferred by the fit to the anisotropy (Galactic longitudes $-45^\circ\lesssim l\lesssim 110^\circ$, Fig.~\ref{fig:fit_sibyll}).

The direction of the dipole of all the Galactic sources we tested is only slightly displaced by the GMF even for the smallest rigidities of interest in our study, as can be seen in Fig.~\ref{fig:amp_ra} (\textit{bottom}) and exemplarily for two sources in Fig.~\ref{fig:cygx1_westerlund_direc} in the Appendix. All coherent GMF models we used (all variations of the \texttt{UF23} suite and the \texttt{KST24} model) preferably deflect particles with rigidities between $\sim0.1$\,EV and $\sim1$\,EV originating in the Cygnus region or the GC towards the Galactic south, i.e.~to latitudes $-30^\circ \lesssim b \lesssim 0^\circ$ -- which brings them into good agreement with the preferred direction of the Component B dipole (Fig.~\ref{fig:fit_sibyll})\footnote{That CRs originating in the Galactic plane direction are preferably deflected towards the Galactic south is also visible for extragalactic CRs in Fig.~19 of \citet{Unger:2024uf23} (also for the \texttt{JF12} GMF model), consistently for almost all directions along the Galactic plane.}. The longitude of the dipole of Galactic sources is not much affected by the GMF for sources in the Cygnus region (see Fig.~\ref{fig:cygx1_westerlund_direc}). For Westerlund 1, however, coherent deflections shift the dipole further away from the Galactic center and even towards $b\gtrsim0^\circ$ (see Fig.~\ref{fig:cygx1_westerlund_direc}) between $R\simeq10^{17.3}\,$V and $R=10^{18.5}\,$V. This leads to a disagreement with the right ascension preferred for Component B in the (1-2)\,EeV and (2-4)\,EeV energy bin, as visible in Fig.~\ref{fig:amp_ra}.

The dipole amplitude of individual Galactic sources is larger than that of the source distributions because for source distributions, CRs are injected more isotropically around Earth leading to an overall more isotropic flux. The dipole amplitude of the point sources grows like a power-law with the energy. At a rigidity of $R\gtrsim3\,$EV, propagation is mostly ballistic as visible in Fig.~\ref{fig:time_N}. Hence, the resulting arrival direction distribution resembles a point source more than a dipolar distribution and the dipole amplitudes can become larger than unity. For these larger rigidities, the dipole direction stays very close to the source direction (see Fig.~\ref{fig:cygx1_westerlund_direc}).

The dipole moment of CRs from the Galactic center is the largest of all tested point sources below $R\simeq1\,$EV. This is related to the fact that CRs from the GC hardly diffuse to larger galactocentric radii below $1\,$EV, so that no diffuse flux of CRs from the outer Galaxy counteracts the dipole pointing to the source direction, unlike for sources at larger galactocentric radii (see Fig.~\ref{fig:topview}). 
Due to this, the dipole of CRs from the GC overshoots the 99\% C.L upper limits on the amplitude of Component B in all energy bins. This makes it extremely unlikely that Component B is due to emission from the GC based on arrival-direction arguments. Instead, at most a subdominant flux of $\lesssim0.01 / 0.31 \simeq 4\%$ (upper limit on measured dipole amplitude / amplitude of the GC dipole) of particles from the GC with $R\simeq0.1\,$EV would be in agreement with the measured anisotropy below 4\,EeV.

The dipole moment of CRs from the nearby Cygnus X-1 / Cygnus Cocoon region also overshoots the upper limits in all energy bins. The dipole amplitude of the further away Westerlund 1 and Cygnus X-3 is as expected lower, but is still in significant disagreement with the upper limits from the fit to the anisotropy for all energy bins below 2\,EeV.

An additional caveat for far away sources is that the source age has to be large enough in order for CRs to be able to reach Earth, see middle panel of Fig.~\ref{fig:time_N}. Below 1\,EeV, the UHECR propagation time exceeds Cygnus X-3's estimated lifetime of $\mathcal{O}(10^{5}~\rm yr)$~\citep{Lommen:2005tc}. The source is not old enough for nitrogen injected early in its active phase to have reached us at these energies. Cygnus X-3 therefore cannot be the source of the nitrogen observed in the $(0.5\text{--}1)$\,EeV bin, independent of the anisotropy constraint already discussed.

Even though we tested only a few promising source candidates, our conclusions are general. For the dipole amplitude to be low enough to be compatible with the limits on Component B, the source has to be very far away, further than Cygnus X-3 at $\sim10\,$kpc, and hence unusually powerful in order for enough UHECRs to reach Earth. However, as we will show in the following section~\ref{sec:energetics}, even Cygnus X-3, which is an unusually bright, one-per-galaxy source, is not powerful enough to overcome the energetic challenge of significantly contributing to the Component B flux. 

Furthermore, a source cannot lie in or behind the Galactic center since, in the diffusive regime, particles reaching this region are far more likely to be transported out of the Galaxy along the poloidal halo field (the X-field) than to continue diffusing across the disk to an observer on the far side. Combined with the distance requirement above, this leaves only sources very close to the edge of the Galaxy, but from there, CRs escape too quickly for a significant fraction to reach Earth (cf.\ Fig.~\ref{fig:time_N}).
Thus, no continuously emitting Galactic source or source population can generate a dipole that is small enough to be compatible with the maximum allowed amplitude of Component B.

\subsection{Anisotropy from a transient Galactic source}   \label{sec:aniso_transient}
Instead of a steady source, the anisotropy of Component B could have been generated by a past transient event. That idea was first investigated in~\citet{Farrar:ICRC2021}, and the supernova related to the nearby SNR G65.3+5.7 was identified as a possible source of Component B based on its direction, distance, and anisotropy under the assumption of isotropic diffusion. 
A transient source has the advantage that the dipole amplitude is expected to decrease with time, which may bring it into agreement with the upper limits on the amplitude of Component B at late times. 

The time evolution of the dipole amplitude $\delta$ of a burst from a single source in a fully turbulent magnetic field in the diffusive regime (neglecting particle escape) is given by~\citep{Shen:1971, Savchenko:2015}:
\begin{equation} \label{eq:transient}
    \delta = \frac{3\,d_s}{2\,c\,t},
\end{equation}
where $d_s$ is the source distance, $c$ the speed of light, and $t$ the time since the burst. Thus, a nearby transient that happened long ago would be expected to reach low enough amplitudes to be compatible with the upper limits on Component B.
Notably, eq.~\ref{eq:transient} does not depend on the diffusion coefficient, and thus also not on the rigidity of the CRs~\citep{Farrar:2000}. Note that we find that an increasing amplitude of Component B with the energy is necessary to describe the measured anisotropy (Fig.~\ref{fig:fit_sibyll}), which already hints that Component B is not generated by fully isotropically diffusing CRs from a past transient burst.

To test whether the isotropic-diffusion approximation without considering particle escape underlying eq.~\ref{eq:transient} provides an accurate description of nitrogen propagation in the energy range 0.5\,EeV to 4\,EeV relevant for our analysis, we conducted further simulations with $R=10^{17.3}\,$EV, using the same setup as for the continuous sources but now for a burst from the location of SNR G65.3+5.7 ($d_s=0.8$\,kpc, $l=65.3^\circ$, $b=5.7^\circ$). In addition to the GMF models used before, we employ a uniform turbulent magnetic field with $B_\mathrm{rms}=3\,\mu\mathrm{G}$ and $L_c=54\,$pc (similar values as the \texttt{JF12+Planck} field nearby Earth).

\begin{figure}[ht]
\includegraphics[width=0.49\textwidth]{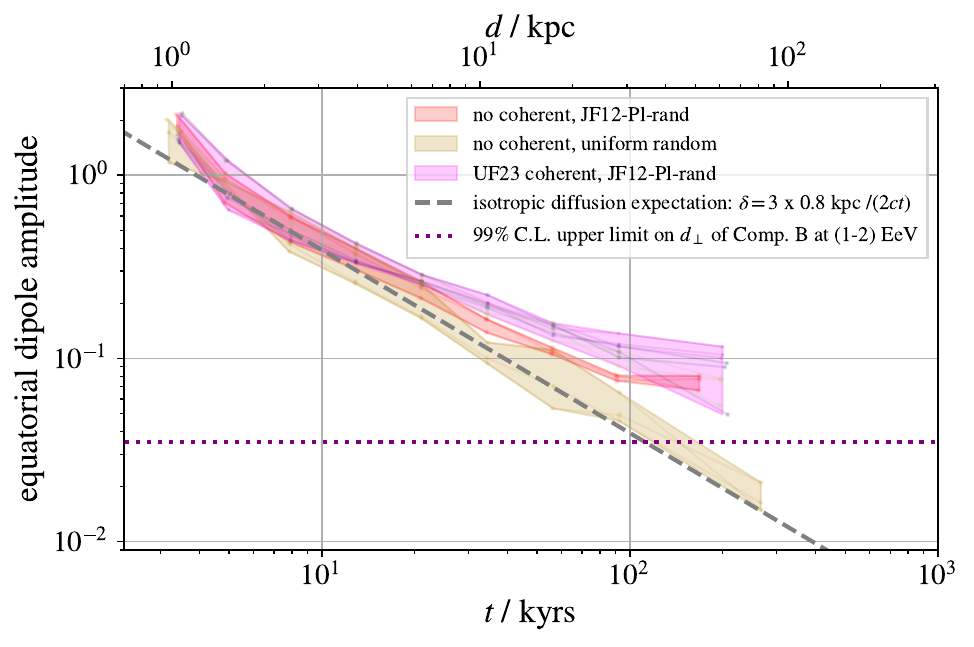}
\includegraphics[width=0.49\textwidth]{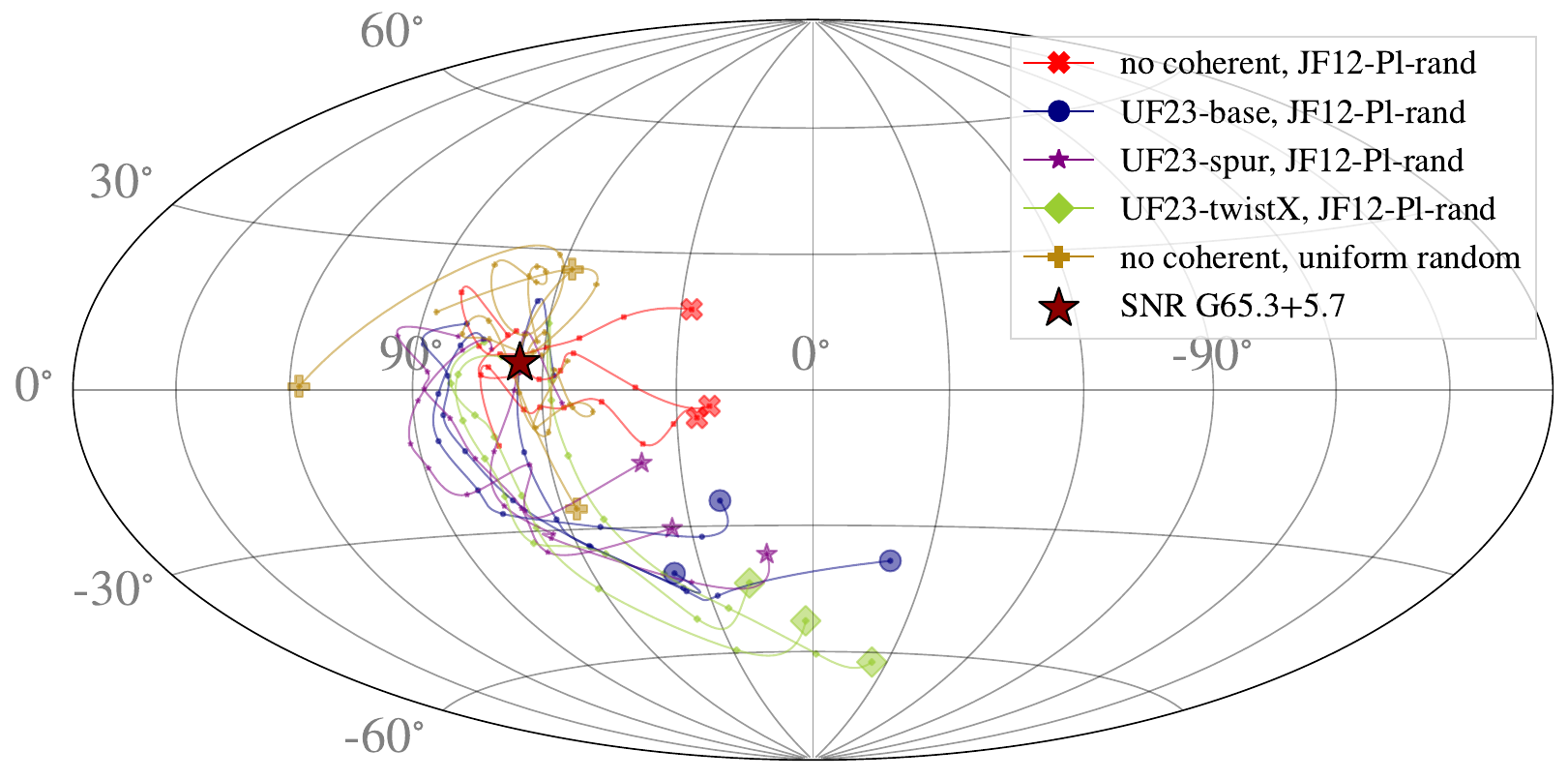}
\caption{Dipole amplitude and direction as a function of time $t$ since injection from SNR G65.3+5.7 for different GMF models and rigidity $R=10^{17.3}\,$EV. The data are binned into 9 logarithmic bins and one cumulative bin at late times containing $\lesssim10\%$ of all recorded CRs (denoted by a larger marker in the lower figure). The bins are centered on the median time/distance of the respective bin in the upper figure. We always use three random realizations of each random field, and the error bars in the upper figure show the minimum and maximum. For the coherent field, 3 coherent field models (\texttt{UF23-base}, \texttt{UF23-spur}, \texttt{UF23-twistX}) are used, and the error bars in the upper figure again show the minimum and maximum over all combinations with the turbulent field realizations.
}
\label{fig:transient}
\end{figure}

The dipole amplitude and direction at Earth as functions of time are shown in Fig.~\ref{fig:transient}. Only for the uniform turbulent field, the dipole amplitude follows the expectation from eq.~\ref{eq:transient} for all times after the initial ballistic phase. 
When using the \texttt{JF12+Planck} random field, the dipole amplitude starts deviating from eq.~\ref{eq:transient} at late times, and a residual dipole moment of $\mathcal{O}(10\%)$ persist until all CRs have left the Galaxy. The same effect is also visible when a coherent GMF is added, in addition to a slightly larger dipole at all times for that case.

The reason for this residual dipole at late times is the radial decay of the field strength of the turbulent GMF which makes it easier for CRs to leave the Galaxy at larger radii. Thus, CRs are only present in the inner Galaxy at later times in the diffusive regime, as is also visible in Fig.~\ref{fig:topview}. 
Due to this, the dipole direction always points towards the Galactic center at later times (plus some displacement towards the Galactic south if coherent fields are included, see above). Note that the effect is observed without fail also for other rigidities and source locations.

The residual dipole for a realistic turbulent field with radial decay is too large to be compatible with the upper limits on the equatorial dipole amplitude of Component B of 3.5\% at 1-2\,EeV (Table~\ref{tab:fit}), so that a past transient event can be excluded as the source of Component B.

\subsection{Parametric validation of the dipole anisotropy from Galactic sources}   \label{sec:aniso_theo}
While the numerical results robustly confirm that, even under the conservative assumption of a rather strong turbulent field such as \texttt{JF12+Planck}, the dipole from a Galactic source is too strong compared to the measurements, it is instructive to also develop an intuition for the anisotropies we find. 

The dipole depends rather strongly on the diffusion coefficient of CRs. The physics of CR diffusion is rather involved, especially in regards to the anisotropy of the process with respect to the direction of the regular field. On the other hand, our choice of relying on the Planck-tuned JF12 field somewhat simplifies our task. The random field therein is so strong that diffusion can be considered to a good approximation isotropic at $R\simeq10^{17.3}\,$eV. We have explicitly verified this by tracking the trajectories of CRs emitted from point sources at varying positions in the Galactic plane. In all such cases we find that all three coordinates shifts $\langle \Delta x^2\rangle$, $\langle \Delta y^2\rangle$, and $\langle \Delta z^2\rangle$ grow linearly with time with the same rate (although $\langle \Delta z^2\rangle$ stops growing after some time due to vertical escape).

For the root-mean-square random field at $z=0$, an approximate fit to the combined halo and disk random field from the model is $B_{\rm rand}\simeq 8\,\mathrm{\mu G}\, e^{-r/12\,\mathrm{kpc}}$ in terms of the galactocentric radius $r$. Taking as a reference rigidity $R=10^{17.3}\,\mathrm{V}$, the gyroradius is $r_g\simeq 27\,\mathrm{pc}/B_{8}$ with $B_8=B/8\,\mathrm{\mu G}$. By comparison with the correlation length $L_c \approx 50\,\mathrm{pc}$, we can estimate the diffusion coefficient e.g.~from the numerical simulations shown in Fig.~8 of \citet{Kuhlen:2022tov}. A good fit to the curve shown there for a purely random field is\begin{equation}\label{eq:diffusion_approximation}
    D\simeq \frac{cL_c}{3}\left[\left(\frac{r_g}{L_c}\right)^{0.6}+3\left(\frac{r_g}{L_c}\right)^{2}\right].
\end{equation}
At a radius $r=8\,\mathrm{kpc}$ (which corresponds roughly to the galactocentric radius of Earth and Cygnus X-1), CRs are just at the transition to the small-angle diffusion regime, and we find $D\simeq 6\times 10^{30}\,\mathrm{cm}^2/\mathrm{s}$. This estimated diffusion coefficient is completely consistent with the escape timescales we retrieve from our tracking simulations in the central panel of Fig.~\ref{fig:time_N}: a rough value for the height over which the field decreases is $h\sim 3\,\mathrm{kpc}$ and the typical timescale for CRs to leave the Galaxy is $t_{\rm esc}\simeq h^2/2D$, corresponding to a traversed distance of $d_{\rm esc}\simeq c h^2/2D\simeq 70\,\mathrm{kpc}$. Similarly, the typical distances traversed from the source before reaching Earth is also consistent.  Taking as an example Cygnus X-1, the closest source to Earth at a distance of $d_{\rm Cyg-X1}=2.2\,\mathrm{kpc}$, a dimensional estimate of the expected traversed distance, based on the standard diffusion equation, is $d\simeq cd_{\rm Cyg\,X-1}^2/4D\simeq 40\,\mathrm{kpc}$, roughly consistent with the corresponding marker in Fig.~\ref{fig:time_N}.

Having validated our diffusion approach against the propagation physics of CRs deduced from our CRPropa simulations, we now turn to the anisotropy. For a steady CR population in the fully diffusive regime, the dipole amplitude expected at Earth is
\begin{equation}
    \delta=\frac{3 D |\boldsymbol{\nabla} n|}{nc},
\end{equation}
where $n$ is the spatial distribution of CRs. If all CRs come from a single, stationary point source at a distance $d_s$, this gives the simple result
\begin{equation}
    \delta=\frac{3D}{d_s \, c}.
\end{equation}
For the case of Cyg X-1, we find $\delta \simeq 0.09$, lower by roughly a factor 3 than the dipole amplitude obtained through the CRPropa simulations. 
Of course this formula is only very approximate, since it neglects the inhomogeneous diffusion coefficient sampled by CRs across the Galaxy, as well as the effect of the subdominant coherent field which can enhance the dipole amplitude through coherent deflections. We tend to attribute the discrepancy to these factors, especially since the dipole obtained for a purely homogeneous random field in Fig.~\ref{fig:transient} matches the diffusive prediction well.

If a population of CR sources is assumed, e.g.~in the pulsar or cylinder model, a somewhat lower anisotropy is found, as already commented. Nevertheless, our dimensional estimates make it immediately clear why we still obtain a relatively large dipole amplitude. Even with a Galactic population, the typical length scale over which CRs are inhomogeneously distributed is still comparable with a few kpc. With the diffusion coefficient inferred above, we still recover an anisotropy at the level of a few percent from these simple dimensional estimates; the tracking simulations, including coherent deflections and inhomogeneous diffusion coefficients, reinforce this conclusion, showing that the dipole is in fact even slightly larger.

We finally turn to the case of an old transient source, for which, as we have observed, the dipole amplitude saturates to a constant value at late times. Within the diffusion treatment, this effect is easy to understand. While the diffusion tensor is isotropic, its spatial variation induced by the random field causes the CR spatial distribution from a transient burst to not simply expand, but rather approach a constant shape, determined by the longest-surviving eigenmode of the diffusion equation. Physically, as CRs escape faster at outer galactocentric radii, their distribution becomes more peaked towards the GC. One may determine this spatial distribution explicitly by considering the least suppressed eigenmode of the diffusion operator
\begin{equation}
    \boldsymbol{\nabla}\cdot(D\boldsymbol{\nabla}n)=-\frac{n}{\tau_0},
\end{equation}
i.e.~the eigenmode with the longest $\tau_0$. We do not need to go through a complex procedure, since the dipole amplitude can be easily estimated. The inhomogeneity length scale induced by diffusion can be estimated directly by combining Eq.~\ref{eq:diffusion_approximation} with our parameterized random field amplitude, obtaining
\begin{equation}
    \ell_{\rm inh}\simeq\frac{D}{|\boldsymbol{\nabla}D|}\simeq 7.5\,\mathrm{kpc}.
\end{equation}
Due to the inhomogeneous diffusion coefficient, CRs will inherit an inhomogeneity on a comparable length scale, so that their dipole amplitude even for very old sources saturate to a typical value
\begin{equation}
    \delta_{\rm res}\simeq \frac{3 D}{c\ell_{\rm inh}}\simeq 0.03.
\end{equation}
Once again, the tracking simulations retrieve a value slightly larger, by a factor $\sim2-3$, compared to this simplified dimensional estimate. A similar effect, although in a different energy regime and with a different spatial dependence of the diffusion coefficient, was also noted in~\cite{Evoli:2012ha}.

\section{Energetic constraints on Galactic sources of Component B} \label{sec:energetics}
Whether a particular Galactic source can account for a sizable fraction of Component B is not just a question of the induced anisotropy -- which depends on details of the GMF model -- but also a question of energetics. In what follows we show that the energetic requirements provide an additional, independent constraint on the possible contribution of powerful Galactic accelerators to the subankle nitrogen. 

\paragraph{Cygnus X-3}
If Cygnus X-3 supplies a sizable fraction, $f_{\rm compB}$, of the CR flux in the energy range of interest, 0.5-4 EeV, with $f_{\rm compB} = 0.6, 0.3$, and 0.2, in the $E = $ 0.5-1, 1-2, 2-4 EeV bins respectively, it must be able to provide $J(E) \cdot f_{\rm CompB} \approx 5.8 \times 10^{-17}\cdot f_{\rm CompB}\cdot~\rm eV^{-1}~km^{-2}~sr^{-1}~yr^{-1}$, where, the quoted differential flux is the total diffuse flux measured by Auger at $E = 10^{18}$~eV~\citep{PierreAuger:2021hun}\footnote{The corresponding integral fluxes $\int E\,J(E)\,dE$ are $6.5\times10^{19}$, $2.6\times10^{19}$, and $1.0\times10^{19}\,\rm eV\,km^{-2}\,sr^{-1}\,yr^{-1}$ in the $(0.5\text{--}1)$, $(1\text{--}2)$, and $(2\text{--}4)$~EeV bins, respectively, so that $L_{\rm bin} \approx (4\pi^2 r_{\rm obs}^2/f_{\rm hit})\,f_{\rm CompB}\int E\,J(E)\,dE$.}.

We can translate this flux requirement into a source luminosity, from the hit rates $f_{\rm hit}$ recorded in the CRPropa3 simulations and shown in Fig.~\ref{fig:time_N}.
For a roughly isotropic population of particles with density $n$, the rate at which particles cross into the sphere of radius $r_{\rm obs}$ is $\dot{N}_{\rm hit} (E) = \pi r_{\rm obs}^2 \cdot c \cdot n(E)$ (the flux of ingoing particles has a geometric factor $1/4$), and we obtain the number density at energy $E$ from the measured differential CR flux, $n(E) = 4\pi \cdot f_{\rm compB} \cdot J(E)/c$. 

The corresponding luminosity, $L \approx E^2 Q(E)$, where $Q(E) = \dot{N}_{\rm hit} (E) \cdot E / f_{\rm hit}$ is the particle injection rate, is 
\begin{equation}
L(E) \approx \frac{4\pi^2 r_{\rm obs}^2}{f_{\rm hit}(E)}
E^2J(E)\, f_{\rm compB}(E).
\end{equation}
Summing the contribution of the UHECR nitrogen in the three energy bins, with their respective values of $f_{\rm hit}$ and $f_{\rm compB}$, gives
$L_{N}^{\rm 0.5-4 EeV} \approx (4 \times10^{35} + 2 \times10^{35} + 10^{35})\, \mathrm{erg\,s^{-1}} \sim 7 \times 10^{35}\ \mathrm{erg\,s^{-1}}.$
Assuming an ${\rm d}N/{\rm d}E \propto E^{-2}$ spectrum at acceleration, the total nitrogen luminosity of the source would be $L_{\rm Tot,N} = L_N^{0.5-4\,\rm EeV} \cdot \ln(E_N^{\rm max}/E_N^{\rm min})/(\ln(4\,{\rm EeV}/0.5\,{\rm EeV})) \approx 10^{37}~{\rm erg\,s^{-1}}$, where $E_N^{\rm max} = 4$~EeV conservatively and $E_N^{\rm min} = m_p c^2$. 

The surface composition of the donor star in Cyg X-3, predicted by Wolf-Rayet atmosphere models, corresponds to a number ratio between nitrogen and helium $N_{\rm He}/N_N = (X_{\rm He}/A_{\rm He})/(X_{\rm N}/A_{\rm N}) = (0.98\times 14)/(0.02 \times 4) = 171.5$~\citep{2015A&A...579A..75T}. Such an amount of helium, if accelerated with the same efficiency as nitrogen, would exceed the total diffuse CR flux by a factor of 6, 3, and 2, at energy 0.15-0.3, 0.3-0.6, and 0.6-1 EeV respectively.    

The corresponding UHECR helium luminosity of the source would be $L_{\rm He}^{\rm 0.15-0.6 EeV} = (Z_{\rm He}/Z_{\rm N})(N_{\rm He}/N_{\rm N})\,L_{N}^{\rm 0.5-4 EeV} \approx 49\,L_{N}^{\rm 0.5-4 EeV}$. The corresponding total helium luminosity is $L_{\rm He} = L_{\rm He}^{\rm 0.15-0.6 EeV} \cdot \ln(E_{\rm max}/E_{\rm min})/\ln(1.5\,{\rm EeV}/0.15\,{\rm EeV})
= 4.2\times10^{38}~{\rm erg\,s^{-1}}$.

This luminosity is comparable to the upper limit on the kinetic power of the jet/outflow $\sim 10^{39}~\rm erg~s^{-1}$~\citep{2013MNRAS.429L.104Z,2021A&A...649A.176V} and over five orders of magnitude higher than the observed UHE $\gamma$-ray luminosity of $L_{\gamma,\rm UHE} = 10^{33}~\rm erg~s^{-1}$ measured by LHAASO, making it unlikely that Cyg X-3 is the true source of a Galactic Component B on energetic grounds. The presence of some hydrogen in the wind of the WR donor is not ruled out~\citep{Koljonen:2017gvy}, but would exacerbate the energetics problem.

\paragraph{Galactic Center}
The Galactic Center has been proposed as the locus of extreme accelerators in our Galaxy, motivated, in part, by the observation of very-high-energy gamma-ray emission, suggesting particle acceleration to at least PeV energies~\citep{HESS:2016pst,Albert:2024aaa}. The radial profile of the observed gamma-ray emission, follows an $\sim1/r$ decrease in flux, suggesting continuous acceleration of hadrons to PeV energy over at least $\Delta t \gtrsim t_{\rm diff} \sim 2000 (D/10^{30}{\rm cm^2 \, s^{-1}})$ yr~\citep{HESS:2016pst}. 

As for Cygnus X-3, a Galactic Center source supplying the same flux fraction $f_{\rm compB}$ in the three relevant energy bins must be able to provide the same flux $J(E) \cdot f_{\rm CompB}$ (see above).
In this case, the hit rates $f_{\rm Hit}$ are much smaller than for Cygnus X-3
as shown in the bottom panel of Fig.~\ref{fig:time_N} and we obtain a corresponding required nitrogen luminosity 
$L_{N}^{\rm 0.5-4 EeV} \approx 1.2 \times10^{37} + 2 \times10^{37} + 6 \times 10^{37} \sim 10^{38}\ \mathrm{erg\,s^{-1}}$.
Assuming again an ${\rm d}N/{\rm d}E \propto E^{-2}$ spectrum at acceleration, the total nitrogen luminosity of the source would be $L_{\rm Tot,N}\sim 10^{39}\mathrm{erg\,s^{-1}}$.

In the Galactic Center, Nuclear Star Cluster metallicities span roughly solar to a few times solar, $\rho_{\rm metals}\approx1\text{--}2\,\rho_{\rm metals,\odot}$, where $\rho_i$ denotes the mass fraction of species $i$~\citep{2017MNRAS.464..194F}. To obtain an estimate of the required source luminosity, we adopt the mean metallicity of stars in the nuclear star cluster in the sample of~\citet{2015ApJ...809..143D}, $\rho_{\rm metals}=2.5 \rho_{\rm metals,\odot}$. Scaling the adopted solar composition, where $\rho_{\rm H}$ = 0.7381, $\rho_{\rm He
} = 0.2485$,
and $\rho_{\rm metals} = 0.0134$ are the dimensionless mass fractions of hydrogen, helium, and elements heavier than helium respectively, satisfying $\rho_H+\rho_\mathrm{He}+\rho_\mathrm{metals}=1$~\citep{2009ARA&A..47..481A}, to $\rho_{\rm metals} = 2.5\rho_{\rm metals,\odot}$ at fixed $\rho_{\rm He}/\rho_{\rm H}$ ratio gives $\rho_{\rm H} = 0.7231, \rho_{\rm He} = 0.2434, \rho_{\rm metals} = 0.0335$. 
Splitting the metal budget into a CNO+Ne group and an iron group using the solar relative abundance pattern ($f_{\rm CNO+Ne}=0.75$, $f_{\rm Fe\text{-}group}=0.25$ of $\rho_{\rm metals}$, following Table 1 of~\citealp{2009ARA&A..47..481A}) and converting to number fractions via $N_i\propto \rho_i/A_i$, gives $N_H:N_{\rm He}:N_{\rm CNO}:N_{\rm Fe} \approx 4864:409:12:1$. Summing the four species gives a required CR luminosity $L_{\rm UHECR}^{\rm 0.07-0.6 EV} = 6 \times 10^{39}~\rm erg~s^{-1}$, and extrapolating each species down to $E_{\rm min} \approx m_p c^2 \approx 1$~GeV assuming an ${\rm d}N/{\rm d}E \propto E^{-2}$ spectrum gives a total CR luminosity $L_\mathrm{CR} = 6\times10^{40}~\rm erg~s^{-1}$. Under this assumed source composition, the inferred H+He (Fe) flux exceeds the observed diffuse UHECR flux in the 0.15--0.3 (1.9--3.7) EeV energy bin by a factor of 2.66 (1.1). The flux excess argument is independent of the assumed GMF model, since it depends only on the assumed composition and $f_{\rm compB}$, not on the propagation efficiency $f_{\rm hit}$.

The required CR luminosity is four orders of magnitude higher than the present-day bolometric luminosity of Sgr A$^*$, $L_{\rm bol}=6.8\text{--}9.2\times10^{35}~\rm erg~s^{-1}$~\citep{EventHorizonTelescope:2022urf}. However, it is more relevant to compare the past activity of the source to the required CR luminosity, because, as shown in Fig~\ref{fig:time_N}, activity $\sim 10^5-10^6$~yr ago could have produced the Component B CRs observed today. 
The inferred $L_{\rm CR}=6\times10^{40}~\rm erg~s^{-1}$ is comparable to upper limit of $\sim10^{41}~\rm erg~s^{-1}$ on the mechanical power of the Galactic Center's past activity thought to have produced the Fermi Bubbles~\citep{2017MNRAS.467.3544S}. This would require the shock to convert $\gtrsim60\%$ of its mechanical power into CRs, well above the $\sim10\%$ efficiency typically invoked for diffusive shock acceleration. We conclude that a Galactic Center origin of Component B is disfavored not only by the dipole anisotropy and diffuse flux-overshoot constraints but also due to the severe energetic requirements. 

\paragraph{Cygnus X-1}
Among the PeVatron microquasars detected by LHAASO~\citep{LHAASO:2024psv}, Cygnus X-1 has the advantage of being the nearest, and of possibly being magnetically connected to Earth by lying along the same coherent Galactic disk field line (see discussion in~\citealp{Zhang:2026igt} and Fig.~\ref{fig:topview}). Indeed, the fraction of UHECRs that hit us from this source as shown in Fig.~\ref{fig:time_N} is $\sim20$--$30\times$ larger than in the case of Cygnus X-3. 
Following the same procedure as above, we obtain 
$L_N^{\rm 0.5-4\,EeV}\approx1.7\times10^{34}+6\times10^{33}+3.2\times10^{33}\sim2.6\times10^{34}~\rm erg\,s^{-1}$ for Cygnus X-1. Assuming a ${\rm d}N/{\rm d}E\propto E^{-2}$ spectrum at acceleration, the total nitrogen luminosity of the source would be $L_{\rm Tot,N}\approx2.7\times10^{35}~\rm erg\,s^{-1}$.

As the donor of Cygnus X-1 (HDE~226868, O9.7~Iab) shows no evidence of Wolf-Rayet composition, we
adopt a solar composition $\rho_{\rm metals}=1\,\rho_{\rm metals,\odot}$. This gives $\rho_{\rm H}=0.7381,\rho_{\rm He}=0.2485,\rho_{\rm metals}=0.0134$~\citep{2009ARA&A..47..481A}, and splitting into a CNO+Ne group and an iron group as above
gives $N_H:N_{\rm He}:N_{\rm CNO}:N_{\rm Fe}\approx12413:1045:12:1$. 

Summing the four species and extrapolating as above gives a required UHECR luminosity $L_{\rm UHECR}^{R=0.07-0.6\,\rm EV}=4.4\times10^{36}~\rm erg\,s^{-1}$, and $L_{\rm CR}=4.3\times10^{37}~\rm erg\,s^{-1}$. Such CR luminosity would be 3-10 times higher than the time-averaged power of the jet of Cygnus X-1 $L_{\rm jet}= 4-14 \times 10^{36}~\rm erg~s^{-1}$~\citep{Russell:2007ig}. Furthermore, the inferred H+He flux exceeds the observed UHECR flux by factors of 6.8 and 3.7 in the 0.14 -- 0.29 and 0.29 -- 0.57~EeV energy bins respectively, and the inferred Fe flux exceeds the observed flux by a factor of 1.1 in the 1.9 -- 3.7~EeV bin. We thus conclude that Cygnus X-1 cannot be the source of the Component B CRs.

\paragraph{Cygnus Cocoon}
Although our dipole simulations use the position and distance of Cygnus X-1, they can be regarded more broadly as representative of a source in the Cygnus region since sources within the Cygnus region cannot be significantly distinguished  at the angular resolution meaningful for our analysis considering GMF deflections. Here we focus specifically on Cygnus OB2, the massive stellar association embedded within it, as a possible particle accelerator powering the Cocoon. 
The stellar wind luminosity of Cygnus OB2 has been estimated as $L_w = 1.5\text{--}3 \times 10^{38}~\rm erg~ s^{-1}$~\citep{Menchiari:2024tqw,Vieu:2024qjx}. Therefore, unlike Cygnus X-1, the available mechanical wind luminosity of Cygnus OB2 comfortably
exceeds the required UHECR luminosity of $4.3\times10^{37}~\rm erg\,s^{-1}$. However, as with Cygnus X-1 before,
assuming the same source metalicity and UHECR composition, the necessary accompanying
H+He and Fe flux would overshoot the total diffuse UHECR flux at lower energies by the same amount as above. Finally, dedicated theoretical studies of particle acceleration in Cygnus OB2 estimate a proton cutoff energy of $\lesssim1$~PeV~\citep{Blasi:2023quw,Haerer:2025ull,Li:2026kxd}. Even allowing for the maximum energy 
boost afforded by the heaviest nuclei considered here (Fe, $Z=26$), this corresponds to a
cutoff energy of $\lesssim26$~PeV, more than an order of magnitude below the sub-ankle
energies relevant for Component B.

\paragraph{Westerlund 1}
For completeness we briefly comment on Westerlund 1, although it does not correctly reproduce the direction of the Component B dipole within the \texttt{UF23} GMF model suite. The stellar wind luminosity of Westerlund 1 has been estimated as $L_w = 1.6\text{--}2.8 \times 10^{39}~\rm erg~ s^{-1}$~\citep{Haerer:2025ull}. With the same calculations as above and a solar composition as for Cyg X-1, we find that the required CR luminosity is $1.4\times 10^{39}~\rm erg~s^{-1}$, which is uncomfortably close to, but within the total mechanical budget inferred from the wind. As with Cygnus X-1 and Cygnus OB2 above, assuming the same source metallicity and UHECR composition, the necessary accompanying H+He and Fe flux would overshoot the total diffuse UHECR flux by the same amount as above. 

\paragraph{Summary of energetic requirements}
The recent LHAASO detection of PeV particle acceleration in microquasars and the Galactic Center observations of HESS and HAWC have generated considerable excitement about their potential contribution to the CR spectrum beyond the second knee. In this work, we have shown that this excitement is difficult to reconcile with the observed UHECR dipole anisotropy: the amplitude and direction required of a Galactic Component B, under the GMF models considered here, are too constraining to be accommodated by Cygnus X-3, the Galactic Center, or Cygnus X-1 (Sect.~\ref{sec:fit}), while Westerlund~1 additionally fails to reproduce the required dipole direction. Independent of the dipole constraints, and even if the specific GMF models adopted here are not taken at face value, we have shown in this section that all candidates are also excluded on energetic grounds. Cygnus X-3, the Galactic Center, Cygnus X-1 and Westerlund~1 require a luminosity comparable to, or in some cases orders of magnitude in excess of, independently inferred power budgets for the these systems. Cygnus OB2 does possess sufficient power, but as for the other tested sources, given the source composition, if it were to produce the subankle nitrogen needed to explain Component B, it would produce a H+He, and Fe flux that exceeds the total measured diffuse flux in the relevant energy range, independent of the source's power budget. Because this flux-consistency argument depends only on the assumed source composition and the measured diffuse flux, and not on the propagation efficiency $f_{\rm hit}$, and hence not on the choice of GMF model, it provides a robust, model-independent constraint that applies even to the most energetically favorable candidates. We therefore conclude that, despite their promise as PeV accelerators, none of the five Galactic sources considered here can plausibly supply the Component B flux both by the dipole anisotropy and, independently, by the severe energetic requirements implied by their inferred composition and luminosity.

\section{Summary and discussion}
\label{sec:discussion}
In this work, we have revisited the question of the transition energy from Galactic to extragalactic CRs. Pinpointing this transition precisely is fundamental to understand the acceleration requirements on Galactic sources and to characterize the extragalactic UHECR source population(s). To search for observational signatures of this transition, we computed the dipole anisotropy expected from candidate Galactic scenarios and compared it to current anisotropy measurements and upper limits.

We find that the observed transition from a dipole pointing to extragalactic space at the highest energies to one with right ascension compatible with the Galactic Center direction below 4\,EeV cannot be reproduced by extragalactic sources following the large-scale structure with any model of the GMF (Fig.~\ref{fig:XGal}). Additionally, the measured dipole amplitude of $\sim0.5\%$ is too large to be reproduced by extragalactic sources below 1\,EeV due to the large energy loss length of CRs at those energies, even when the Compton-Getting effect is considered (Fig.~\ref{fig:fit_sibyll}). For these two reasons, at least some Galactic contribution is required in the energy range we investigated (0.5\,EeV to 4\,EeV). It is however difficult to reconcile the low measured dipole amplitude of $\sim0.5\%$ in this energy range with a predominantly Galactic origin, so the only viable Galactic contribution is a subdominant ($\lesssim 20\%$) iron tail from a source population which is presumably dominant at lower energies below our fitting range. 

\paragraph{Galactic Component B: too large anisotropy}
Keeping this constraint aside, two scenarios are in principle able to explain the dipole amplitude and the observed phase change. In the first, the Galactic iron is completely depleted in the energy range considered in our work, and an \emph{anisotropic} Galactic Component B (contributing around 60, 30, 20$\%$ of the total flux in the 0.5--1, 1--2, 2--4~EeV bins following the intermediate-mass fractions measured by~\citet{PierreAuger:2026MassComposition} interpreted using the Sybill2.3e hadronic interaction model mean, see Fig.~\ref{fig:comp}) is preferred over a purely isotropic origin for Component B ($\chi^2/\mathrm{ndf}=1.44/1$, compared to $\chi^2/\mathrm{ndf}\simeq10$ for the isotropic case). The preferred dipole amplitude of Component B is around 1\% below 2\,EeV, increasing to $\sim10$\% in the 2--4\,EeV bin, at a preferred right ascension of $-60^{+17}_{-16}$ degrees (Fig.~\ref{fig:fit_sibyll}). The declination is not well constrained, leading to a broad range of allowed directions that includes the Galactic Center.

Motivated by this allowed anisotropic component, we simulated the dipole anisotropy induced by diffuse Galactic source populations, individual long-lived Galactic sources, and short-lived Galactic transients. However, none of these source classes can reproduce a sufficiently low dipole amplitude. 
We find that at 99\% C.L., no Galactic source population or individual source candidate should contribute more than 25\% of the total CR flux $\lesssim1\,$EeV if it emits intermediate elements in this energy range (Fig.~\ref{fig:amp_ra}). 
Stronger limits apply to individual source candidates because they produce larger dipole moments. For a source at the Galactic Center -- which produces the largest dipole amplitude of all sources we tested at rigidities $\lesssim1\,\mathrm{EV}$ -- this leads to an upper limit of $\sim4\%$ of the total flux. 
Nearby transient sources are also excluded because we find that a radially decaying turbulent magnetic field always leads to an accumulation of CRs in the inner Galaxy, and thereby to a too large residual dipole pointing in the Galactic Center direction at late times (Fig.~\ref{fig:transient}). We have validated our simulation results with analytical estimates which we presented in Sec.~\ref{sec:aniso_theo}.

Note that also a dominant contribution from Galactic sources of \textit{heavy} elements directly below the ankle is forbidden by anisotropy constraints now (as well as by mass composition measurements), contrary to results of previous works based on old GMF models and smaller data sets~\citep{Giacinti:2011Transition, Auger:2012_anisotropy}.

Thus, although a Galactic Component B (i.e.~a transition from Galactic to extragalactic sources at the ankle) is allowed by current mass-composition estimates and by our fit to the dipole amplitude and direction in a three-component model, we found no viable realization among the Galactic source scenarios explored here. This is also supported by the severe energetics that would be required of such a source, and the associated fluxes of lighter elements from the source that would overshoot the energy spectrum at lower energies, as discussed in Sec.~\ref{sec:energetics}.

\paragraph{Extragalactic Component B}
The second viable scenario that explains the large-scale anisotropy measurements is a tail of Galactic iron extending above 1\,EeV (Fig.~\ref{fig:fit_epos}). Both the dipole amplitude and the change in dipole phase can be well reproduced by this iron-tail of a Galactic source population following the Galactic matter distribution with flux fractions of around 15, 10, 5$\%$ at 0.5--1, 1--2, 2--4~EeV. These fractions are compatible with both the mean iron fractions predicted using the EPOS-LHC-R hadronic interaction model, as well as the upper limits using Sibyll 2.3e~\citep{PierreAuger:2026MassComposition}.
Component B, which accounts for $\sim$45, 40, 25$\%$ of the total flux under the EPOS-LHC-R model, is compatible with a negligible dipole amplitude, and hence extragalactic in this scenario.

Note that the Galactic Component B scenario is not intrinsically less economical than the extragalactic one. It substitutes an additional extragalactic component for an additional Galactic intermediate-mass population. The two scenarios are therefore not distinguishable on the grounds that one or the other invokes a smaller number of populations. Within the source and magnetic-field scenarios explored here, and given that our forward-tracking simulations find no Galactic scenario with a small enough dipole amplitude in the no-iron scenario, we therefore consider the most economical joint interpretation of the composition and anisotropy measurements to be a sub-ankle extragalactic CNO component accompanied by a tail of Galactic iron. Future diffuse $\gamma$-ray observations as proposed by~\citet{Prevotat:2025qdp} would confirm this scenario.

\paragraph{Choice of Galactic Magnetic Field}
In our analysis, we used the Planck-tuned JF12 random field. This is a comparatively strong random field. For example, \citet{Giacinti:2017dgt} reduced the turbulent component of the original JF12 random field by a factor of $\sim8$--$10$ relative to its nominal normalization in order to correctly reproduce the observed B/C ratio. Since a stronger turbulent field isotropizes CRs more efficiently, our use of the JF12+Planck random field yields a robust lower limit to the expected dipole anisotropy from the Galactic Component B scenarios we studied here. 

While this manuscript was being finalized, the new random field model of~\citet{Unger:2026yau} appeared on arXiv. Both representative versions of the model have similar or somewhat weaker field strengths as the \texttt{JF12Planck} model we used here, so its adoption would strengthen our constraints on a Galactic origin of the Component B cosmic-ray flux from the dipole anisotropy, while leaving our overall conclusions unchanged. The vertical extension of the field is significantly smaller than that of \texttt{JF12Planck}, which would lead to faster particle escape and thus even stronger requirements on the source luminosity to power a Galactic Component B.

In this analysis we assumed that the coherence length of the random GMF is $L_{\rm coh} = 56$~pc, i.e.~the outer scale of the turbulence is $L = 5 L_{\rm coh} \sim 280$~pc. This outer scale is somewhat above the $\mathcal{O}(100~\rm pc)$ scale typically associated with injection by supernova remnants and hence again leads to a conservative estimate of the dipole of Galactic sources. Also, our adopted coherence length is close to the independent estimate of \citet{Mertsch:2013pua} from Galactic synchrotron emission, and within the allowed range (although well below the best-fit value) of~\cite{Beck:2014pma}. In stark contrast,~\citet{Haverkorn:2008tb} find coherence lengths of $\mathcal{O}(1~\rm pc)$. Were we to assume the latter value we would find a much larger dipole amplitude. Our choice is therefore conservative with regard to the amplitude of the anisotropy.

\paragraph{Effects of a Giant Circumgalactic Halo on the Origin of Component B} There is evidence for a hot, ionized halo surrounding the Milky Way~\citep{Gupta:2012rh}, which may also be magnetized to some degree, e.g.~\citet{Becker:2024aqj}. For an estimate of the effect such a Giant Halo could have on the Component B UHECRs, we can use the values adopted by~\citet{Li:2026xat} whose halo size and diffusion coefficient were shown to be consistent with current secondary-to-primary CR measurements. 

\citet{Li:2026xat} adopted a halo radius of 250~kpc and diffusion coefficient $D(R)= D_0\left(\mathcal{R}/\mathrm{GV}\right)^\delta$ with $D_0 = 10^{28.6}~\rm cm^2/s$ at 1~GV and $\delta~=~0.5$. The adopted diffusion coefficient fixes the scattering length $\lambda = 3D/c \sim 1.29\ {\rm pc}\ \left(\mathcal{R}/\rm GV\right)^{\delta}$ but does not fix $B$ and $L_{\rm c}$ separately, only their ratio. To relate $\lambda$ to the field, we use the test-particle simulations
of~\citet{Kuhlen:2022tov}, who tabulate the scattering mean free path in isotropic
turbulence as a function of reduced rigidity $r_g/L_{\rm c}$. We adopt $\lambda/L_{\rm c} = A\,(r_g/L_{\rm c})^{\delta}$, taking the
exponent to be $\delta = 0.5$ as assumed by~\citet{Li:2026xat}, so that
$D\propto\mathcal{R}^{0.5}$ is reproduced at every rigidity, and fixing the
normalization $A = 0.687$ on their $\eta = 1$ (purely turbulent) tabulated values, matching the
assumption of~\citet{Li:2026xat}.\footnote{Our Eq.~\ref{eq:diffusion_approximation} uses an exponent $0.6$, fitted to the
Kolmogorov turbulence of~\citet{Kuhlen:2022tov}, whereas here we adopt $\delta = 0.5$ for
consistency with the Kraichnan turbulence assumed by~\citet{Li:2026xat}. Over the range of
reduced rigidities simulated by~\citet{Kuhlen:2022tov} the two estimates agree to within
$30\%$. Beyond it we extrapolate with the $\delta$ of~\citet{Li:2026xat}. This yields a stronger field at
large $L_{\rm c}$, namely $B = 0.31\,{\rm nG}\,(L_{\rm coh}/{\rm kpc})$ against
$B = 0.10\,{\rm nG}\,(L_{\rm coh}/{\rm kpc})^{2/3}$ and is therefore the conservative choice.} Finally, substituting the gyroradius $r_g = 27\ {\rm pc}\,(\mathcal{R}/10^{17.3}\,{\rm V})
(B/8\ \mu{\rm G})^{-1}$ gives $B = 0.31\ {\rm nG}\,(L_{\rm c}/1\ {\rm kpc})$ for the giant halo field. For a purely random field as proposed by~\citet{Li:2026xat}, requiring at least $\sim10$
coherence cells across the halo gives $L_{\rm c}\lesssim25$~kpc. For the Component~B UHECRs, the
Larmor radius is then $r_g = 704\ {\rm kpc}\,(\mathcal{R}/10^{17.3}\,{\rm V})
(L_{\rm c}/1\ {\rm kpc})^{-1}$, and transport is ballistic. Hence a giant halo field of such strength has no effect on our conclusions.

\paragraph{Other caveats}
In principle, the dipoles produced by several Galactic sources could partially cancel. However, reproducing the small observed amplitude across multiple energy bins would require a finely tuned combination of source directions, spectra, compositions, and maximum rigidities. Such a cancellation would not be possible for transient sources, for which the residual dipole amplitude towards the inner Galaxy persists at late times (cf.\ Fig.~\ref{fig:transient}). Although we cannot exclude an accidental cancellation of continuously emitting sources, we regard it as a very fine-tuned realization of the no-iron scenario.

Our results cannot be extrapolated to lower energies where there are no published mass fractions with up-to-date hadronic interaction models, and where tracking simulations become computationally prohibitive. We therefore do not constrain the Galactic sources below $\mathcal{R}\sim 10^{17}$~eV. Microquasars, e.g.~\citep{Kaci:2025gyb,Zhang:2025tew,Zhang:2026igt,Vecchiotti:2026okk}, and young stellar clusters, e.g.~\citep{Morlino1:2021zwu} remain viable as contributors at lower energies closer to the knee, although there is open debate as to whether either of them can saturate the flux at the knee. PeV $\gamma$-ray observations in the Southern sky, for example, with the proposed SWGO~\citep{SWGO:2025taj} and PEPS~\citep{Maris:2023anl} detectors, will shed more light to the most extreme accelerators in our Galaxy. 

\paragraph{Possible Extragalactic Sources of the Component B}
Several source classes have been proposed as possible sources of an extragalactic Component B flux, including Binary Neutron Star mergers~\citep{Rodrigues:2018bjg}, Galaxy Clusters~\citep{Murase:2008yt,Fang:2017zjf,Zhang:2025tew}, Filament Accretion Shocks~\citep{Simeon:2025gxd}, Hypernovae~\citep{Wang:2007ya}, and ultra-fast outflows in AGN~\citep{Ehlert:2024lji}. 
The CNO dominated composition of this component suggests environments unusually rich in CNO nuclei, with hydrogen and helium depleted. 
Low luminosity gamma-ray bursts or engine-driven supernovae of stars that have lost their hydrogen and helium envelopes~\citep{Zhang:2017moz,Zhang:2018agl}, tidal disruption of CNO rich progenitors~\citep{Zhang:2017hom,Plotko:2024gop}, and environments unusually enriched in massive star ejecta such as Galactic Centers~\citep{Ehlert:2024lji}, may be suitable environments for such composition to be accelerated and emerge. Finally, the recently proposed binary-neutron star merger scenario of~\citet{Farrar:2024zsm,Farrar:2025kpk}, which offers an explanation to the narrow maximum ridigity distribution of UHECR sources observed above the ankle~\citep{Ehlert:2022jmy} could lead to intermediate-mass fragments below the ankle through the in-source processing of ultraheavy elements. In such a scenario, the CNO component below the ankle would also originate from the same extragalactic population dominant above the ankle, just like the sub-ankle protons in~\citet{unger:2015_ufa} and as we assume in out three-component model.

\paragraph{Observations and Hadronic Interaction Model uncertainties}
The principal remaining uncertainty in interpreting our results is the mass composition in the Component B energy range, and particularly whether a subdominant iron fraction persists below the ankle. Future improved hadronic interaction models and composition measurements with high $X_{\rm max}$ resolution extending to energies below $0.5$~EeV with HEAT~\citep{Porcelli:2015jli}, IceCube~\citep{IceCube:2019hmk} and upcoming instruments will be required to definitively distinguish between the no-iron and iron-tail scenarios.
In addition, measurements of the full dipole direction (not just its right ascension) below 4\,EeV would allow to constrain the possible contribution of individual Galactic sources way more stringently depending on their direction. Even more information could be gained through mass-composition dependent arrival direction studies such as~\citet{Golup_ICRC2025}.

Taken together, our results identify a subdominant Galactic iron component and an intermediate-mass (CNO-like) extragalactic population as the simplest interpretation of the observations in the energy range between the second knee and the ankle. In this picture, the Galactic iron tail accounts for the observed evolution of the amplitude and direction of the dipole and the transition to extragalactic UHECRs starts already at the second knee.

\appendix
\section{More details on the 3-component model}
In Fig.~\ref{fig:XGal}, the mass composition fractions as measured by Auger, along with our interpretation in the three-component model, are shown. Fig.~\ref{fig:XGal} shows the dipole directions expected for the extragalactic source population following the LSS~\citep{Bister:2024Anisotropies, Bister:2024GMF}, extended to energies between 0.5\,EeV and 8\,EeV. In Fig.~\ref{fig:topview}, we show the time evolution of Galactic CRs of different rigidities and sources in the GMF, and in Fig.~\ref{fig:cygx1_westerlund_direc} the resulting dipole directions for Cygnus X-1 and Westerlund 1.
\begin{figure}[ht]
\centering
\includegraphics[width=0.65\textwidth]{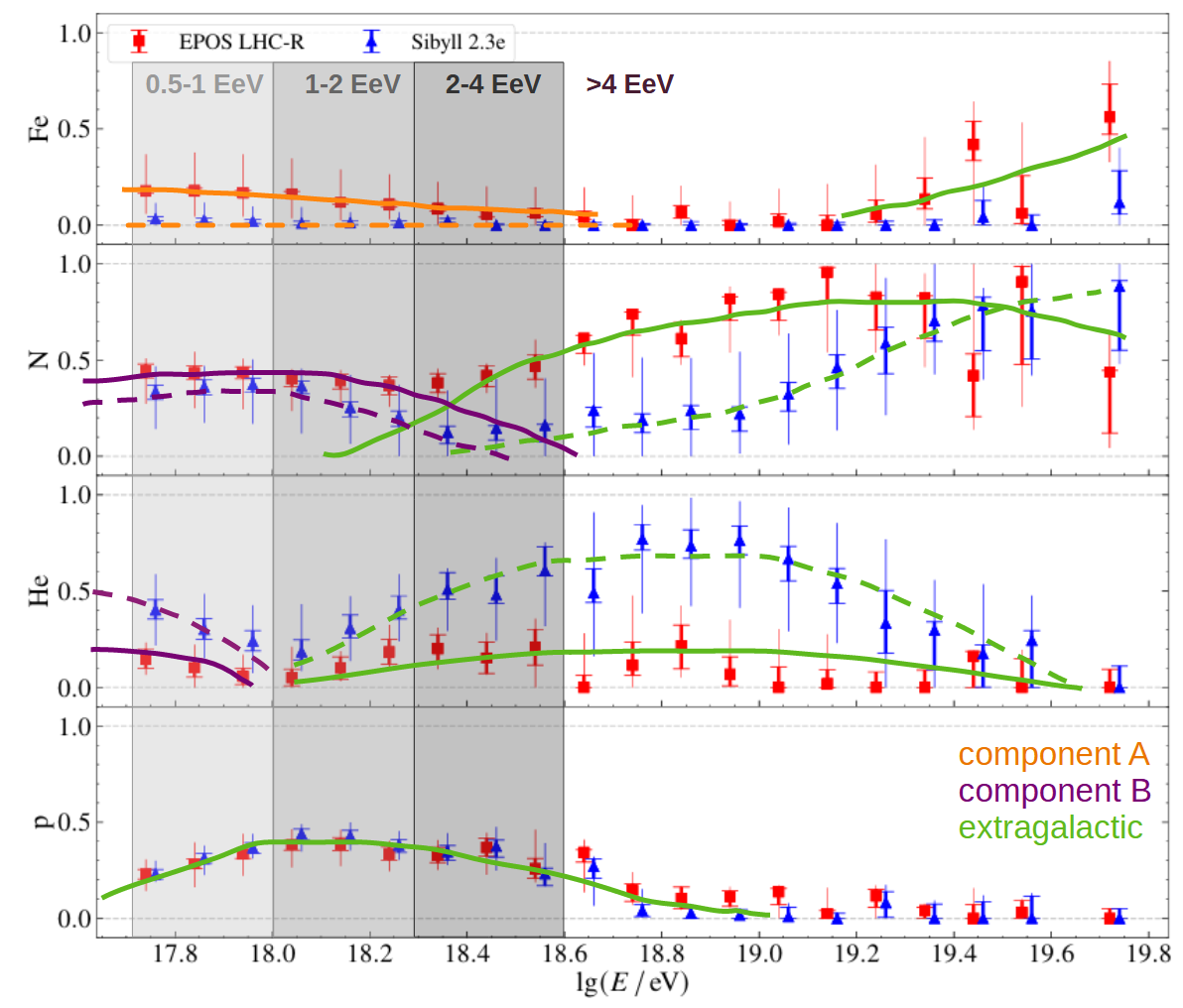}
\caption{Model for the relative contributions of the three components: a possible tail of Galactic iron (Component A, orange), an intermediate component that could be either Galactic or extragalactic (Component B, purple), and an extragalactic mixed composition following a Peters cycle that dominates above the ankle (extragalactic, green). The dashed (solid) lines are meant to guide the eye and are oriented on the median values for the Sibyll2.3e (EPOS-LHC-R) hadronic interaction model. 
The mass composition fractions as derived by Auger from~\citet{PierreAuger:2026MassComposition} for the two hadronic interaction models are shown as markers. The three energy bins used in the above analysis (guided by the binning used for the dipole in~\citet{PierreAuger:2024Anisotropy19Y}) are indicated by the grey boxes.}
\label{fig:comp}
\end{figure}

\begin{figure}[ht]
\centering
\includegraphics[width=0.4\textwidth]{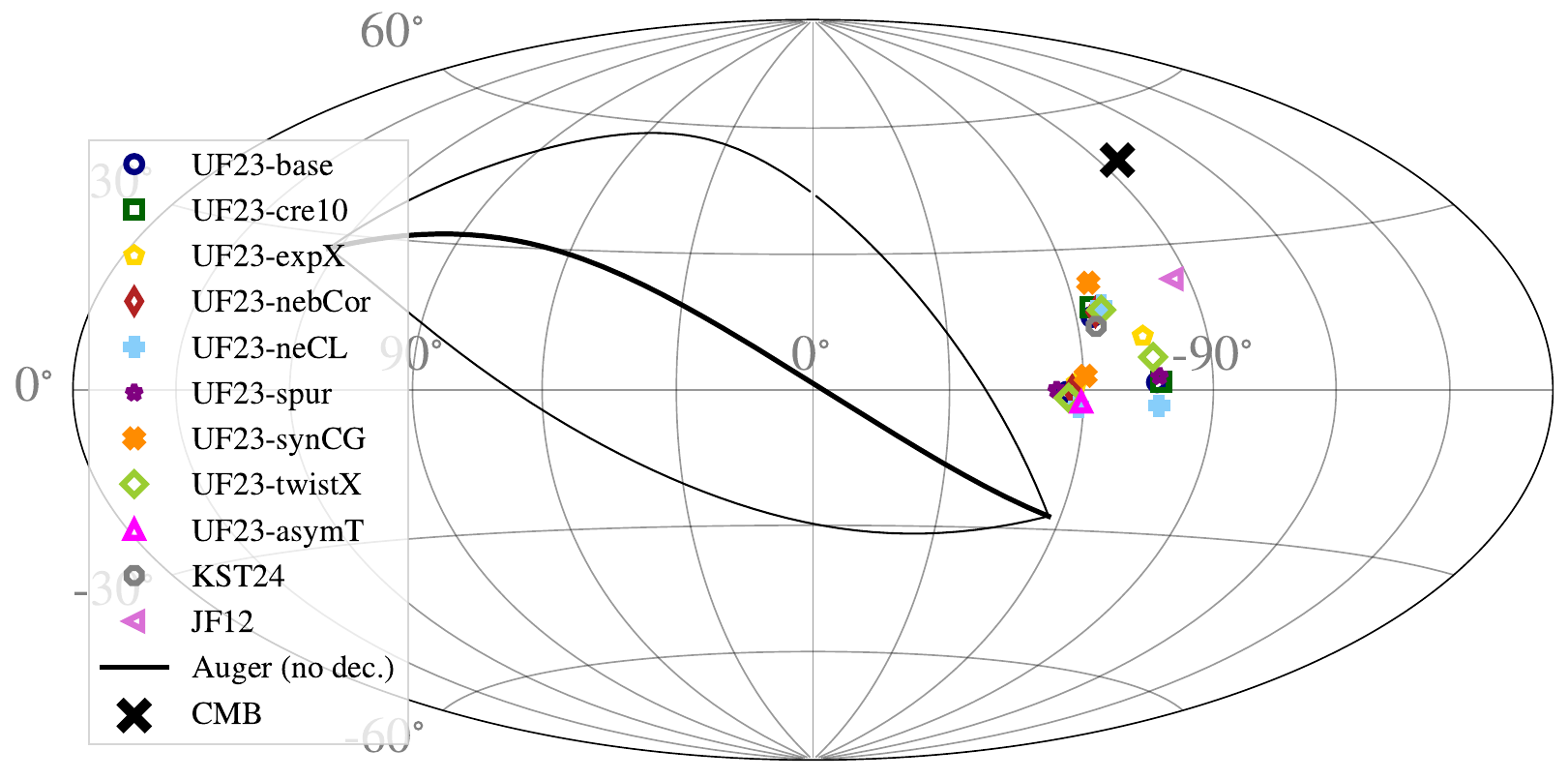}
\includegraphics[width=0.4\textwidth]{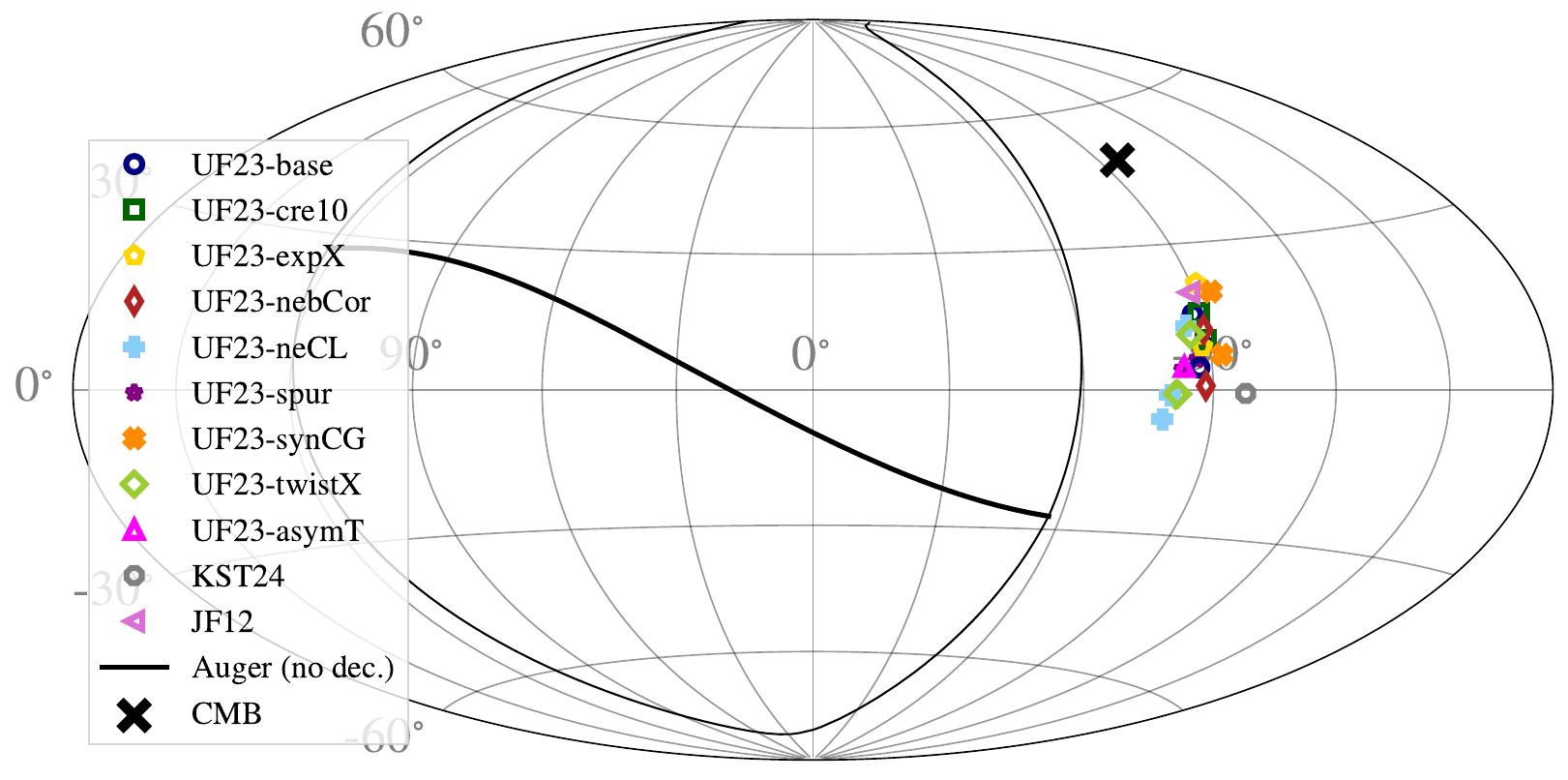}\\
\includegraphics[width=0.4\textwidth]{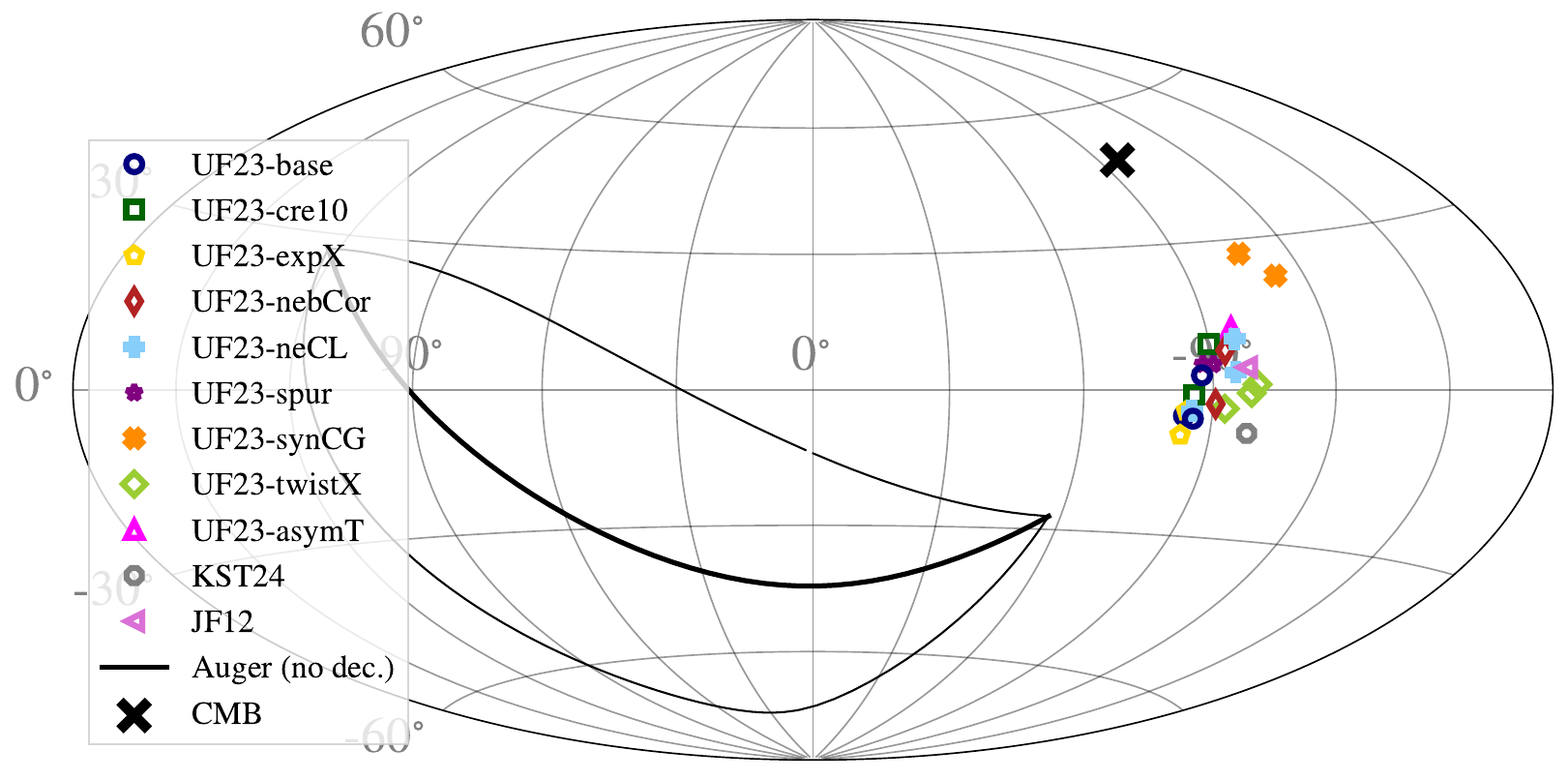}
\includegraphics[width=0.4\textwidth]{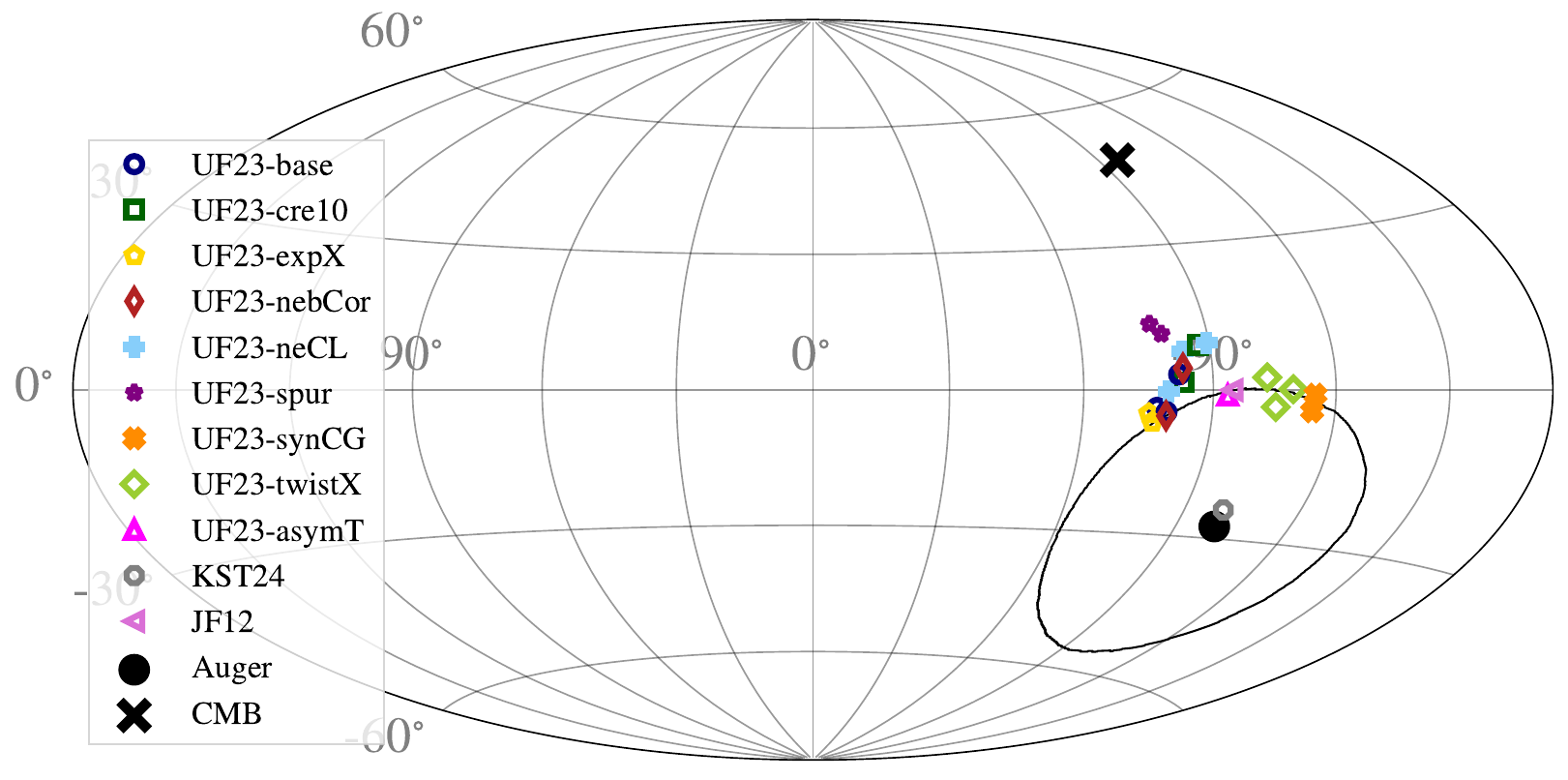}
\caption{Predicted dipole directions for extragalactic sources following the LSS from CosmicFlows4~\citep{Valade:2026_CF4} in four energy bins, (0.5-1)\,EeV (\textit{upper left}), (1-2)\,EeV (\textit{upper right}), (2-4)\,EeV (\textit{lower left}), and (4-8)\,EeV (\textit{lower right}). The dipole direction is calculated taking into account the limited exposure of the Pierre Auger Observatory. The dipole expected from the Compton-Getting effect (0.6\% before GMF pointing towards the CMB dipole direction~\citep{Compton_Getting:1935, Kachelriess:2006}) is included in the model and dominates over the LSS dipole below $\simeq1\,$EeV.
The markers show different coherent GMF models, combined with up to three different realizations of the random field from \texttt{JF12+Planck}~\citep{Jansson:2012pc, Jansson:2012b_Random, Planck:2016GMF}.
The measured dipole directions from Auger~\citep{PierreAuger:2024Anisotropy19Y} are either shown as a central marker and $1\sigma$ contour for (4-8)\,EeV, or as a thick line representing the measured right ascension, and thinner lines visualizing the uncertainty (below 4\,EeV where the declination is not measured).}
\label{fig:XGal}
\end{figure}

\begin{figure}[ht]
\centering
\includegraphics[width=0.3\textwidth]{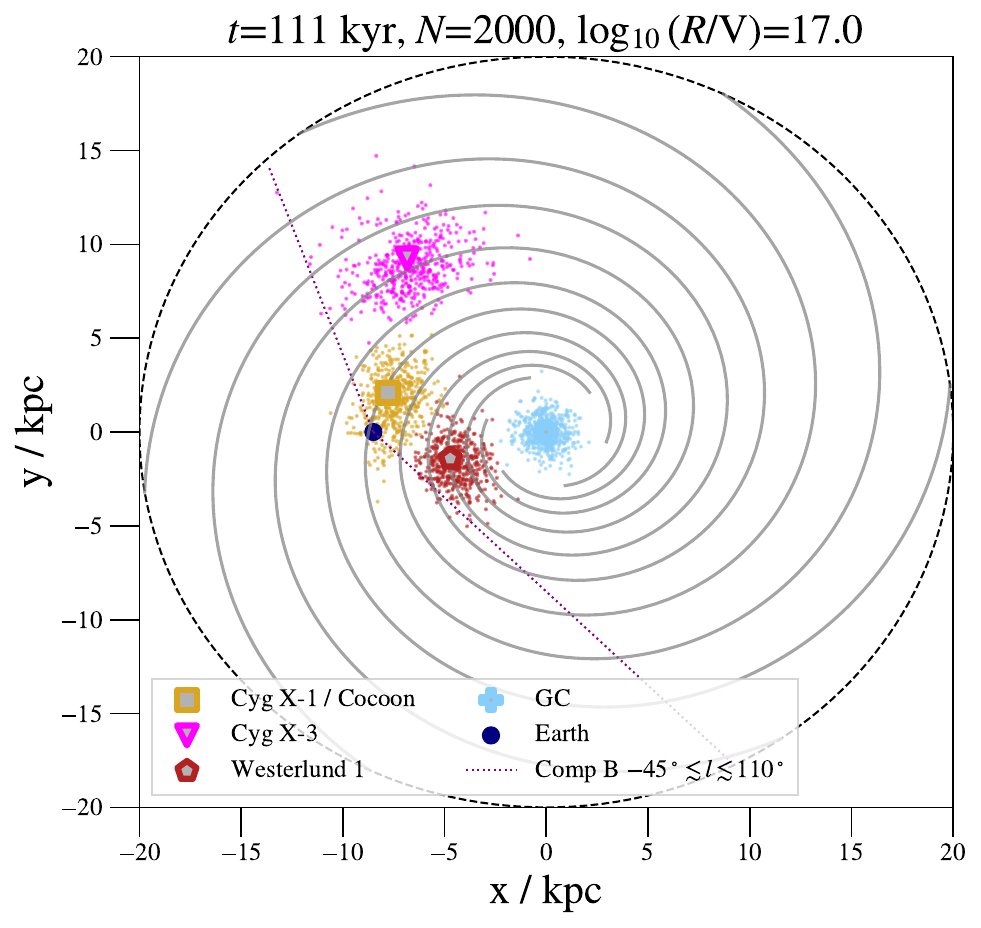}
\includegraphics[width=0.3\textwidth]{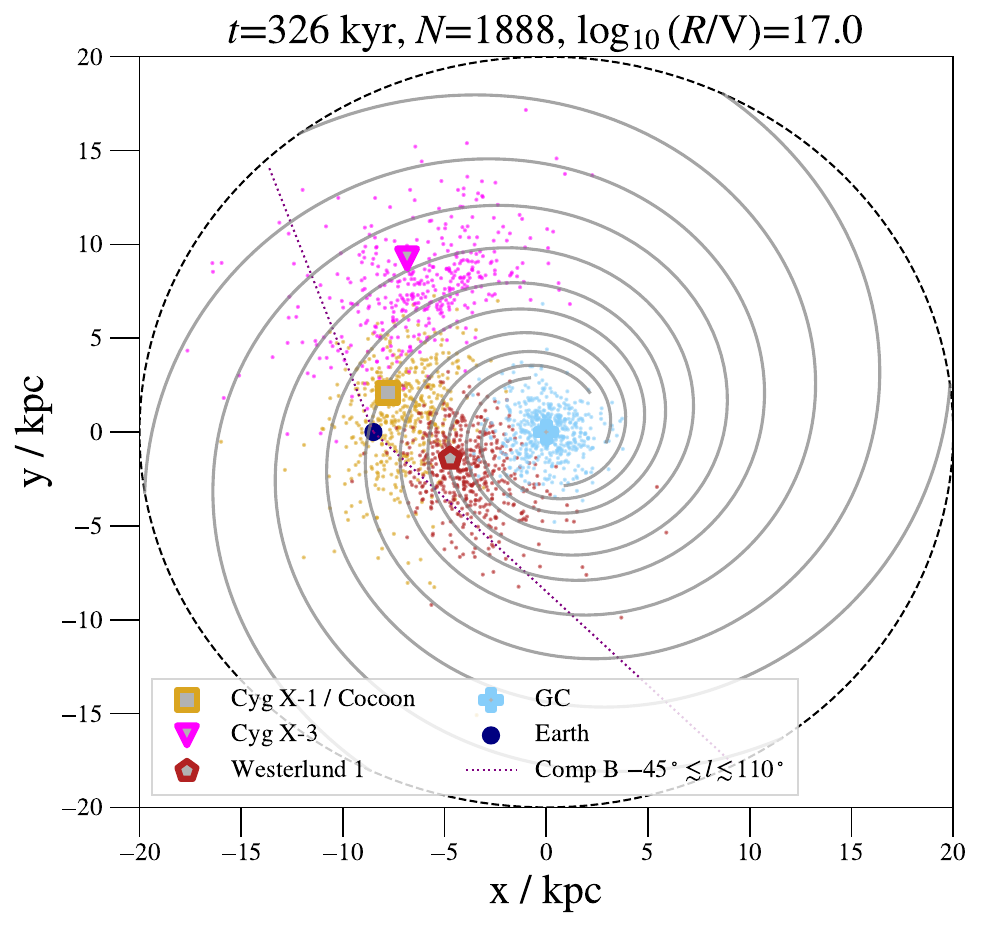}
\includegraphics[width=0.3\textwidth]{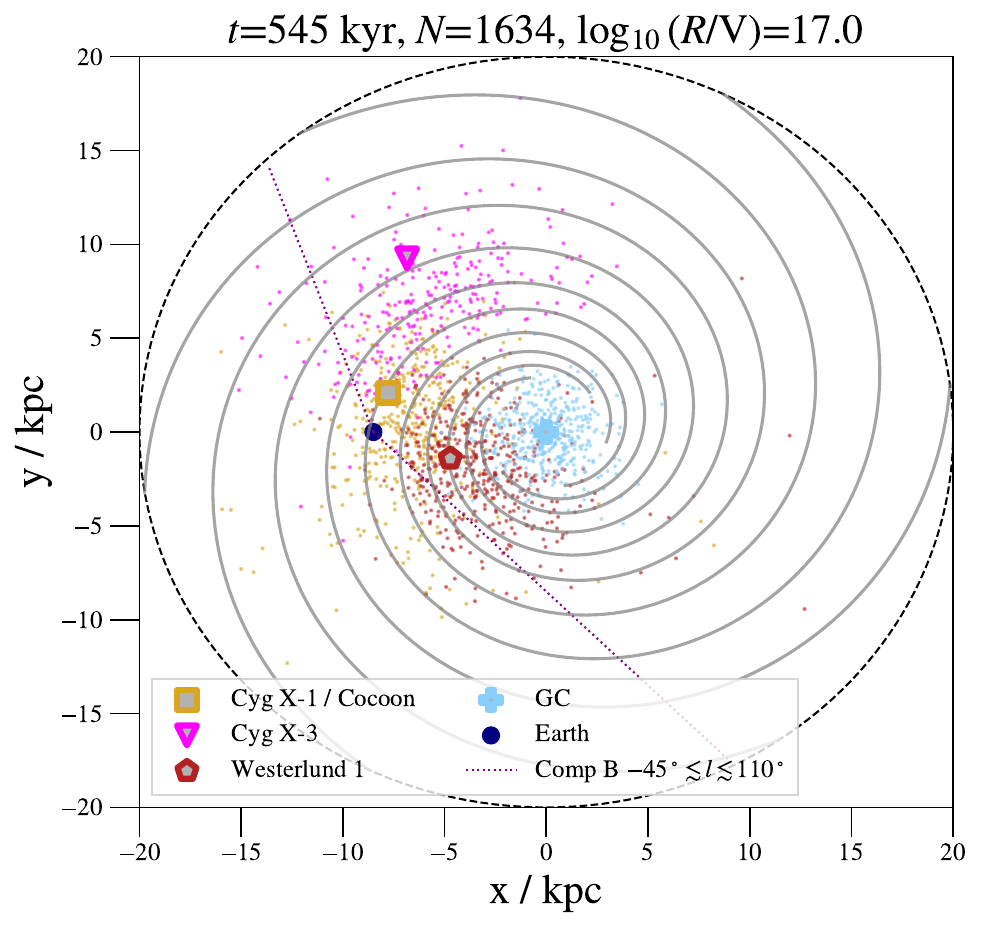}\\
\includegraphics[width=0.3\textwidth]{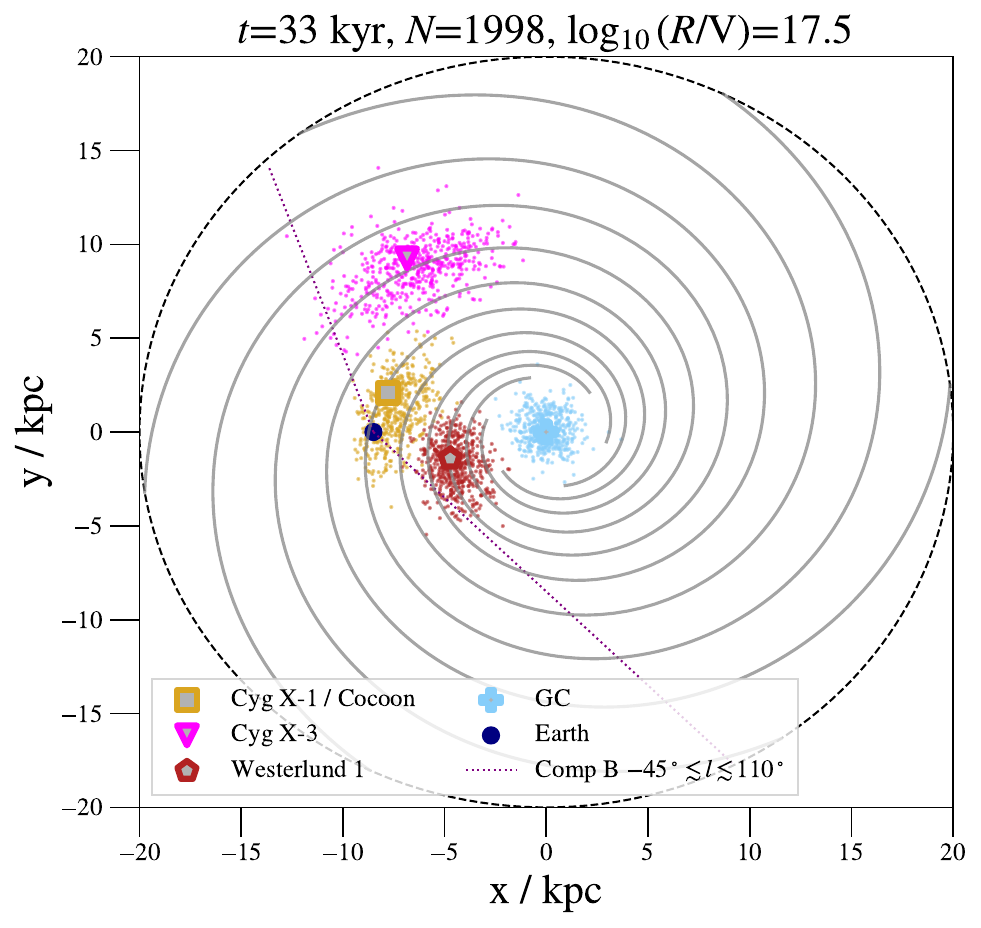}
\includegraphics[width=0.3\textwidth]{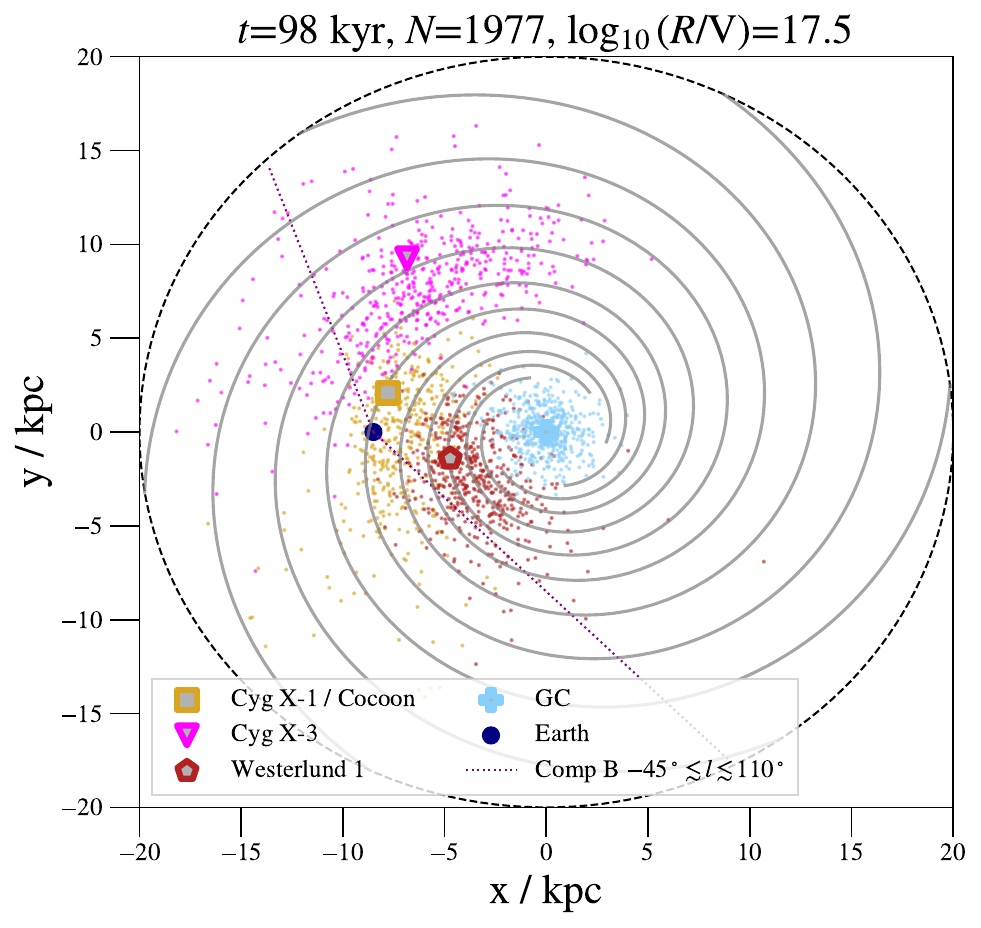}
\includegraphics[width=0.3\textwidth]{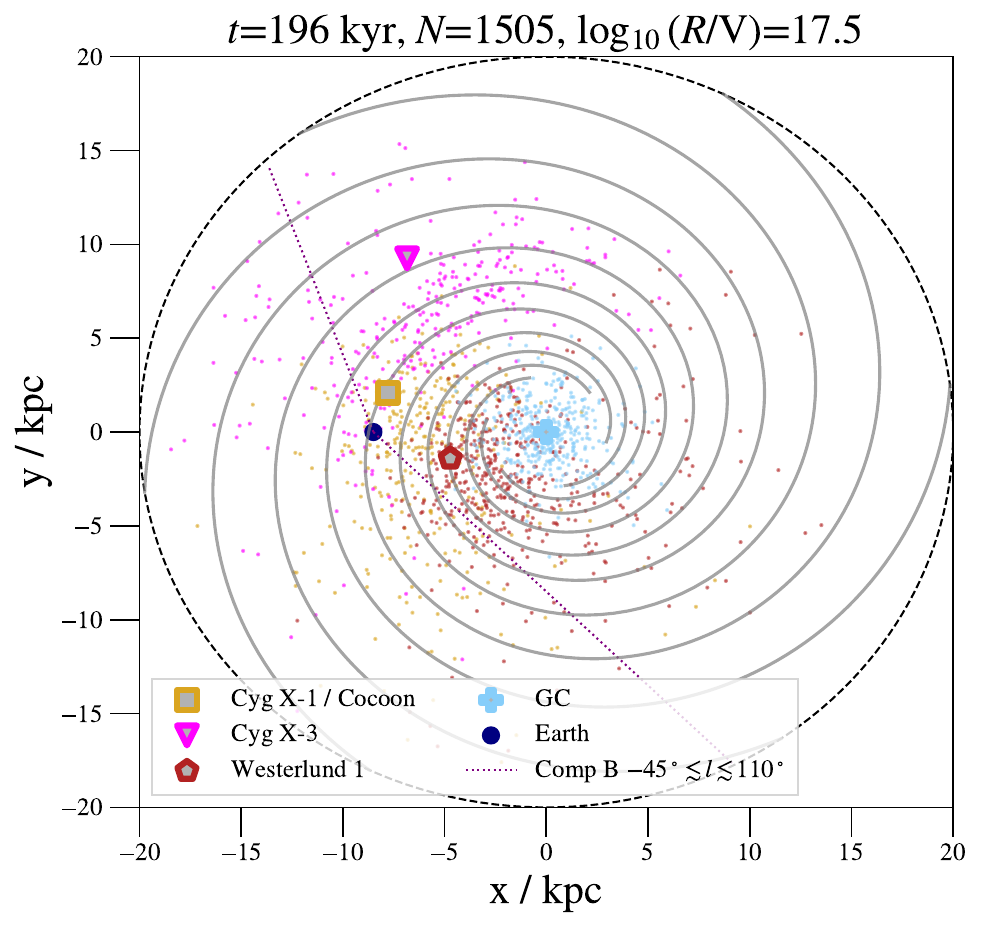}\\
\includegraphics[width=0.3\textwidth]{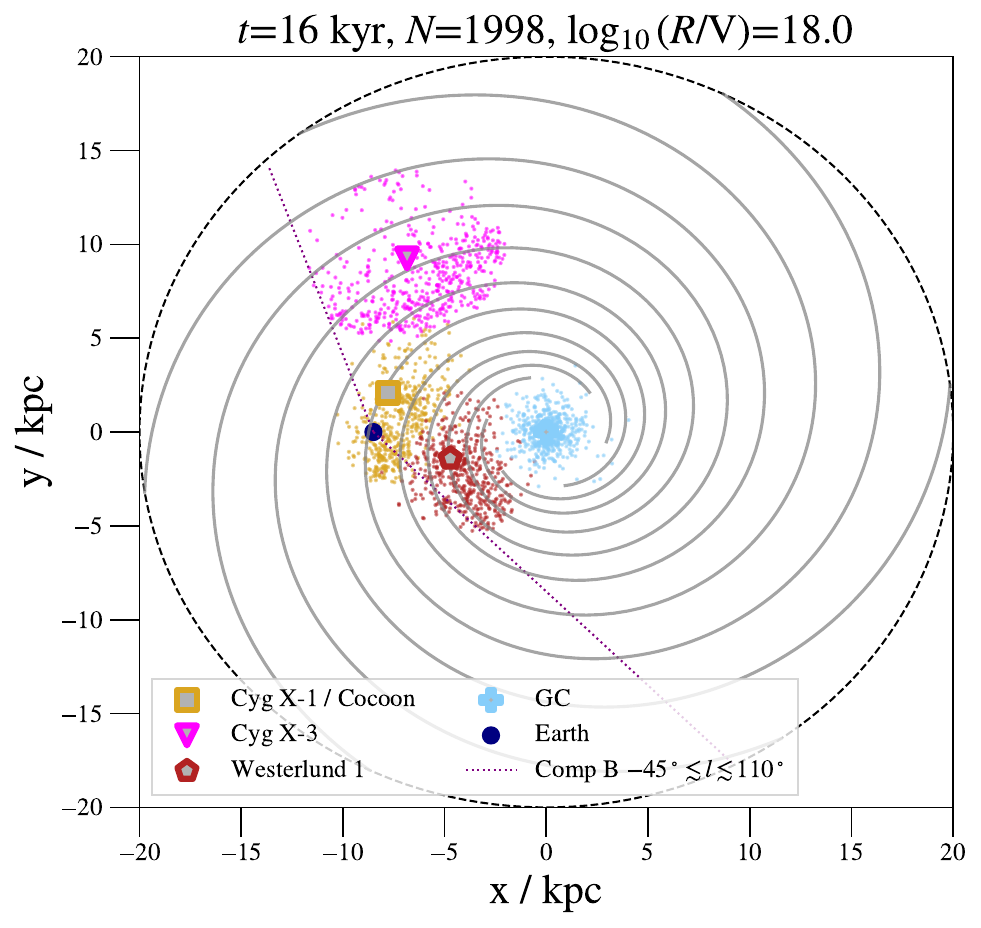}
\includegraphics[width=0.3\textwidth]{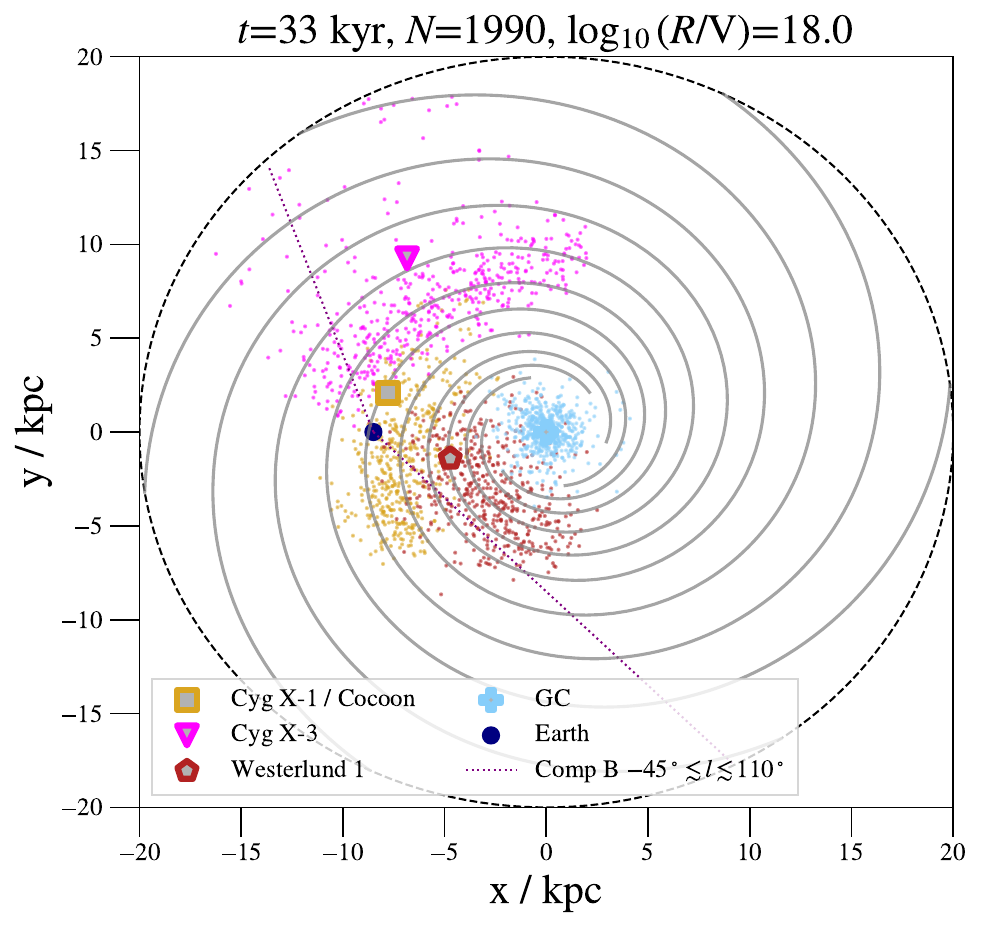}
\includegraphics[width=0.3\textwidth]{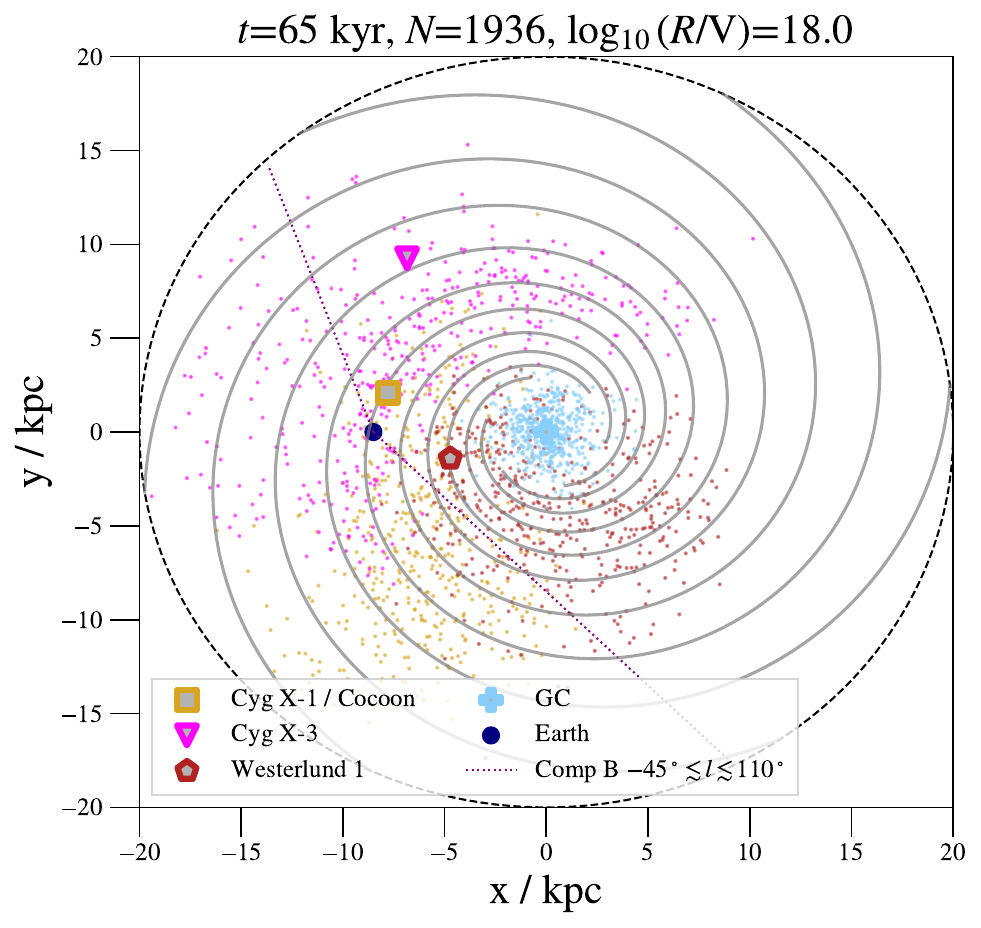}\\
\includegraphics[width=0.3\textwidth]{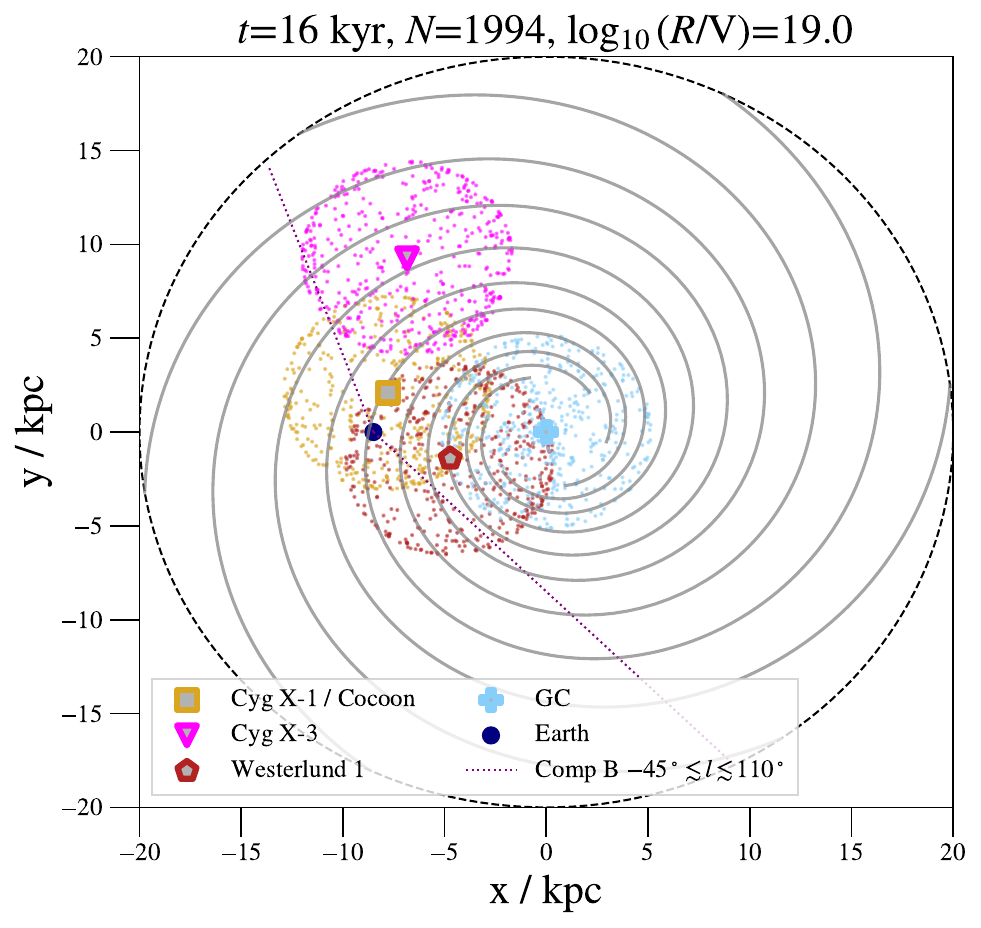}
\includegraphics[width=0.3\textwidth]{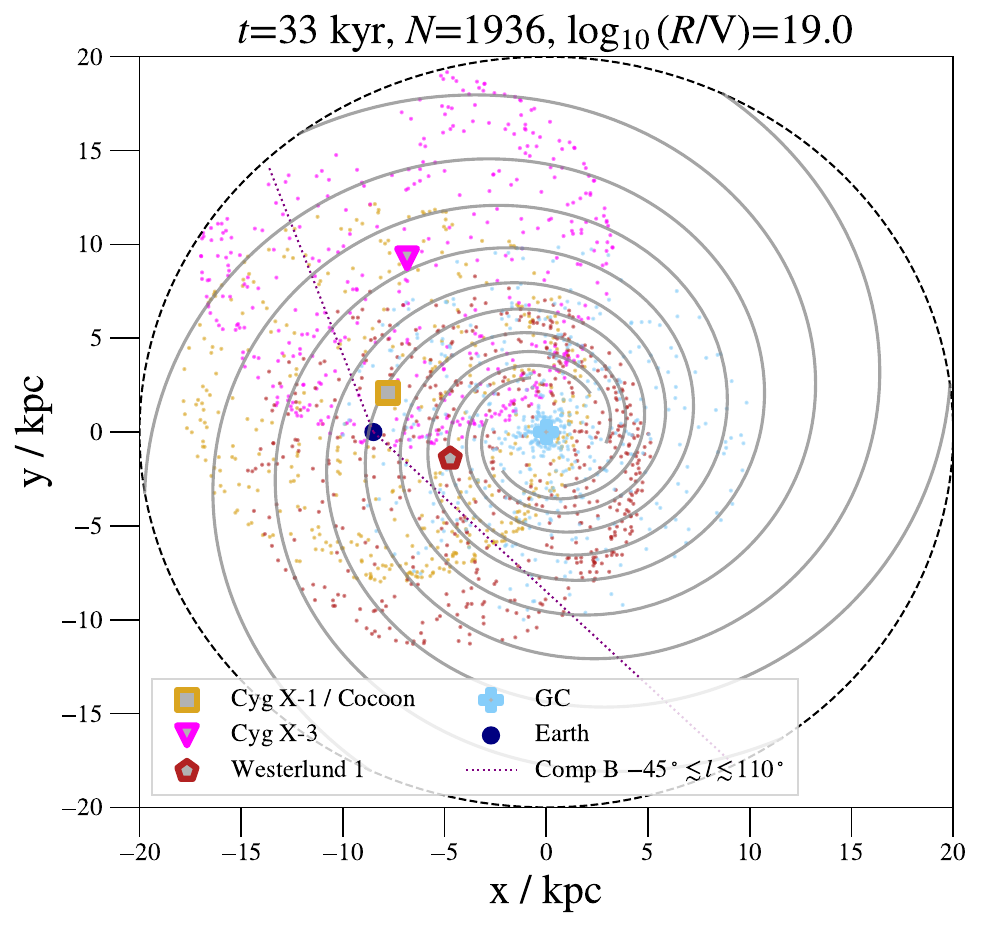}
\includegraphics[width=0.3\textwidth]{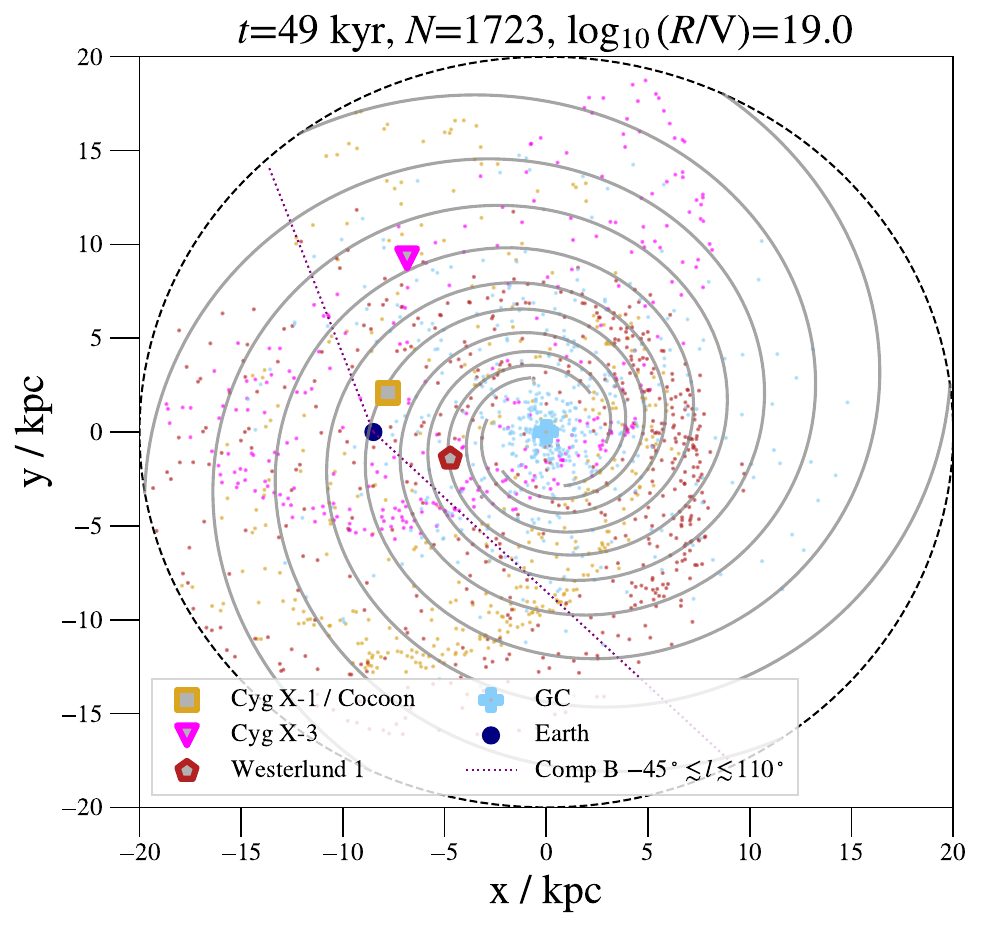}
\caption{CRPropa3 simulations of particles with $R=0.1\,$EV (\textit{top row}), $R=0.3\,$EV (\textit{second row}), $R=1\,$EV (\textit{third row}), and $R=10\,$EV (\textit{lower row}) propagating over time $t$ (columns, see figure titles) in the \texttt{UF23-base} GMF model (field lines shown as gray spiral) with the \texttt{JF12+Planck} random field. In total, $N(t=0)=2000$ CRs are injected, 500 each from Cygnus X-3, Cygnus X-1 / Cocoon, Westerlund 1, and the Galactic Center. Over time, CRs spread from the sources and some leave the Galaxy so that $N$ decreases with time. The purple dotted lines show the edges of the preferred direction of the dipole of Component B, $-45^\circ\lesssim l\lesssim 110^\circ$ (see Fig.~\ref{fig:fit_sibyll}).}
\label{fig:topview}
\end{figure}

\begin{figure}[ht]
\centering
\includegraphics[width=0.32\textwidth]{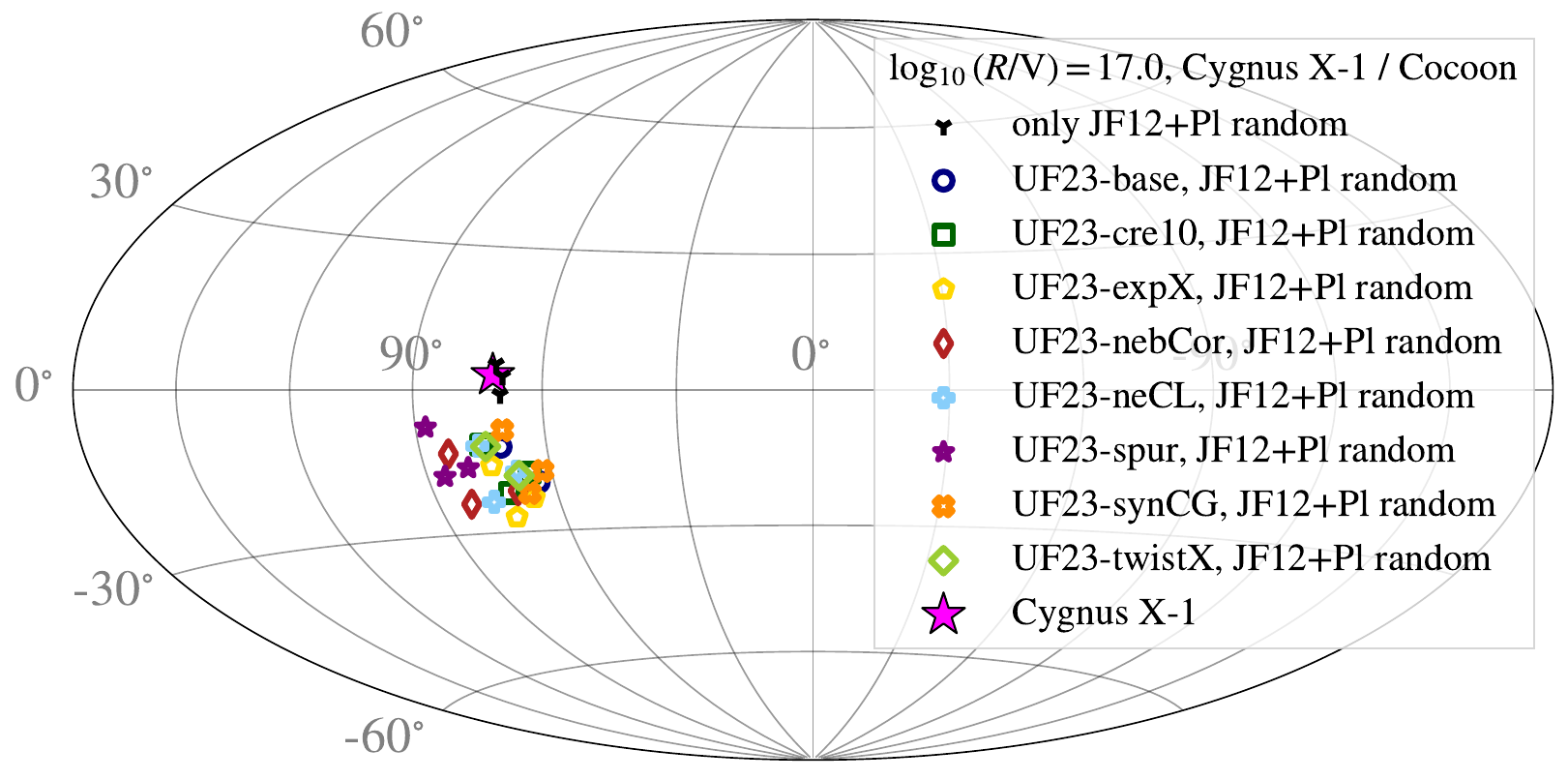}
\includegraphics[width=0.32\textwidth]{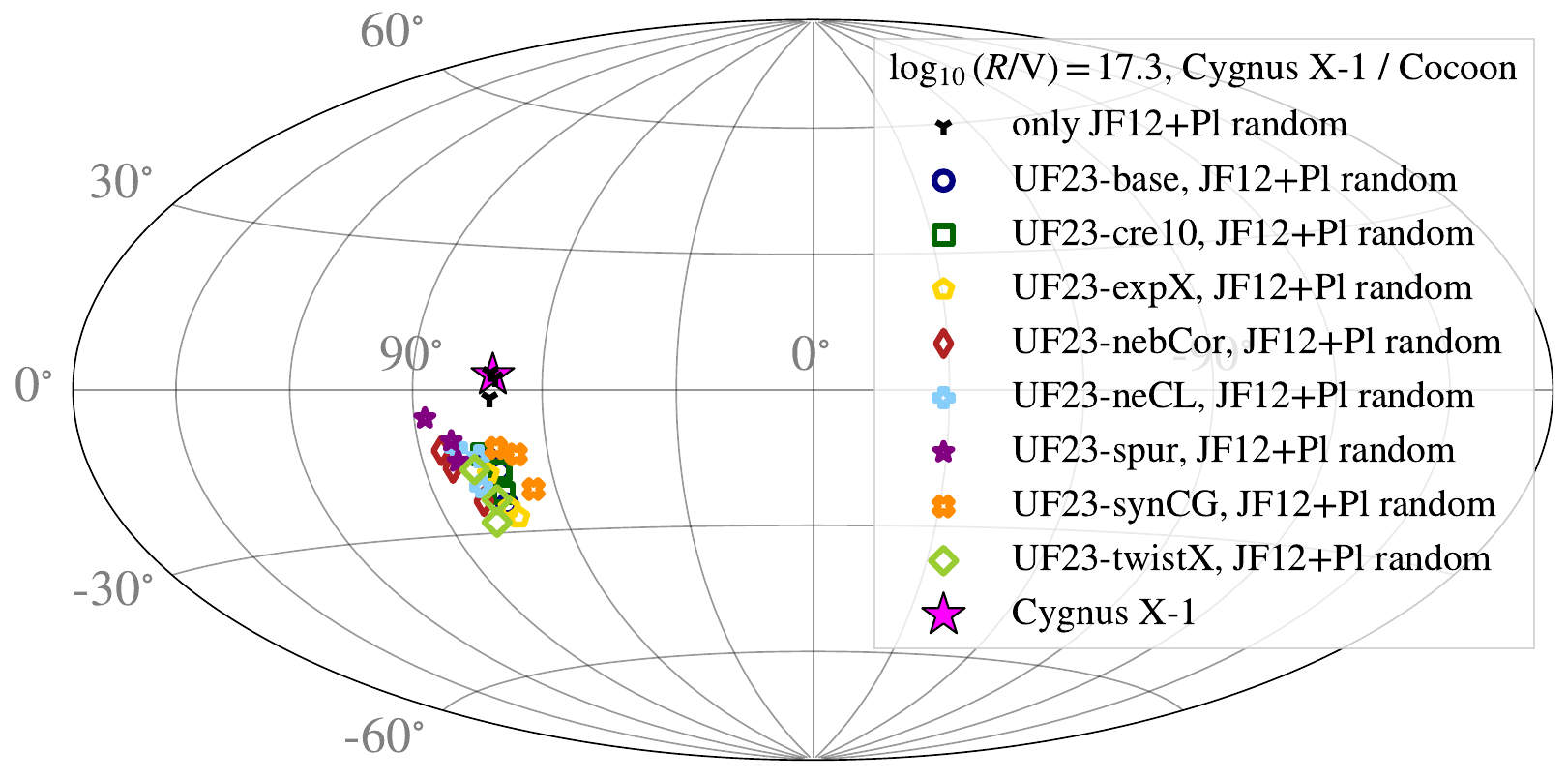}
\includegraphics[width=0.32\textwidth]{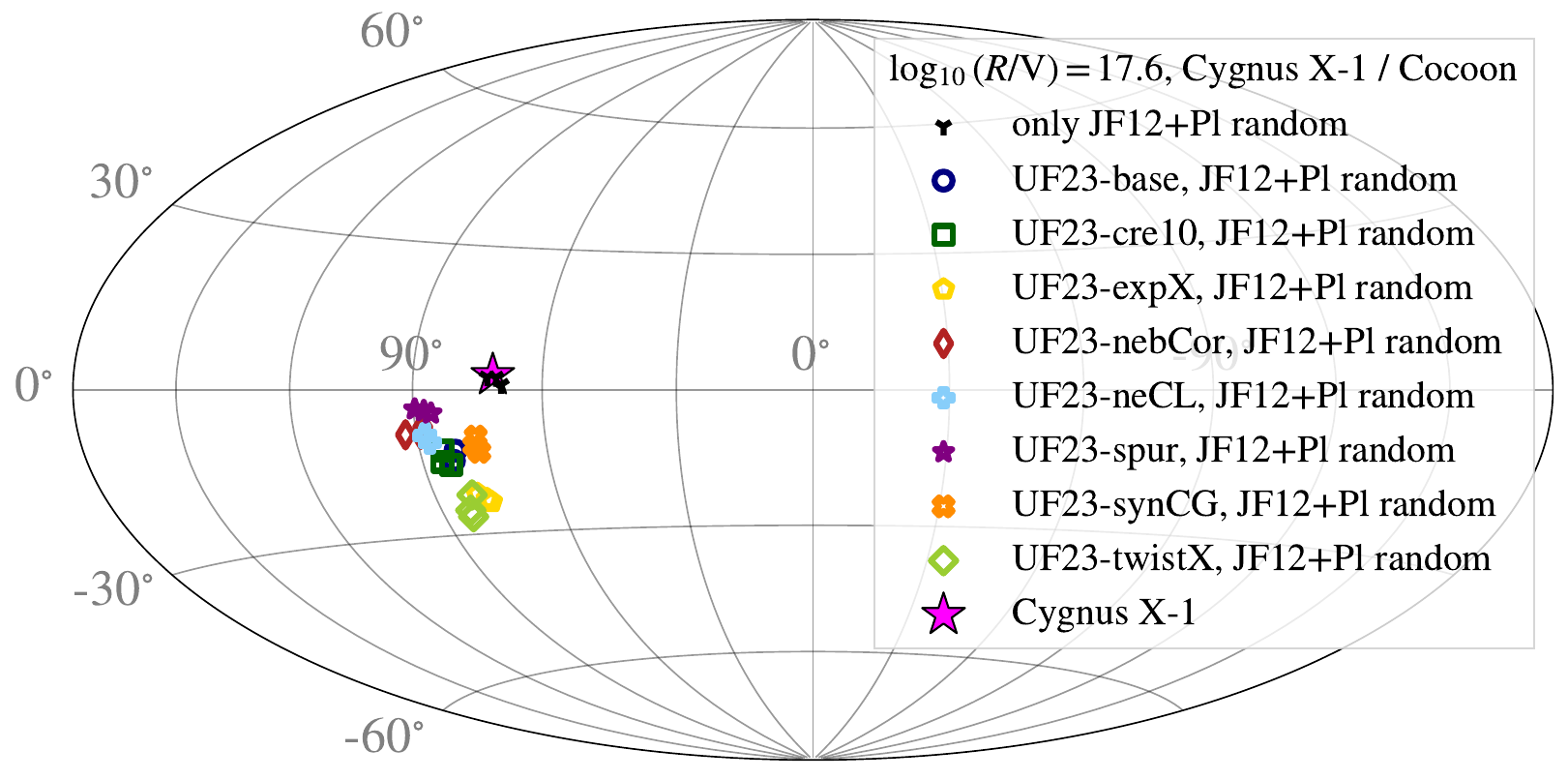}\\
\includegraphics[width=0.32\textwidth]{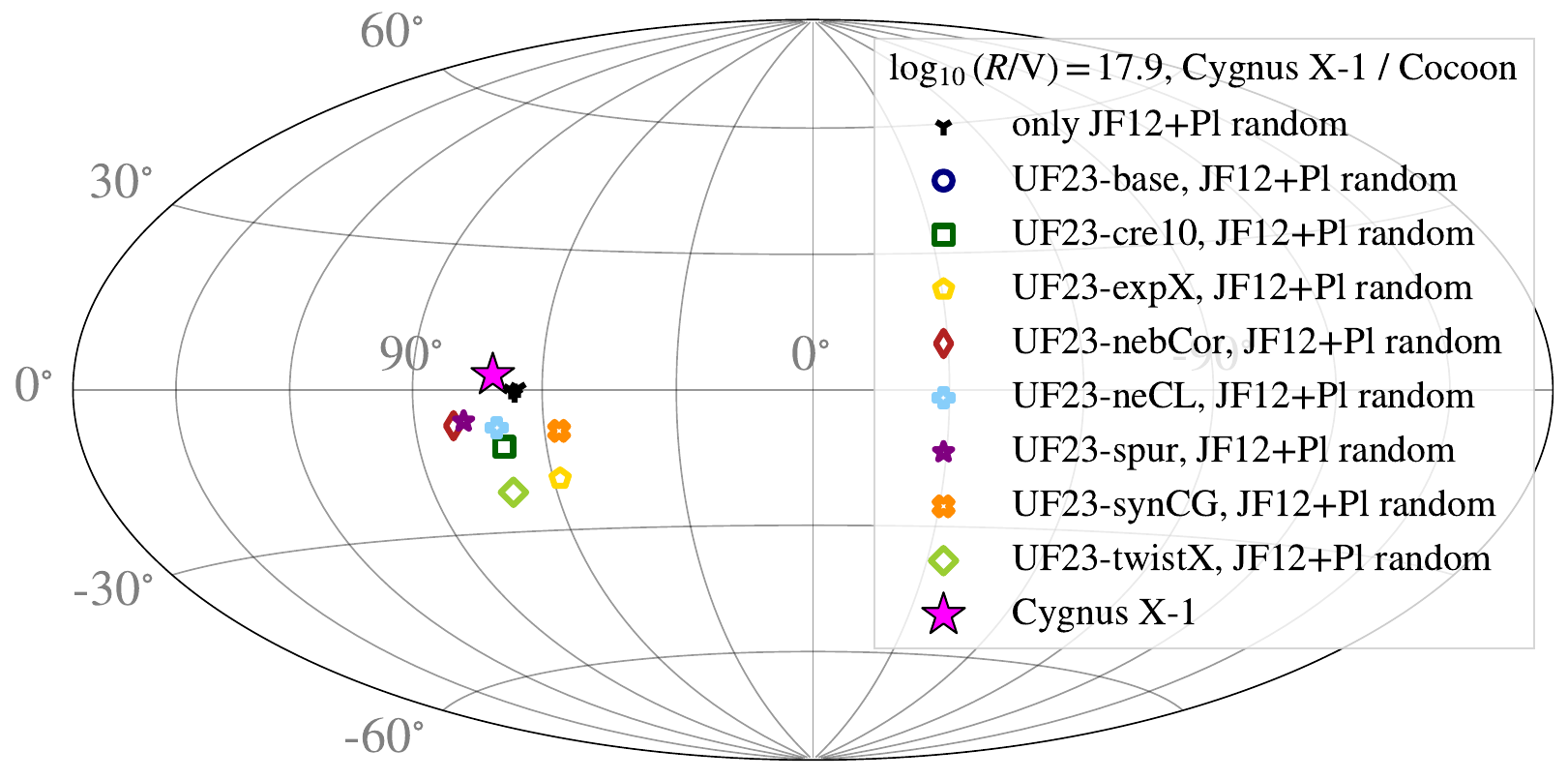}
\includegraphics[width=0.32\textwidth]{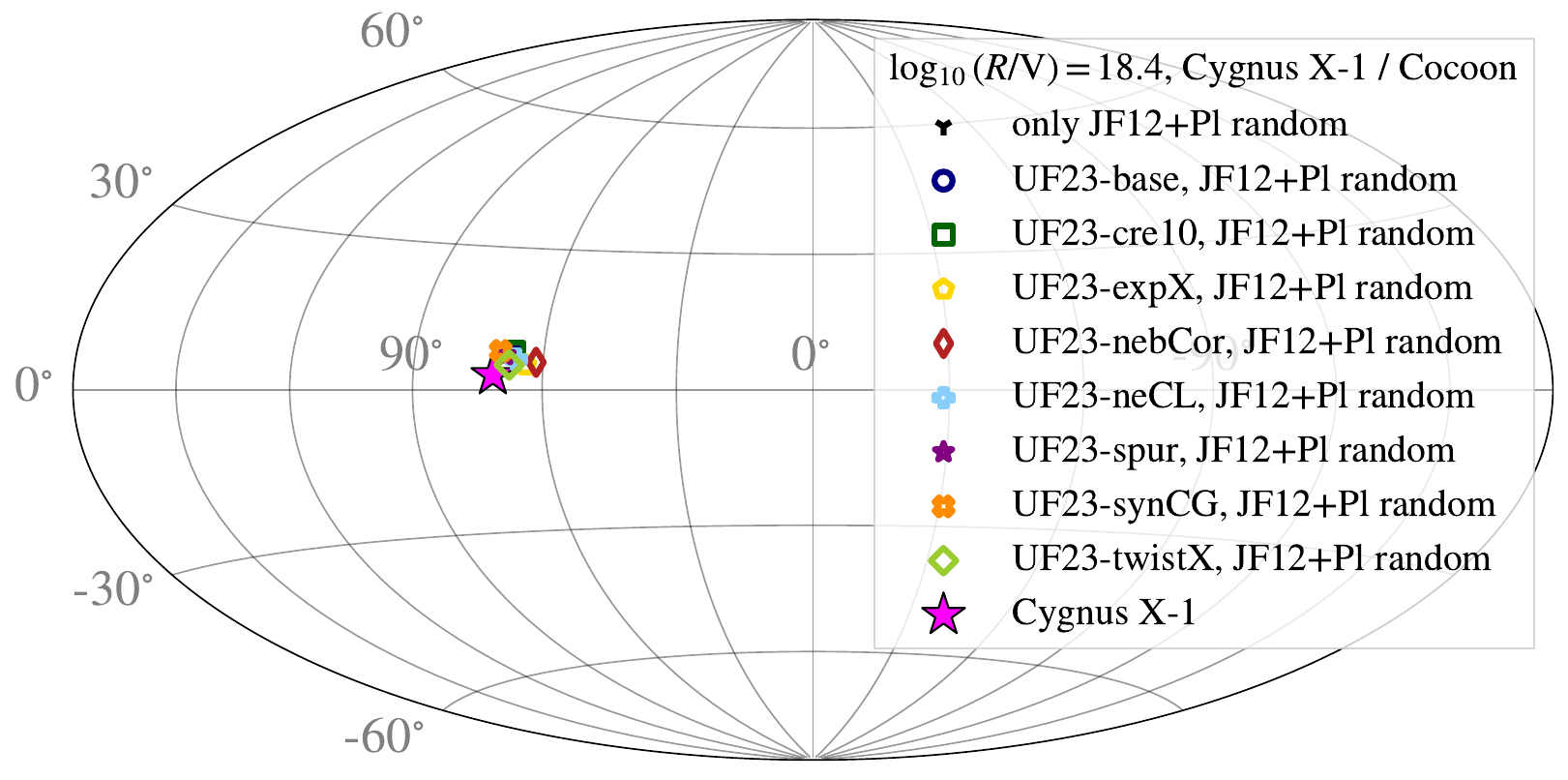}
\includegraphics[width=0.32\textwidth]{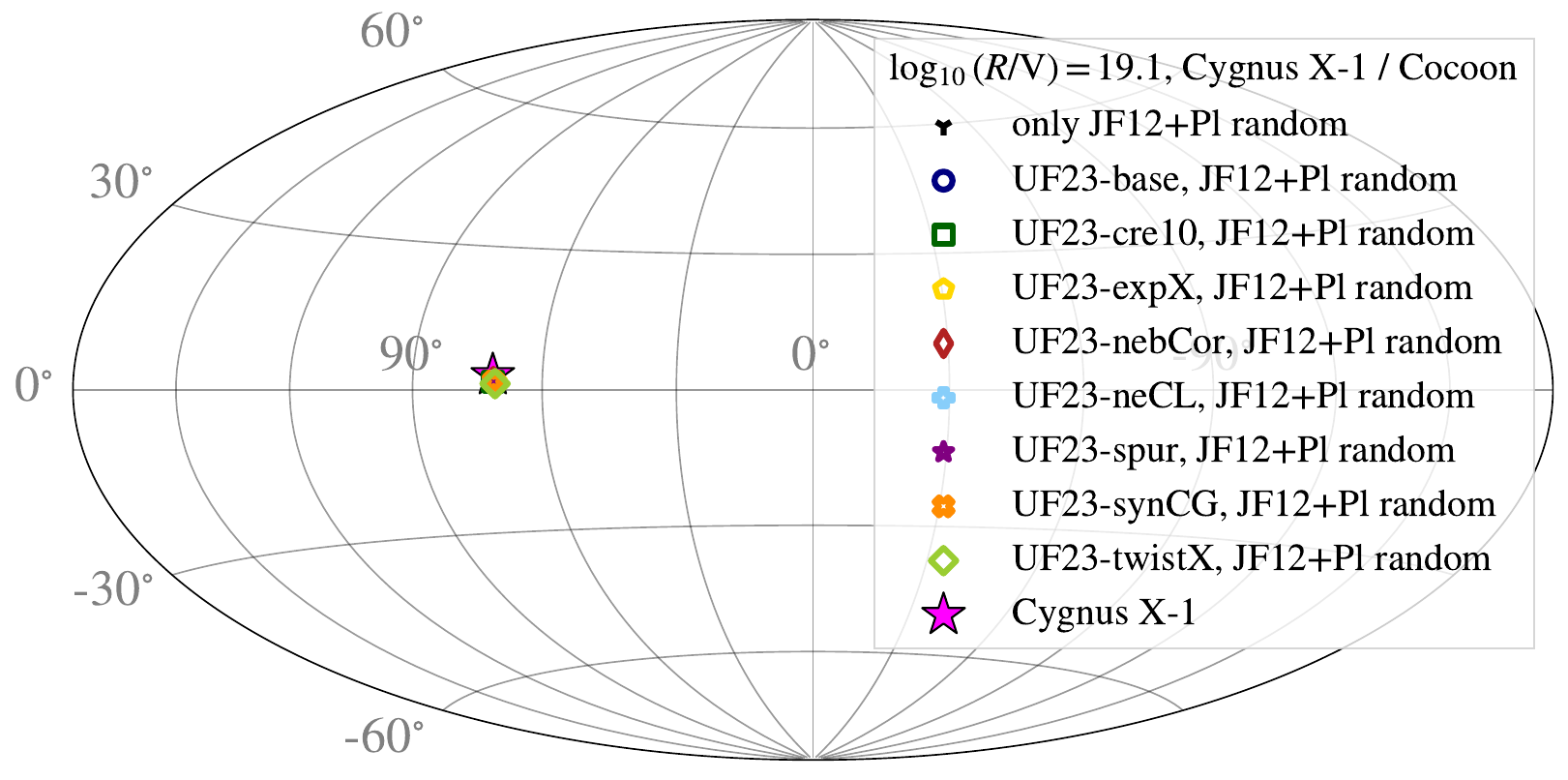}\\
\includegraphics[width=0.32\textwidth]{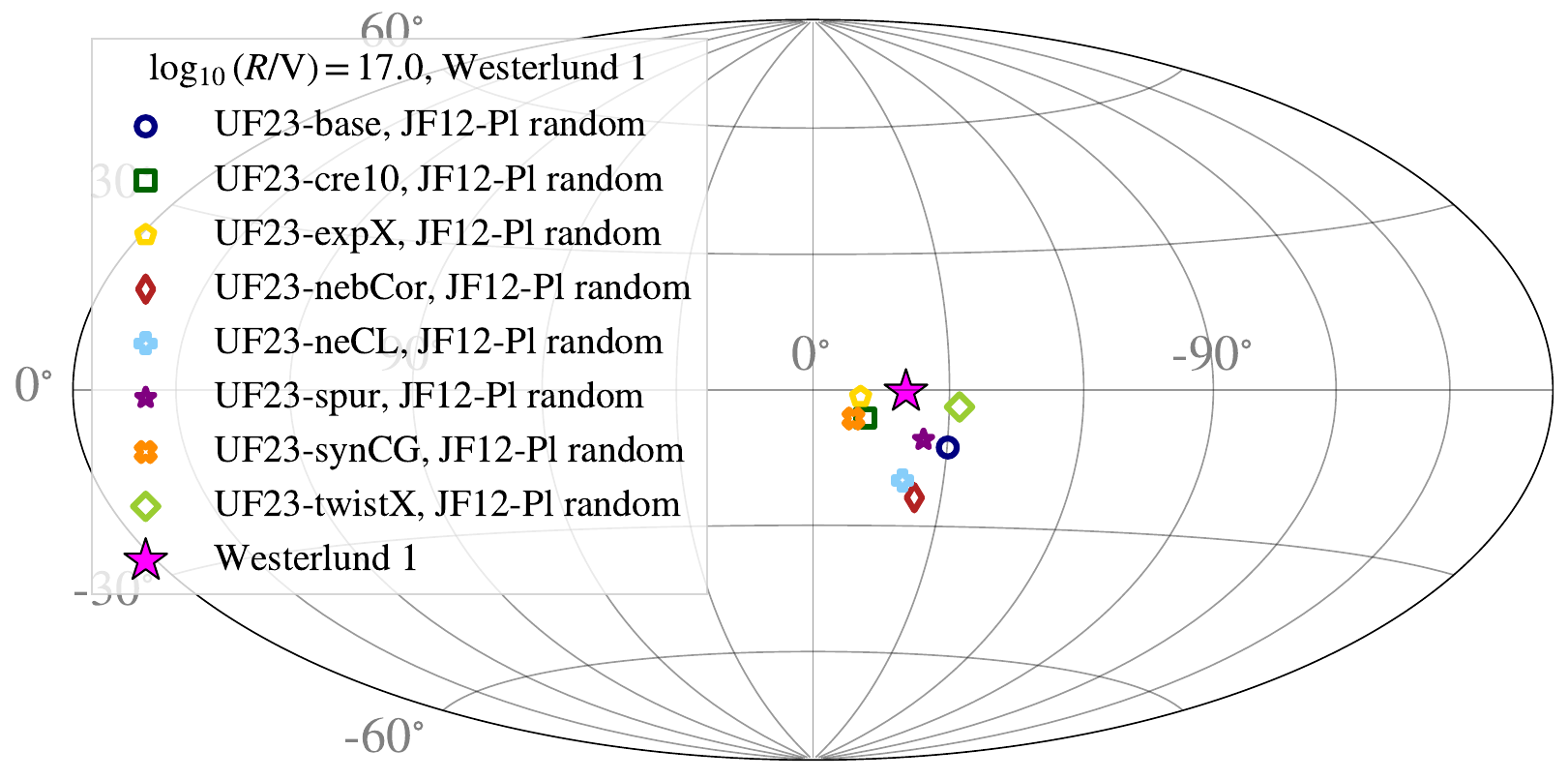}
\includegraphics[width=0.32\textwidth]{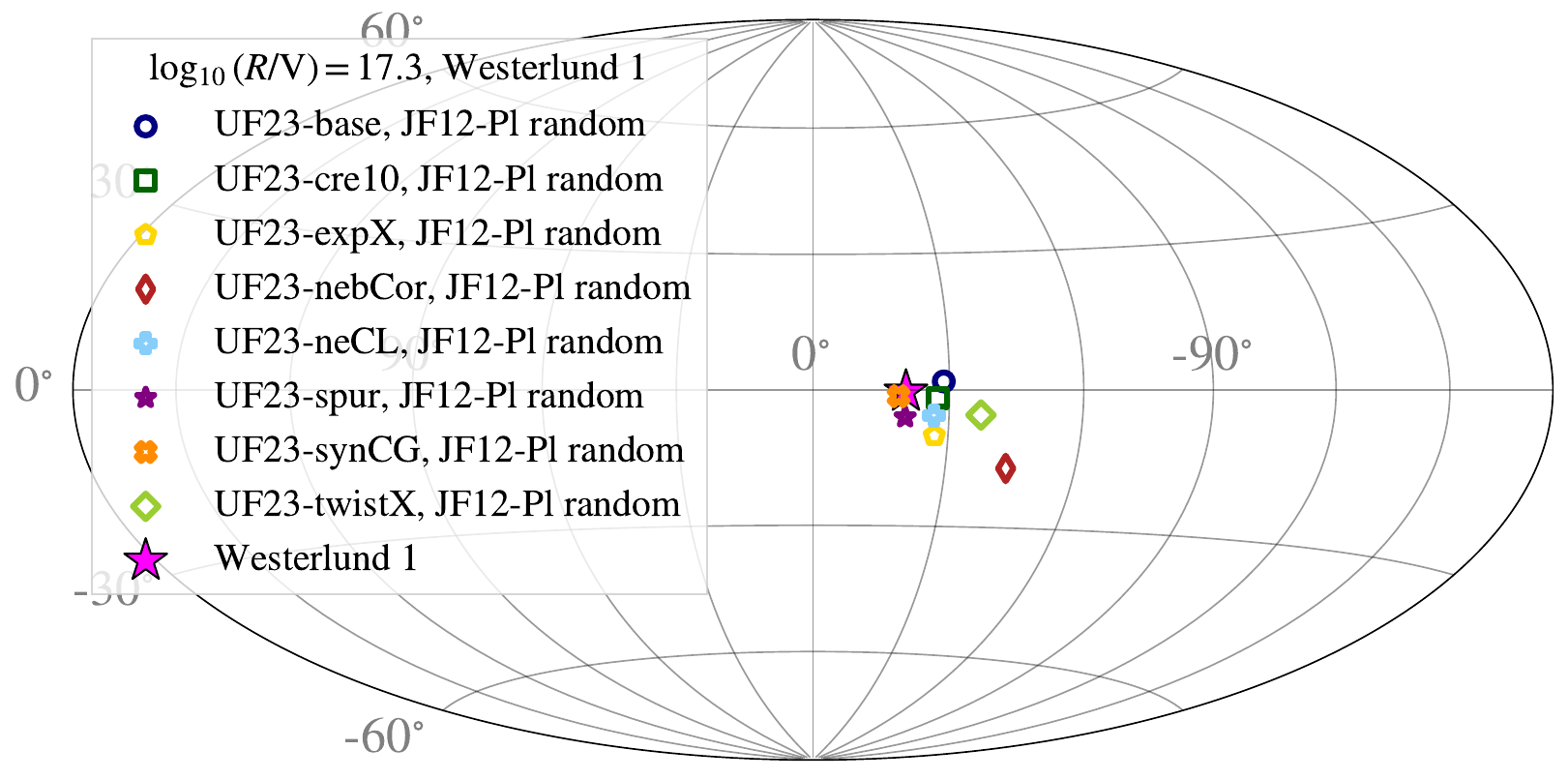}
\includegraphics[width=0.32\textwidth]{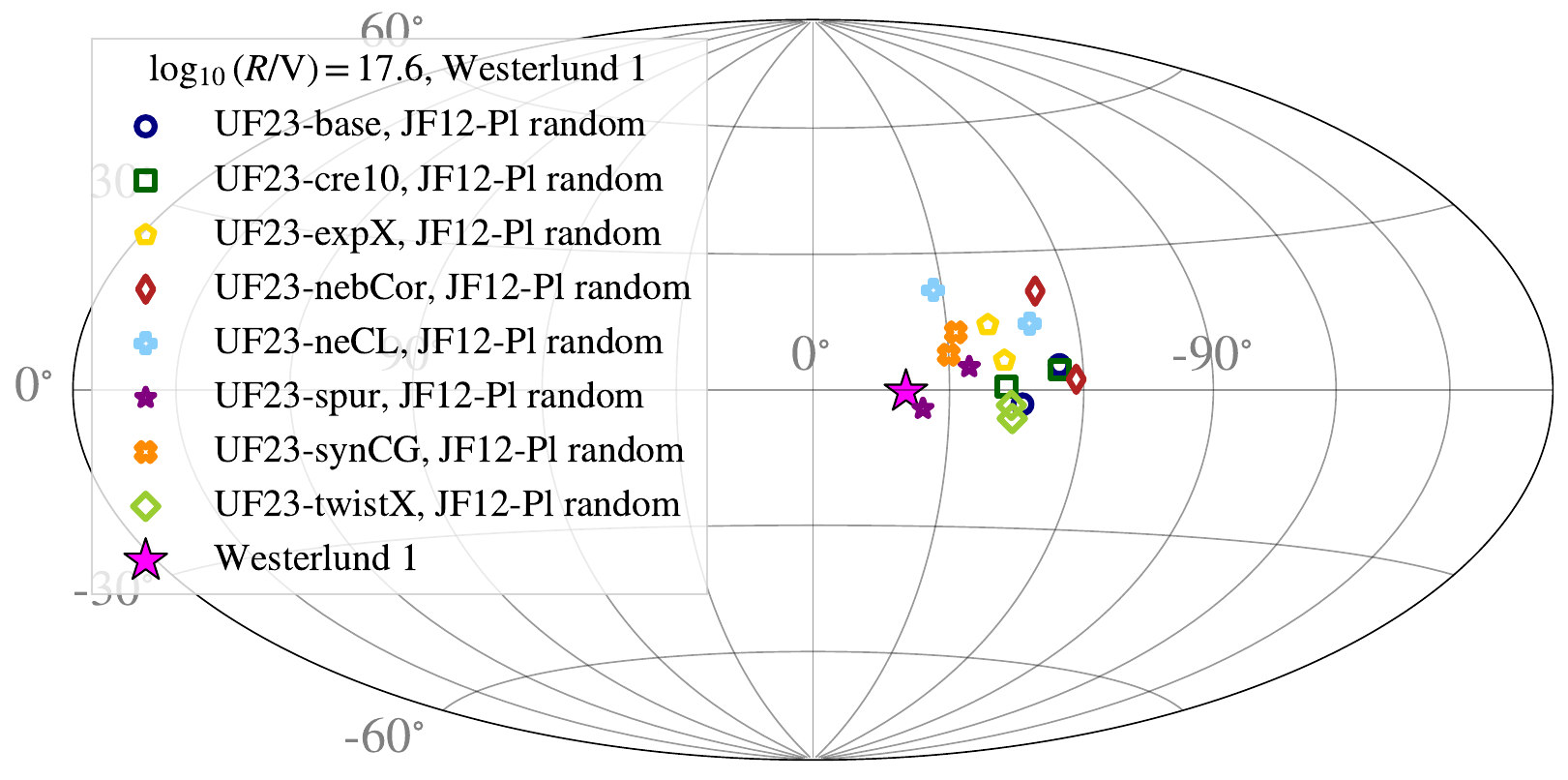}\\
\includegraphics[width=0.32\textwidth]{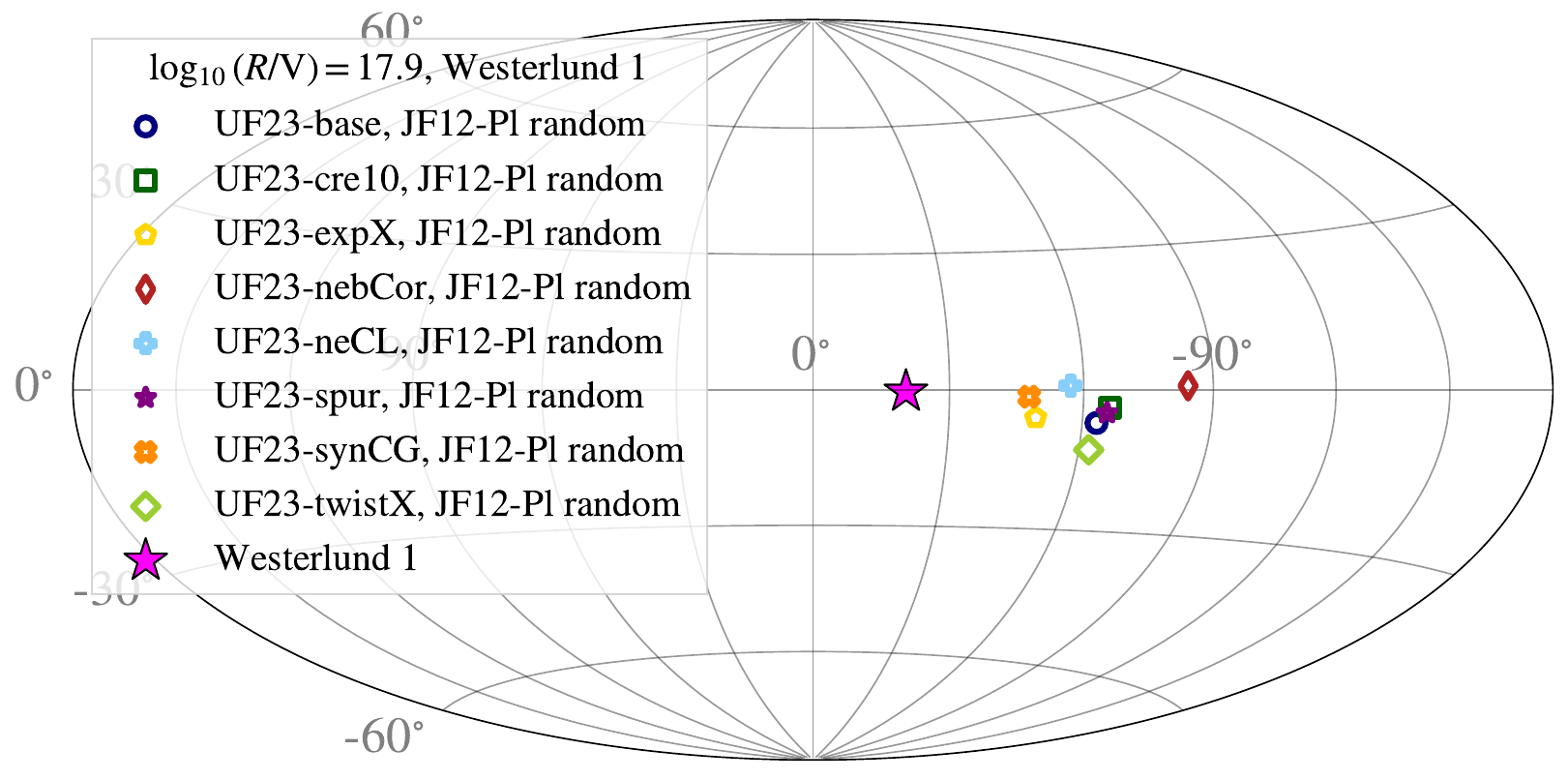}
\includegraphics[width=0.32\textwidth]{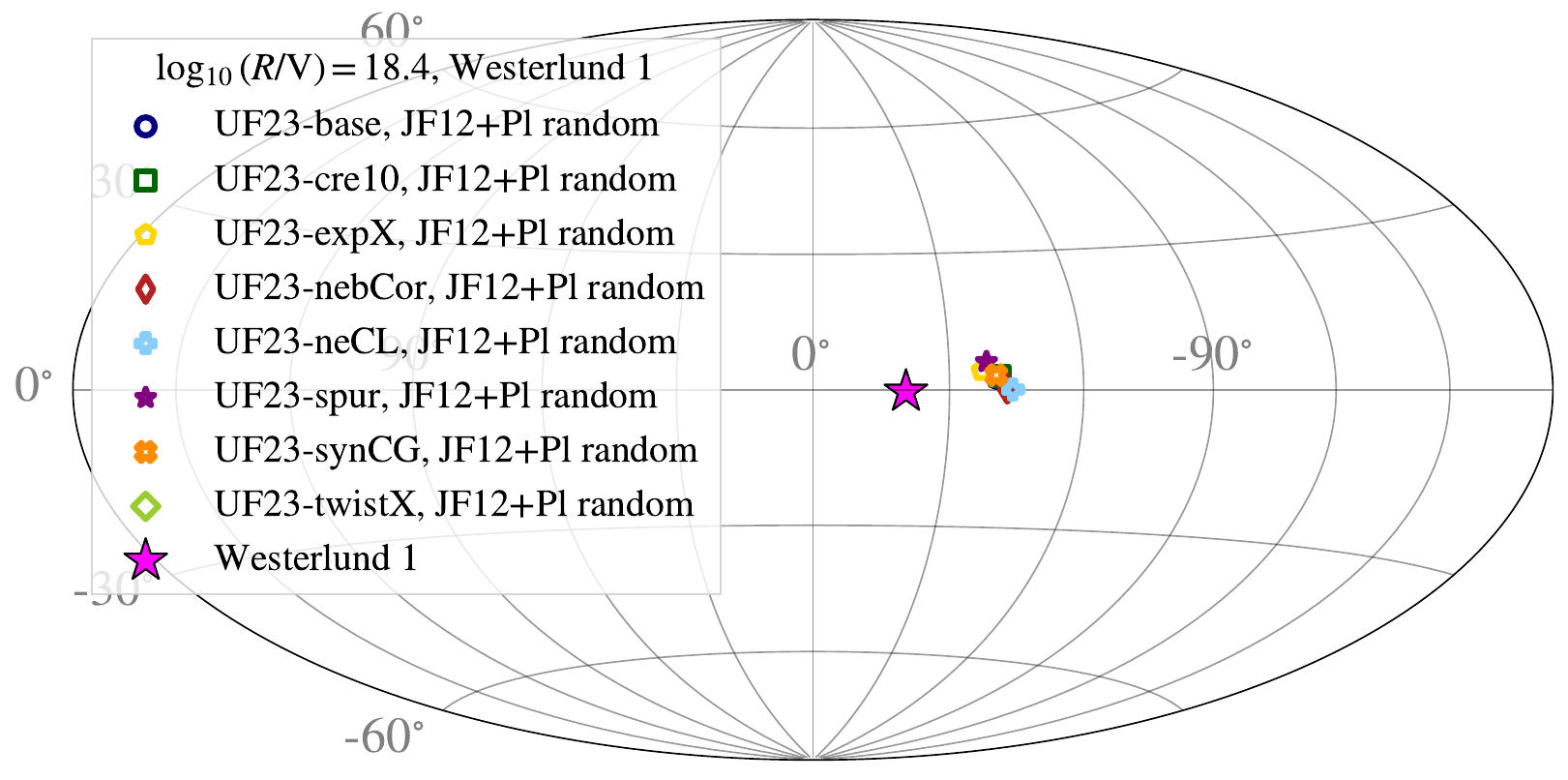}
\includegraphics[width=0.32\textwidth]{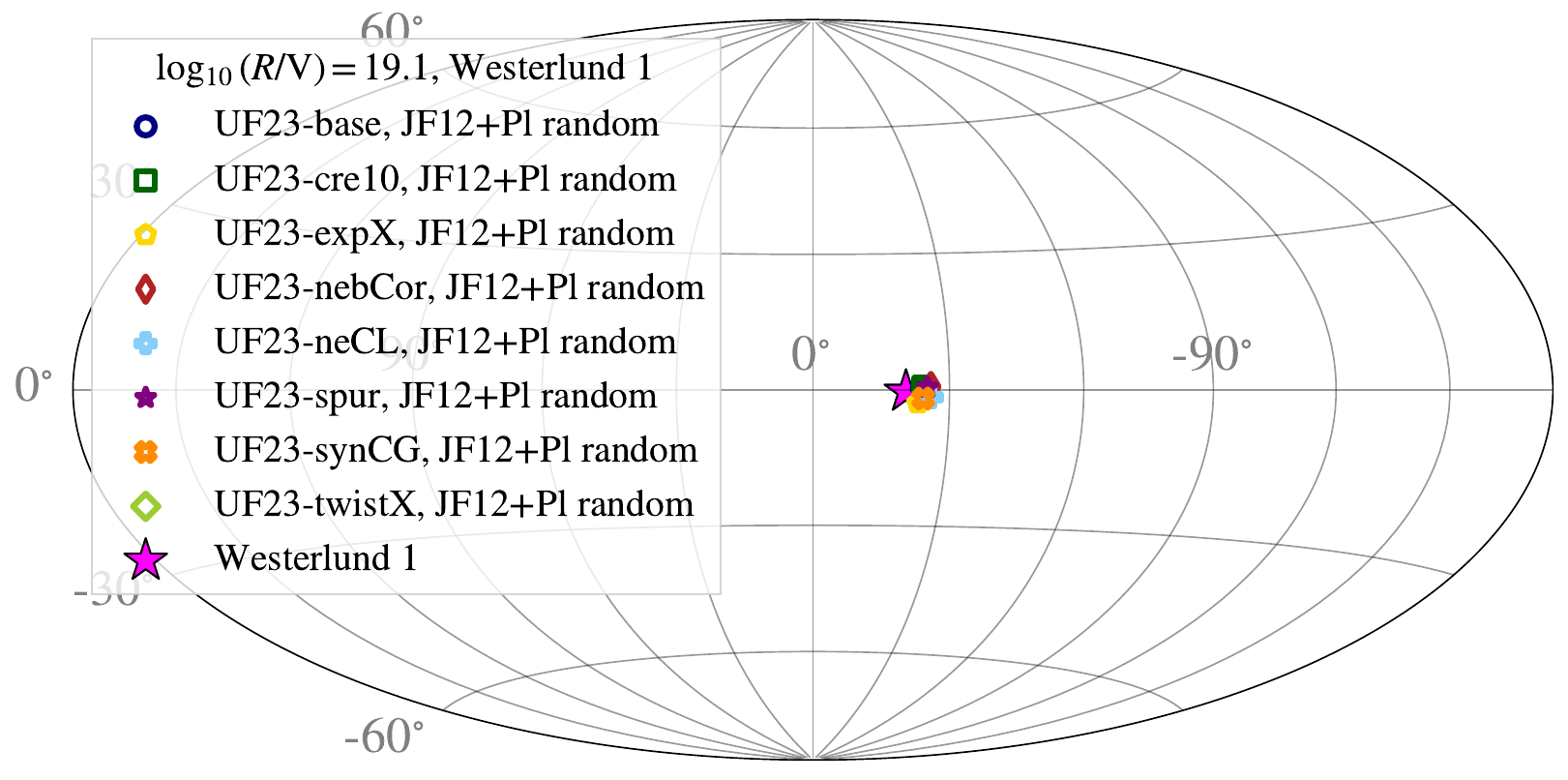}
\caption{Dipole directions for CRs from Cygnus X-1 (first two rows) and Westerlund 1 (last two rows) using the \texttt{JF12+Planck} random GMF model (up to 3 different random seeds) in combination with the \texttt{UF23} coherent field models or no coherent model, for different rigidities as stated in the legends. The pink star represents the direction of the source.}
\label{fig:cygx1_westerlund_direc}
\end{figure}

\begin{acknowledgments}
This research was supported by the Center for Advanced Study in Oslo via a Young CAS Grant awarded to F.O. The authors gratefully acknowledge the funding, administrative support, and hospitality provided during the research stay at the Center. T.B. acknowledges funding by the Dutch Research Council (NWO) under the grant VI.Veni.232.023 (\textit{Search for the sources of ultra-high-energy cosmic rays}).  DFGF acknowledges support from the Italian Ministero dell'Università e della Ricerca through the FIS 3 project FIS-2024-03087 (DD n. 18010 12-11-2025, CUP E53C25002700001), and  by the TAsP (Theoretical Astroparticle Physics) project. \\

We thank everyone who joined us at CAS, Mainak Mukhopadhyay, Michael Unger, Bing Theodore Zhang, Michael Kachelriess, Karri Koljonen, Glennys Farrar, Walter Winter, and Egor Podlesnyi for fruitful discussions and valuable comments. We are also grateful to Roberto Aloisio, Pasquale Blasi, Carmelo Evoli, and Julien Dörner for useful conversations. 

\end{acknowledgments}

\bibliography{bibliography}{}

@article{Li:2026xat,
    author = "Li, Chao-Ming and Taylor, Andrew M.",
    title = "{Galactic Cosmic Ray Transport in the Giant Circumgalactic Medium Halo}",
    eprint = "2606.23317",
    archivePrefix = "arXiv",
    primaryClass = "astro-ph.HE",
    month = "6",
    year = "2026"
}

@article{Gupta:2012rh,
    author = "Gupta, A. and Mathur, S. and Krongold, Y. and Nicastro, F. and Galeazzi, M.",
    title = "{A huge reservoir of ionized gas around the Milky Way: Accounting for the Missing Mass?}",
    eprint = "1205.5037",
    archivePrefix = "arXiv",
    primaryClass = "astro-ph.HE",
    doi = "10.1088/2041-8205/756/1/L8",
    journal = "Astrophys. J. Lett.",
    volume = "756",
    pages = "L8",
    year = "2012"
}

@ARTICLE{2013MNRAS.429L.104Z,
       author = {{Zdziarski}, A.~A. and {Mikolajewska}, J. and {Belczynski}, K.},
        title = "{Cyg X-3: a low-mass black hole or a neutron star.}",
      journal = {\mnras},
         year = 2013,
        month = feb,
       volume = {429},
        pages = {L104-L108},
          doi = {10.1093/mnrasl/sls035},
archivePrefix = {arXiv},
       eprint = {1208.5455},
       adsurl = {https://ui.adsabs.harvard.edu/abs/2013MNRAS.429L.104Z},
}

@article{Zhang:2026igt,
    author = "Zhang, Bing Theodore and Yu, Shiqi",
    title = "{Microquasar Remnants as Pevatrons Illuminating the Galactic Cosmic Ray Knee}",
    eprint = "2602.08940",
    archivePrefix = "arXiv",
    primaryClass = "astro-ph.HE",
    month = "2",
    year = "2026"
}

@article{Russell:2007ig,
    author = "Russell, David M. and Fender, R. P. and Gallo, E. and Kaiser, C. R.",
    title = "{The jet-powered optical nebula of Cygnus X-1}",
    eprint = "astro-ph/0701645",
    archivePrefix = "arXiv",
    doi = "10.1111/j.1365-2966.2007.11539.x",
    journal = "Mon. Not. Roy. Astron. Soc.",
    volume = "376",
    pages = "1341--1349",
    year = "2007"
}

@ARTICLE{2021A&A...649A.176V,
       author = {{Vilhu}, O. and {Kallman}, T.~R. and {Koljonen}, K.~I.~I. and {Hannikainen}, D.~C.},
        title = "{Wind suppression by X-rays in Cygnus X-3}",
      journal = {\aap},
         year = 2021,
        month = may,
       volume = {649},
          eid = {A176},
        pages = {A176},
          doi = {10.1051/0004-6361/202140620},
archivePrefix = {arXiv},
       eprint = {2104.02305},
 primaryClass = {astro-ph.HE},
       adsurl = {https://ui.adsabs.harvard.edu/abs/2021A&A...649A.176V}
}

@ARTICLE{2017MNRAS.467.3544S,
       author = {{Sarkar}, Kartick C. and {Nath}, Biman B. and {Sharma}, Prateek},
        title = "{Clues to the origin of Fermi bubbles from O viii/O vii line ratio}",
      journal = {\mnras},
         year = 2017,
        month = may,
       volume = {467},
       number = {3},
        pages = {3544-3555},
          doi = {10.1093/mnras/stx314},
archivePrefix = {arXiv},
       eprint = {1610.00719},
 primaryClass = {astro-ph.GA},
       adsurl = {https://ui.adsabs.harvard.edu/abs/2017MNRAS.467.3544S}
}

@inproceedings{Hillas:2006ms,
    author = "Hillas, Anthony M.",
    title = "{Cosmic Rays: Recent Progress and some Current Questions}",
    booktitle = "{Conference on Cosmology, Galaxy Formation and Astro-Particle Physics on the Pathway to the SKA}",
    eprint = "astro-ph/0607109",
    archivePrefix = "arXiv",
    month = "7",
    year = "2006"
}

@ARTICLE{2015ApJ...809..143D,
       author = {{Do}, Tuan and {Kerzendorf}, Wolfgang and {Winsor}, Nathan and {St{\o}stad}, Morten and {Morris}, Mark R. and {Lu}, Jessica R. and {Ghez}, Andrea M.},
        title = "{Discovery of Low-metallicity Stars in the Central Parsec of the Milky Way}",
      journal = {\apj},
         year = 2015,
        month = aug,
       volume = {809},
       number = {2},
          eid = {143},
        pages = {143},
          doi = {10.1088/0004-637X/809/2/143},
archivePrefix = {arXiv},
       eprint = {1506.07891},
 primaryClass = {astro-ph.GA},
       adsurl = {https://ui.adsabs.harvard.edu/abs/2015ApJ...809..143D}
}

@article{Evoli:2012ha,
    author = "Evoli, Carmelo and Gaggero, Daniele and Grasso, Dario and Maccione, Luca",
    title = "{A common solution to the cosmic ray anisotropy and gradient problems}",
    eprint = "1203.0570",
    archivePrefix = "arXiv",
    primaryClass = "astro-ph.HE",
    reportNumber = "LMU-ASC-13-12, MPP-2012-50, LAPTH-008-12",
    doi = "10.1103/PhysRevLett.108.211102",
    journal = "Phys. Rev. Lett.",
    volume = "108",
    pages = "211102",
    year = "2012"
}

@article{HESS:2016pst,
    author = "Abramowski, A. and others",
    collaboration = "H.E.S.S.",
    title = "{Acceleration of petaelectronvolt protons in the Galactic Centre}",
    eprint = "1603.07730",
    archivePrefix = "arXiv",
    primaryClass = "astro-ph.HE",
    doi = "10.1038/nature17147",
    journal = "Nature",
    volume = "531",
    pages = "476",
    year = "2016"
}

@article{Kuhlen:2022tov,
    author = "Kuhlen, Marco and Mertsch, Philipp and Phan, Vo Hong Minh",
    title = "{Diffusion of Relativistic Charged Particles and Field Lines in Isotropic Turbulence. I. Numerical Simulations}",
    eprint = "2211.05881",
    archivePrefix = "arXiv",
    primaryClass = "astro-ph.HE",
    reportNumber = "TTK-22-35",
    doi = "10.3847/1538-4357/adee9a",
    journal = "Astrophys. J.",
    volume = "992",
    number = "1",
    pages = "10",
    year = "2025"
}

@ARTICLE{2017MNRAS.464..194F,
       author = {{Feldmeier-Krause}, A. and {Kerzendorf}, W. and {Neumayer}, N. and {Sch{\"o}del}, R. and {Nogueras-Lara}, F. and {Do}, T. and {de Zeeuw}, P.~T. and {Kuntschner}, H.},
        title = "{KMOS view of the Galactic Centre - II. Metallicity distribution of late-type stars}",
      journal = {\mnras},
         year = 2017,
        month = jan,
       volume = {464},
       number = {1},
        pages = {194-209},
          doi = {10.1093/mnras/stw2339},
archivePrefix = {arXiv},
       eprint = {1610.01623},
 primaryClass = {astro-ph.GA},
       adsurl = {https://ui.adsabs.harvard.edu/abs/2017MNRAS.464..194F}
}

@article{Albert:2024aaa,
    author = "Albert, A. and others",
    title = "{Observation of the Galactic Center PeVatron beyond 100 TeV with HAWC}",
    eprint = "2407.03682",
    archivePrefix = "arXiv",
    primaryClass = "astro-ph.HE",
    doi = "10.3847/2041-8213/ad772e",
    journal = "Astrophys. J. Lett.",
    volume = "973",
    number = "1",
    pages = "L34",
    year = "2024"
}

@article{EventHorizonTelescope:2022urf,
    author = "Akiyama, Kazunori and others",
    collaboration = "Event Horizon Telescope",
    title = "{First Sagittarius A* Event Horizon Telescope Results. V. Testing Astrophysical Models of the Galactic Center Black Hole}",
    eprint = "2311.09478",
    archivePrefix = "arXiv",
    primaryClass = "astro-ph.HE",
    reportNumber = "FERMILAB-PUB-22-419-PPD",
    doi = "10.3847/2041-8213/ac6672",
    journal = "Astrophys. J. Lett.",
    volume = "930",
    number = "2",
    pages = "L16",
    year = "2022"
}

@book{Gaisser:2016uoy,
    author = "Gaisser, Thomas K. and Engel, Ralph and Resconi, Elisa",
    title = "{Cosmic Rays and Particle Physics}: {2nd Edition}",
    isbn = "978-0-521-01646-9",
    publisher = "Cambridge University Press",
    month = "6",
    year = "2016"
}

@article{Koljonen:2017gvy,
    author = "Koljonen, Karri I. I. and Maccarone, Thomas J.",
    title = "{Gemini/GNIRS infrared spectroscopy of the Wolf{\textendash}Rayet stellar wind in Cygnus X-3}",
    eprint = "1708.04050",
    archivePrefix = "arXiv",
    primaryClass = "astro-ph.HE",
    doi = "10.1093/mnras/stx2106",
    journal = "Mon. Not. Roy. Astron. Soc.",
    volume = "472",
    number = "2",
    pages = "2181--2195",
    year = "2017"
}

@ARTICLE{2015A&A...579A..75T,
       author = {{Todt}, H. and {Sander}, A. and {Hainich}, R. and {Hamann}, W.-R. and {Quade}, M. and {Shenar}, T.},
        title = "{Potsdam Wolf-Rayet model atmosphere grids for WN stars}",
      journal = {\aap},
         year = 2015,
        month = jul,
       volume = {579},
          eid = {A75},
        pages = {A75},
          doi = {10.1051/0004-6361/201526253},
       adsurl = {https://ui.adsabs.harvard.edu/abs/2015A&A...579A..75T}
}

@article{PierreAuger:2012egl,
    author = "Abreu, Pedro and others",
    collaboration = "Pierre Auger",
    title = "{Measurement of the proton-air cross-section at $\sqrt{s}=57$ TeV with the Pierre Auger Observatory}",
    eprint = "1208.1520",
    archivePrefix = "arXiv",
    primaryClass = "hep-ex",
    reportNumber = "FERMILAB-PUB-12-327-AD-AE-CD-TD",
    doi = "10.1103/PhysRevLett.109.062002",
    journal = "Phys. Rev. Lett.",
    volume = "109",
    pages = "062002",
    year = "2012"
}

@article{Reid:2023ksq,
    author = "Reid, M. J. and Miller-Jones, J. C. A.",
    title = "{On the Distances to the X-Ray Binaries Cygnus X-3 and GRS 1915+105}",
    eprint = "2309.15027",
    archivePrefix = "arXiv",
    primaryClass = "astro-ph.HE",
    doi = "10.3847/1538-4357/acfe0c",
    journal = "Astrophys. J.",
    volume = "959",
    number = "2",
    pages = "85",
    year = "2023"
}

@article{PierreAuger:2021hun,
    author = "Abreu, P. and others",
    collaboration = "Pierre Auger",
    title = "{The energy spectrum of cosmic rays beyond the turn-down around $10^{17}$ eV as measured with the surface detector of the Pierre Auger Observatory}",
    eprint = "2109.13400",
    archivePrefix = "arXiv",
    primaryClass = "astro-ph.HE",
    reportNumber = "FERMILAB-PUB-21-474-AD-AE-SCD-TD",
    doi = "10.1140/epjc/s10052-021-09700-w",
    journal = "Eur. Phys. J. C",
    volume = "81",
    number = "11",
    pages = "966",
    year = "2021"
}

@article{Berezinsky:2002nc,
    author = "Berezinsky, V. and Gazizov, A. Z. and Grigorieva, S. I.",
    title = "{On astrophysical solution to ultrahigh-energy cosmic rays}",
    eprint = "hep-ph/0204357",
    archivePrefix = "arXiv",
    doi = "10.1103/PhysRevD.74.043005",
    journal = "Phys. Rev. D",
    volume = "74",
    pages = "043005",
    year = "2006"
}

@article{Lommen:2005tc,
    author = "Lommen, Dave and Yungelson, Lev and van den Heuvel, Ed and Nelemans, Gijs and Portegies Zwart, Simon",
    title = "{Cygnus X-3 and the problem of the missing Wolf-Rayet x-ray binaries}",
    eprint = "astro-ph/0507304",
    archivePrefix = "arXiv",
    doi = "10.1051/0004-6361:20052824",
    journal = "Astron. Astrophys.",
    volume = "443",
    pages = "231",
    year = "2005"
}

@article{Prevotat:2025qdp,
    author = "Pr{\'e}votat, Cl{\'e}ment and Alves Batista, Rafael and Koldobskiy, Sergey and Semikoz, Dmitri and Di Marco, Gaetano",
    title = "{Probing the transition between Galactic and extragalactic cosmic rays with a multi-messenger approach}",
    doi = "10.22323/1.501.0365",
    journal = "PoS",
    volume = "ICRC2025",
    pages = "365",
    year = "2025"
}

@article{Kaci:2025gyb,
    author = "Kaci, Samy and Giacinti, Gwenael and Aharonian, Felix and Wang, Jie-Shuang",
    title = "{Microquasars as the major contributors to Galactic cosmic rays around the ''knee''}",
    eprint = "2510.01369",
    archivePrefix = "arXiv",
    primaryClass = "astro-ph.HE",
    month = "10",
    year = "2025"
}

@article{Zhang:2025tew,
    author = "Zhang, B. Theodore and Kimura, Shigeo S. and Murase, Kohta",
    title = "{Microquasar jet-cocoon systems as PeVatrons}",
    eprint = "2506.20193",
    archivePrefix = "arXiv",
    primaryClass = "astro-ph.HE",
    doi = "10.1103/p6r1-qg5q",
    journal = "Phys. Rev. D",
    volume = "112",
    number = "12",
    pages = "123015",
    year = "2025"
}

@article{Murase:2008yt,
    author = "Murase, Kohta and Inoue, Susumu and Nagataki, Shigehiro",
    title = "{Cosmic Rays Above the Second Knee from Clusters of Galaxies and Associated High-Energy Neutrino Emission}",
    eprint = "0805.0104",
    archivePrefix = "arXiv",
    primaryClass = "astro-ph",
    doi = "10.1086/595882",
    journal = "Astrophys. J. Lett.",
    volume = "689",
    pages = "L105",
    year = "2008"
}

@article{Wang:2007ya,
    author = "Wang, Xiang-Yu and Razzaque, Soebur and Meszaros, Peter and Dai, Zi-Gao",
    title = "{High-energy Cosmic Rays and Neutrinos from Semi-relativistic Hypernovae}",
    eprint = "0705.0027",
    archivePrefix = "arXiv",
    primaryClass = "astro-ph",
    doi = "10.1103/PhysRevD.76.083009",
    journal = "Phys. Rev. D",
    volume = "76",
    pages = "083009",
    year = "2007"
}

@article{Simeon:2025gxd,
    author = "Simeon, Paul and Globus, No{\'e}mie and Barrow, Kirk S. S. and Blandford, Roger",
    title = "{A Hierarchical Shock Model of Ultra-high-energy Cosmic Rays}",
    eprint = "2503.10795",
    archivePrefix = "arXiv",
    primaryClass = "astro-ph.HE",
    doi = "10.3847/1538-4357/ae2852",
    journal = "Astrophys. J.",
    volume = "998",
    number = "1",
    pages = "164",
    year = "2026"
}

@article{Fang:2017zjf,
    author = "Fang, Ke and Murase, Kohta",
    title = "{Linking High-Energy Cosmic Particles by Black Hole Jets Embedded in Large-Scale Structures}",
    eprint = "1704.00015",
    archivePrefix = "arXiv",
    primaryClass = "astro-ph.HE",
    doi = "10.1038/s41567-017-0025-4",
    journal = "Nature Phys.",
    volume = "14",
    number = "4",
    pages = "396--398",
    year = "2018"
}

@article{Porcelli:2015jli,
    author = "Porcelli, Alessio",
    title = "{Measurements of the first two moments of the depth of shower maximum over nearly three decades of energy, combining data from}",
    doi = "10.22323/1.236.0420",
    journal = "PoS",
    volume = "ICRC2015",
    pages = "420",
    year = "2016"
}

@article{Rodrigues:2018bjg,
    author = "Rodrigues, Xavier and Biehl, Daniel and Boncioli, Denise and Taylor, Andrew M.",
    title = "{Binary neutron star merger remnants as sources of cosmic rays below the {\textquotedblleft}Ankle{\textquotedblright}}",
    eprint = "1806.01624",
    archivePrefix = "arXiv",
    primaryClass = "astro-ph.HE",
    reportNumber = "DESY-18-097",
    doi = "10.1016/j.astropartphys.2018.10.007",
    journal = "Astropart. Phys.",
    volume = "106",
    pages = "10--17",
    year = "2019"
}

@article{IceCube:2019hmk,
    author = "Aartsen, M. G. and others",
    collaboration = "IceCube",
    title = "{Cosmic ray spectrum and composition from PeV to EeV using 3 years of data from IceTop and IceCube}",
    eprint = "1906.04317",
    archivePrefix = "arXiv",
    primaryClass = "astro-ph.HE",
    doi = "10.1103/PhysRevD.100.082002",
    journal = "Phys. Rev. D",
    volume = "100",
    number = "8",
    pages = "082002",
    year = "2019"
}

@article{Maris:2023anl,
    author = "Maris, Ioana Codrina and Gonzalez, Nicolas Martin",
    collaboration = "Pierre Auger",
    title = "{On the possibility to measure galactic photons at the altitude of the Pierre Auger Observatory}",
    doi = "10.22323/1.444.0718",
    journal = "PoS",
    volume = "ICRC2023",
    pages = "718",
    year = "2023"
}

@article{SWGO:2025taj,
    author = "Abreu, P. and others",
    collaboration = "SWGO",
    title = "{Science Prospects for the Southern Wide-field Gamma-ray Observatory: SWGO}",
    eprint = "2506.01786",
    archivePrefix = "arXiv",
    primaryClass = "astro-ph.HE",
    month = "6",
    year = "2025"
}

@article{Vecchiotti:2026okk,
    author = "Vecchiotti, V. and Amato, E. and Giacinti, G. and Morlino, G. and Peron, G.",
    title = "{Constraints on Hadronic Emission from Microquasars Detected by LHAASO}",
    eprint = "2606.29830",
    archivePrefix = "arXiv",
    primaryClass = "astro-ph.HE",
    month = "6",
    year = "2026"
}

@article{Morlino1:2021zwu,
    author = "Morlino 1, G. and Blasi, P. and Peretti, E. and Cristofari, P.",
    title = "{Particle acceleration in winds of star clusters}",
    eprint = "2102.09217",
    archivePrefix = "arXiv",
    primaryClass = "astro-ph.HE",
    doi = "10.1093/mnras/stab690",
    journal = "Mon. Not. Roy. Astron. Soc.",
    volume = "504",
    number = "4",
    pages = "6096--6105",
    year = "2021"
}

@article{Giacinti:2011Transition,
    author        = "Giacinti, G. and Kachelrie{\ss}, M. and Semikoz, D. V. and Sigl, G.",
    title         = "{Cosmic ray anisotropy as signature for the transition from Galactic to extragalactic cosmic rays}",
    journal       = "Journal of Cosmology and Astroparticle Physics",
    volume        = "2012",
    number        = "07",
    pages         = "031",
    year          = "2012",
    doi           = "10.1088/1475-7516/2012/07/031",
    eprint        = "1112.5599",
    archivePrefix = "arXiv",
    primaryClass  = "astro-ph.HE",
    url           = "https://arxiv.org/abs/1112.5599"
}

@article{Haverkorn:2008tb,
    author = "Haverkorn, M. and Brown, J. C. and Gaensler, B. M. and McClure-Griffiths, N. M.",
    title = "{The outer scale of turbulence in the magneto-ionized Galactic interstellar medium}",
    eprint = "0802.2740",
    archivePrefix = "arXiv",
    primaryClass = "astro-ph",
    doi = "10.1086/587165",
    journal = "Astrophys. J.",
    volume = "680",
    pages = "362",
    year = "2008"
}

@article{Beck:2014pma,
    author = "Beck, Marcus C. and Beck, Alexander M. and Beck, Rainer and Dolag, Klaus and Strong, Andrew W. and Nielaba, Peter",
    title = "{New constraints on modelling the random magnetic field of the MW}",
    eprint = "1409.5120",
    archivePrefix = "arXiv",
    primaryClass = "astro-ph.GA",
    doi = "10.1088/1475-7516/2016/05/056",
    journal = "JCAP",
    volume = "05",
    pages = "056",
    year = "2016"
}

@article{Mertsch:2013pua,
    author = "Mertsch, Philipp and Sarkar, Subir",
    title = "{Loops and spurs: The angular power spectrum of the Galactic synchrotron background}",
    eprint = "1304.1078",
    archivePrefix = "arXiv",
    primaryClass = "astro-ph.GA",
    doi = "10.1088/1475-7516/2013/06/041",
    journal = "JCAP",
    volume = "06",
    pages = "041",
    year = "2013"
}

@article{Giacinti:2017dgt,
    author = "Giacinti, G. and Kachelriess, M. and Semikoz, D. V.",
    title = "{Reconciling cosmic ray diffusion with Galactic magnetic field models}",
    eprint = "1710.08205",
    archivePrefix = "arXiv",
    primaryClass = "astro-ph.HE",
    doi = "10.1088/1475-7516/2018/07/051",
    journal = "JCAP",
    volume = "07",
    pages = "051",
    year = "2018"
}

@article{Apel:2013ura,
    author = "Apel, W. D. and others",
    title = "{Ankle-like Feature in the Energy Spectrum of Light Elements of Cosmic Rays Observed with KASCADE-Grande}",
    eprint = "1304.7114",
    archivePrefix = "arXiv",
    primaryClass = "astro-ph.HE",
    doi = "10.1103/PhysRevD.87.081101",
    journal = "Phys. Rev. D",
    volume = "87",
    pages = "081101",
    year = "2013"
}

@article{KASCADE:2005ynk,
    author = "Antoni, T. and others",
    collaboration = "KASCADE",
    title = "{KASCADE measurements of energy spectra for elemental groups of cosmic rays: Results and open problems}",
    eprint = "astro-ph/0505413",
    archivePrefix = "arXiv",
    doi = "10.1016/j.astropartphys.2005.04.001",
    journal = "Astropart. Phys.",
    volume = "24",
    pages = "1--25",
    year = "2005"
}

@article{Peters:1961mxb,
    author = "Peters, B.",
    title = "{Primary cosmic radiation and extensive air showers}",
    doi = "10.1007/bf02783106",
    journal = "Nuovo Cim.",
    volume = "22",
    number = "4",
    pages = "800--819",
    year = "1961"
}

@article{LHAASO:2024knt,
    author = "Cao, Zhen and others",
    collaboration = "LHAASO",
    title = "{Measurements of All-Particle Energy Spectrum and Mean Logarithmic Mass of Cosmic Rays from 0.3 to 30~PeV with LHAASO-KM2A}",
    eprint = "2403.10010",
    archivePrefix = "arXiv",
    primaryClass = "astro-ph.HE",
    doi = "10.1103/PhysRevLett.132.131002",
    journal = "Phys. Rev. Lett.",
    volume = "132",
    number = "13",
    pages = "131002",
    year = "2024"
}

@article{Apel:2013uni,
    author = "Apel, W. D. and others",
    title = "{KASCADE-Grande measurements of energy spectra for elemental groups of cosmic rays}",
    eprint = "1306.6283",
    archivePrefix = "arXiv",
    primaryClass = "astro-ph.HE",
    doi = "10.1016/j.astropartphys.2013.06.004",
    journal = "Astropart. Phys.",
    volume = "47",
    pages = "54--66",
    year = "2013"
}

@article{Kulikov:1958zh,
  author    = {Kulikov, G. V. and Khristiansen, G. B.},
  title     = {On the Size Spectrum of Extensive Air Showers},
  journal   = {Zh. Eksper. Teor. Fiz.},
  volume    = {35},
  number    = {3},
  pages     = {635--640},
  year      = {1958}
}

@article{Unger:2026yau,
    author = "Unger, Michael and Farrar, Glennys R.",
    title = "{The Random Magnetic Field of the Milky Way}",
    eprint = "2608.21293",
    archivePrefix = "arXiv",
    primaryClass = "astro-ph.GA",
    month = "8",
    year = "2026"
}

@article{Becker:2024aqj,
    author = "Becker, C. T. and Kachelriess, M.",
    title = "{Polarized synchrotron data and the structure of the Galactic magnetic field}",
    eprint = "2408.02554",
    archivePrefix = "arXiv",
    primaryClass = "astro-ph.GA",
    doi = "10.1103/PhysRevD.111.063057",
    journal = "Phys. Rev. D",
    volume = "111",
    number = "6",
    pages = "063057",
    year = "2025"
}

@article{Zhang:2018agl,
    author = "Zhang, B. Theodore and Murase, Kohta",
    title = "{Ultrahigh-energy cosmic-ray nuclei and neutrinos from engine-driven supernovae}",
    eprint = "1812.10289",
    archivePrefix = "arXiv",
    primaryClass = "astro-ph.HE",
    doi = "10.1103/PhysRevD.100.103004",
    journal = "Phys. Rev. D",
    volume = "100",
    number = "10",
    pages = "103004",
    year = "2019"
}

@article{Plotko:2024gop,
    author = "Plotko, Pavlo and Winter, Walter and Lunardini, Cecilia and Yuan, Chengchao",
    title = "{Ultrahigh-energy Cosmic Rays from Neutrino-emitting Tidal Disruption Events}",
    eprint = "2410.19047",
    archivePrefix = "arXiv",
    primaryClass = "astro-ph.HE",
    doi = "10.3847/1538-4357/adef3b",
    journal = "Astrophys. J.",
    volume = "994",
    number = "2",
    pages = "251",
    year = "2025"
}

@article{Farrar:2025kpk,
    author = "Farrar, Glennys R.",
    title = "{Ultra-High-energy Cosmic Ray Production in Binary Neutron Star Mergers}",
    eprint = "2506.22625",
    archivePrefix = "arXiv",
    primaryClass = "astro-ph.HE",
    doi = "10.3847/2041-8213/ae14d5",
    journal = "Astrophys. J. Lett.",
    volume = "994",
    number = "1",
    pages = "L7",
    year = "2025"
}

@article{Farrar:2024zsm,
    author = "Farrar, Glennys R.",
    title = "{Binary Neutron Star Mergers as the Source of the Highest Energy Cosmic Rays}",
    eprint = "2405.12004",
    archivePrefix = "arXiv",
    primaryClass = "astro-ph.HE",
    doi = "10.1103/PhysRevLett.134.081003",
    journal = "Phys. Rev. Lett.",
    volume = "134",
    number = "8",
    pages = "081003",
    year = "2025"
}

@article{Zhang:2017hom,
    author = "Zhang, B. Theodore and Murase, Kohta and Oikonomou, Foteini and Li, Zhuo",
    title = "{High-energy cosmic ray nuclei from tidal disruption events: Origin, survival, and implications}",
    eprint = "1706.00391",
    archivePrefix = "arXiv",
    primaryClass = "astro-ph.HE",
    doi = "10.1103/PhysRevD.96.063007",
    journal = "Phys. Rev. D",
    volume = "96",
    number = "6",
    pages = "063007",
    year = "2017",
    note = "[Addendum: Phys.Rev.D 96, 069902 (2017)]"
}

@article{Zhang:2017moz,
    author = "Zhang, B. Theodore and Murase, Kohta and Kimura, Shigeo S. and Horiuchi, Shunsaku and M{\'e}sz{\'a}ros, Peter",
    title = "{Low-luminosity gamma-ray bursts as the sources of ultrahigh-energy cosmic ray nuclei}",
    eprint = "1712.09984",
    archivePrefix = "arXiv",
    primaryClass = "astro-ph.HE",
    doi = "10.1103/PhysRevD.97.083010",
    journal = "Phys. Rev. D",
    volume = "97",
    number = "8",
    pages = "083010",
    year = "2018"
}

@article{Ehlert:2022jmy,
    author = "Ehlert, Domenik and Oikonomou, Foteini and Unger, Michael",
    title = "{Curious case of the maximum rigidity distribution of cosmic-ray accelerators}",
    eprint = "2207.10691",
    archivePrefix = "arXiv",
    primaryClass = "astro-ph.HE",
    doi = "10.1103/PhysRevD.107.103045",
    journal = "Phys. Rev. D",
    volume = "107",
    number = "10",
    pages = "103045",
    year = "2023"
}

@article{Ehlert:2024lji,
    author = "Ehlert, Domenik and Oikonomou, Foteini and Peretti, Enrico",
    title = "{Ultra-high-energy cosmic rays from ultra-fast outflows of active galactic nuclei}",
    eprint = "2411.05667",
    archivePrefix = "arXiv",
    primaryClass = "astro-ph.HE",
    doi = "10.1093/mnras/staf457",
    journal = "Mon. Not. Roy. Astron. Soc.",
    volume = "539",
    number = "3",
    pages = "2435--2462",
    year = "2025"
}

@article{Gaisser:2013bla,
    author = "Gaisser, Thomas K. and Stanev, Todor and Tilav, Serap",
    title = "{Cosmic Ray Energy Spectrum from Measurements of Air Showers}",
    eprint = "1303.3565",
    archivePrefix = "arXiv",
    primaryClass = "astro-ph.HE",
    doi = "10.1007/s11467-013-0319-7",
    journal = "Front. Phys. (Beijing)",
    volume = "8",
    pages = "748--758",
    year = "2013"
}

@article{Parizot:2014ixa,
    author = "Parizot, Etienne",
    editor = {Tibolla, Omar and Drury, Luke and Persic, Massimo and Kaufmann, Sarah and V{\"o}lk, Heinz and Mannheim, Karl and De Angelis, Alessandro},
    title = "{Cosmic Ray Origin: Lessons from Ultra-High-Energy Cosmic Rays and the Galactic/Extragalactic Transition}",
    eprint = "1410.2655",
    archivePrefix = "arXiv",
    primaryClass = "astro-ph.HE",
    doi = "10.1016/j.nuclphysbps.2014.10.023",
    journal = "Nucl. Phys. B Proc. Suppl.",
    volume = "256-257",
    pages = "197--212",
    year = "2014"
}

@article{Thoudam:2016syr,
    author = {Thoudam, S. and Rachen, J. P. and van Vliet, A. and Achterberg, A. and Buitink, S. and Falcke, H. and H{\"o}randel, J. R.},
    title = "{Cosmic-ray energy spectrum and composition up to the ankle: the case for a second Galactic component}",
    eprint = "1605.03111",
    archivePrefix = "arXiv",
    primaryClass = "astro-ph.HE",
    doi = "10.1051/0004-6361/201628894",
    journal = "Astron. Astrophys.",
    volume = "595",
    pages = "A33",
    year = "2016"
}

@article{Bhadra_2024,
   title={Between the Cosmic-Ray “Knee” and the “Ankle”: Contribution from Star Clusters},
   volume={961},
   ISSN={1538-4357},
   url={http://dx.doi.org/10.3847/1538-4357/ad1605},
   DOI={10.3847/1538-4357/ad1605},
   number={2},
   journal={The Astrophysical Journal},
   publisher={American Astronomical Society},
   author={Bhadra, Sourav and Thoudam, Satyendra and Nath, Biman B and Sharma, Prateek},
   year={2024},
   month=Jan, pages={215} }

@article{Kachelriess:2019oqu,
    author = "Kachelriess, M. and Semikoz, D. V.",
    title = "{Cosmic Ray Models}",
    eprint = "1904.08160",
    archivePrefix = "arXiv",
    primaryClass = "astro-ph.HE",
    doi = "10.1016/j.ppnp.2019.07.002",
    journal = "Prog. Part. Nucl. Phys.",
    volume = "109",
    pages = "103710",
    year = "2019"
}

@article{LHAASO:2024_Cygnus_Cocoon,
   title={An ultrahigh-energy $\gamma$-ray bubble powered by a super PeVatron},
   volume={69},
   ISSN={2095-9273},
   url={http://dx.doi.org/10.1016/j.scib.2023.12.040},
   DOI={10.1016/j.scib.2023.12.040},
   number={4},
   journal={Science Bulletin},
   author = "Cao, Zhen and others",
   collaboration = "LHAASO",
   year={2024},
   month=Feb, pages={449–457} }

@article{LHAASO:2023rpg,
    author = "Cao, Zhen and others",
    collaboration = "LHAASO",
    title = "{The First LHAASO Catalog of Gamma-Ray Sources}",
    eprint = "2305.17030",
    archivePrefix = "arXiv",
    primaryClass = "astro-ph.HE",
    doi = "10.3847/1538-4365/acfd29",
    journal = "Astrophys. J. Suppl.",
    volume = "271",
    number = "1",
    pages = "25",
    year = "2024"
}

@article{Miller-Jones:2021plh,
    author = "Miller-Jones, James C. A. and others",
    title = "{Cygnus X-1 contains a 21{\textendash}solar mass black hole{\textemdash}Implications for massive star winds}",
    eprint = "2102.09091",
    archivePrefix = "arXiv",
    primaryClass = "astro-ph.HE",
    doi = "10.1126/science.abb3363",
    journal = "Science",
    volume = "371",
    number = "6533",
    pages = "1046--1049",
    year = "2021"
}

@article{LHAASO:2024psv,
    author = "Cao, Zhen and others",
    collaboration = "LHAASO",
    title = "{Ultrahigh-energy gamma-ray emission associated with black hole{\textendash}jet systems}",
    eprint = "2410.08988",
    archivePrefix = "arXiv",
    primaryClass = "astro-ph.HE",
    doi = "10.1093/nsr/nwaf496",
    journal = "Natl. Sci. Rev.",
    volume = "12",
    number = "12",
    pages = "nwaf496",
    year = "2025"
}

@article{LHAASO:2025ysm,
    author = "Cao, Zhen and others",
    collaboration = "LHAASO",
    title = "{Cygnus X-3: A variable petaelectronvolt gamma-ray source}",
    eprint = "2512.16638",
    archivePrefix = "arXiv",
    primaryClass = "astro-ph.HE",
    month = "12",
    year = "2025"
}

@article{HES:2022_Westerlund1,
   title={A deep spectromorphological study of the<i>γ</i>-ray emission surrounding the young massive stellar cluster Westerlund 1},
   volume={666},
   ISSN={1432-0746},
   url={http://dx.doi.org/10.1051/0004-6361/202244323},
   DOI={10.1051/0004-6361/202244323},
   journal={Astronomy \& Astrophysics},
   publisher={EDP Sciences},
   author={Aharonian, F. and others},
   collaboration = "HESS",
   year={2022},
   month=Oct, pages={A124} }

@article{Tinyakov:2016,
   title={A signature of EeV protons of Galactic origin},
   volume={460},
   ISSN={1365-2966},
   url={http://dx.doi.org/10.1093/mnras/stw1163},
   DOI={10.1093/mnras/stw1163},
   number={4},
   journal={Monthly Notices of the Royal Astronomical Society},
   publisher={Oxford University Press (OUP)},
   author={Tinyakov, P. G. and Urban, F. R. and Ivanov, D. and Thomson, G. B. and Tirone, A. H.},
   year={2016},
   month=May, pages={3479–3487} }

@article{Golup_ICRC2025,
  author = {{G. Golup for the Pierre Auger Collaboration}},
  title = "{A composition-informed search for large-scale anisotropy with the Pierre Auger Observatory}",
  doi = "10.22323/1.501.0272",
  journal = "PoS",
  year = 2025,
  volume = "ICRC2025",
  pages = "272"
}

@article{Kaapa:2021icrc,
    author        = "K{\"a}{\"a}p{\"a}, Alex and Kampert, Karl-Heinz and Mayotte, Eric William",
    collaboration = "Pierre Auger Collaboration",
    title         = "{Propagation of extragalactic cosmic rays in the Galactic magnetic field}",
    journal       = "Proceedings of Science",
    volume        = "ICRC2021",
    pages         = "004",
    year          = "2021",
    doi           = "10.22323/1.395.0004",
    url           = "https://pos.sissa.it/395/004/pdf"
}

@article{Kaapa:2022gmf,
    author        = "K{\"a}{\"a}p{\"a}, Alex and Kampert, Karl-Heinz and Mayotte, Eric William",
    title         = "{Propagation of extragalactic cosmic rays in the Galactic magnetic field}",
    journal       = "Proceedings of Science",
    volume        = "CRIS2022",
    pages         = "088",
    year          = "2023",
    doi           = "10.22323/1.398.0088",
    url           = "https://sissa.it"
}

@article{CRPropa,
    author        = "Alves Batista, Rafael and others",
    title         = "{CRPropa 3.2: a modular and parallelized framework for cosmic-ray propagation}",
    journal       = "Journal of Cosmology and Astroparticle Physics",
    volume        = "2022",
    number        = "09",
    pages         = "035",
    year          = "2022",
    doi           = "10.1088/1475-7516/2022/09/035",
    eprint        = "2205.11495",
    archivePrefix = "arXiv",
    primaryClass  = "astro-ph.HE",
    url           = "https://arxiv.org"
}

@article{Blasi:2012JCAP,
    author        = "Blasi, Pasquale and Amato, Elena",
    title         = "{Diffusive propagation of cosmic rays from supernova remnants in the Galaxy. II: Anisotropy}",
    journal       = "Journal of Cosmology and Astroparticle Physics",
    volume        = "2012",
    number        = "01",
    pages         = "011",
    year          = "2012",
    doi           = "10.1088/1475-7516/2012/01/011",
    eprint        = "1105.1021",
    archivePrefix = "arXiv",
    primaryClass  = "astro-ph.HE",
    url           = "https://arxiv.org"
}

@article{FaucherGiguere:2006ApJ,
    author        = "Faucher-Gigu{\`e}re, Claude-Andr{\ railway} and Kaspi, Victoria M.",
    title         = "{Birth and Evolution of Isolated Radio Pulsars}",
    journal       = "The Astrophysical Journal",
    volume        = "643",
    number        = "1",
    pages         = "332--355",
    year          = "2006",
    doi           = "10.1086/501516",
    eprint        = "astro-ph/0512585",
}

@article{Bister:2024GMF,
    author        = "Bister, Teresa and Farrar, Glennys R. and Unger, Michael",
    title         = "{The Large-scale Anisotropy and Flux (de)magnification of Ultrahigh-energy Cosmic Rays in the Galactic Magnetic Field}",
    journal       = "The Astrophysical Journal Letters",
    volume        = "975",
    number        = "2",
    pages         = "L21",
    year          = "2024",
    doi           = "10.3847/2041-8213/ad856f",
    eprint        = "2408.00614",
    archivePrefix = "arXiv",
    primaryClass  = "astro-ph.HE",
    url           = "https://arxiv.org/abs/2408.00614"
}

@article{Bister:2024Anisotropies,
    author        = "Bister, Teresa and Farrar, Glennys R.",
    title         = "{Constraints on UHECR Sources and Extragalactic Magnetic Fields from Directional Anisotropies}",
    journal       = "The Astrophysical Journal",
    volume        = "966",
    number        = "1",
    pages         = "71",
    year          = "2024",
    doi           = "10.3847/1538-4357/ad2f3f",
    eprint        = "2312.02645",
    archivePrefix = "arXiv",
    primaryClass  = "astro-ph.HE",
    url           = "https://arxiv.org/abs/2312.02645"
}

@article{Kachelriess:2006,
   title={The Compton–Getting effect on ultra-high energy cosmic rays of cosmological origin},
   volume={640},
   ISSN={0370-2693},
   url={http://dx.doi.org/10.1016/j.physletb.2006.08.006},
   DOI={10.1016/j.physletb.2006.08.006},
   number={5-6},
   journal={Physics Letters B},
   publisher={Elsevier BV},
   author={Kachelrieß, M. and Serpico, P.D.},
   year={2006},
   month=Sept, pages={225–229} }

@article{Cerri:2017,
   title={A signature of anisotropic cosmic-ray transport in the gamma-ray sky},
   volume={2017},
   ISSN={1475-7516},
   url={http://dx.doi.org/10.1088/1475-7516/2017/10/019},
   DOI={10.1088/1475-7516/2017/10/019},
   number={10},
   journal={Journal of Cosmology and Astroparticle Physics},
   publisher={IOP Publishing},
   author={Cerri, Silvio Sergio and Gaggero, Daniele and Vittino, Andrea and Evoli, Carmelo and Grasso, Dario},
   year={2017},
   month=Oct, pages={019–019} }

@ARTICLE{Shen:1971,
       author = {{Shen}, C.~S. and {Mao}, C.~Y.},
        title = "{Anisotropy of High Energy Cosmic-Ray Electrons in the Discrete Source Model}",
      journal = {\aplett},
         year = 1971,
        month = aug,
       volume = {9},
        pages = {169},
       adsurl = {https://ui.adsabs.harvard.edu/abs/1971ApL.....9..169S}
}

@article{Savchenko:2015,
   title={IMPRINT OF A 2 MILLION YEAR OLD SOURCE ON THE COSMIC-RAY ANISOTROPY},
   volume={809},
   ISSN={2041-8213},
   url={http://dx.doi.org/10.1088/2041-8205/809/2/L23},
   DOI={10.1088/2041-8205/809/2/l23},
   number={2},
   journal={The Astrophysical Journal},
   publisher={American Astronomical Society},
   author={Savchenko, V. and Kachelrieß, M. and Semikoz, D. V.},
   year={2015},
   month=Aug, pages={L23} }

@article{Farrar:2000,
   title={Deducing the Source of Ultrahigh Energy Cosmic Rays},
   url={https://arxiv.org/abs/astro-ph/0010370},
   author={Farrar, G. and Piran, T.},
   year={2000},}

@article{Rocha:2022,
   title={Distance and age of the massive stellar cluster Westerlund 1 – II. The eclipsing binary W36},
   volume={517},
   ISSN={1365-2966},
   url={http://dx.doi.org/10.1093/mnras/stac2927},
   DOI={10.1093/mnras/stac2927},
   number={3},
   journal={Monthly Notices of the Royal Astronomical Society},
   publisher={Oxford University Press (OUP)},
   author={Rocha, Danilo F and Almeida, Leonardo A and Damineli, Augusto and Navarete, Felipe and Abdul-Masih, Michael and Mace, Gregory N},
   year={2022},
   month=Oct, pages={3749–3762} }

@article{PierreAuger:2022AcrossAnkle,
    author        = "Abdul Halim, A. and others",
    collaboration = "Pierre Auger Collaboration",
    title         = "{Constraining the sources of ultra-high-energy cosmic rays across and above the ankle with the spectrum and composition data measured at the Pierre Auger Observatory}",
    journal       = "Journal of Cosmology and Astroparticle Physics",
    volume        = "2023",
    number        = "05",
    pages         = "024",
    year          = "2023",
    doi           = "10.1088/1475-7516/2023/05/024",
    eprint        = "2211.02857",
    archivePrefix = "arXiv",
    primaryClass  = "astro-ph.HE",
    url           = "https://arxiv.org/abs/2211.02857"
}

@article{HiRes:2007lra,
    author = "Abbasi, R. U. and others",
    collaboration = "HiRes",
    title = "{First observation of the Greisen-Zatsepin-Kuzmin suppression}",
    eprint = "astro-ph/0703099",
    archivePrefix = "arXiv",
    reportNumber = "RU-PNA-002",
    doi = "10.1103/PhysRevLett.100.101101",
    journal = "Phys. Rev. Lett.",
    volume = "100",
    pages = "101101",
    year = "2008"
}

@article{TIBETIII:2008qon,
    author = "Amenomori, M. and others",
    collaboration = "TIBET III",
    title = "{The All-particle spectrum of primary cosmic rays in the wide energy range from 10**14 eV to 10**17 eV observed with the Tibet-III air-shower array}",
    eprint = "0801.1803",
    archivePrefix = "arXiv",
    primaryClass = "hep-ex",
    doi = "10.1086/529514",
    journal = "Astrophys. J.",
    volume = "678",
    pages = "1165--1179",
    year = "2008"
}

@article{KASCADEGrande:2011kpw,
    author = "Apel, W. D. and others",
    collaboration = "KASCADE Grande",
    title = "{Kneelike structure in the spectrum of the heavy component of cosmic rays observed with KASCADE-Grande}",
    eprint = "1107.5885",
    archivePrefix = "arXiv",
    primaryClass = "astro-ph.HE",
    doi = "10.1103/PhysRevLett.107.171104",
    journal = "Phys. Rev. Lett.",
    volume = "107",
    pages = "171104",
    year = "2011"
}

@article{PierreAuger:2020Spectrum,
    author        = "Aab, Alexander and others",
    collaboration = "Pierre Auger Collaboration",
    title         = "{Features of the energy spectrum of cosmic rays above $2.5 \times 10^{18}$ eV using the Pierre Auger Observatory}",
    journal       = "Physical Review Letters",
    volume        = "125",
    number        = "12",
    pages         = "121106",
    year          = "2020",
    doi           = "10.1103/PhysRevLett.125.121106",
    eprint        = "2008.06486",
    archivePrefix = "arXiv",
    primaryClass  = "astro-ph.HE",
    url           = "https://arxiv.org"
}

@article{Auger:magnetic,
   author        = "Abdul Halim, A. and others",
   collaboration = "Pierre Auger Collaboration",
   title={Impact of the magnetic horizon on the interpretation of the Pierre Auger Observatory spectrum and composition data},
   volume={2024},
   ISSN={1475-7516},
   url={http://dx.doi.org/10.1088/1475-7516/2024/07/094},
   DOI={10.1088/1475-7516/2024/07/094},
   number={07},
   journal={JCAP},
   year={2024},
   month=July, pages={094} }

@article{PierreAuger:2026MassComposition,
    author        = "Abdul Halim, A. and others",
    collaboration = "Pierre Auger Collaboration",
    title         = "{Depth of Maximum of Air-Shower Profiles above $10^{17.7}$ eV Measured with the Fluorescence Detector of the Pierre Auger Observatory}",
    journal = {Phys. Rev. D},
    volume = {114},
    issue = {4},
    pages = {043016},
    numpages = {28},
    year = {2026},
    month = {Aug},
    doi = {10.1103/n616-15v5},
    url = {https://link.aps.org/doi/10.1103/n616-15v5}
}

@article{TelescopeArray:2026rdu,
    author = "Abbasi, R. U. and others",
    collaboration = "Telescope Array",
    title = "{Cosmic ray mass composition measurement in the energy range from \(10^{16.5}\,\,\text{eV}\) to \(10^{18.5}\,\,\text{eV}\) observed with the TALE hybrid detector}",
    eprint = "2603.14804",
    archivePrefix = "arXiv",
    primaryClass = "astro-ph.HE",
    doi = "10.1103/vrky-dxn7",
    journal = "Phys. Rev. D",
    volume = "113",
    number = "6",
    pages = "062003",
    year = "2026"
}

@article{PierreAuger:2017ScienceDipole,
    author        = "Aab, Alexander and others",
    collaboration = "Pierre Auger Collaboration",
    title         = "{Observation of a large-scale anisotropy in the arrival directions of cosmic rays above $8 \times 10^{18}$ eV}",
    journal       = "Science",
    volume        = "357",
    number        = "6357",
    pages         = "1266--1270",
    year          = "2017",
    doi           = "10.1126/science.aan4338",
    eprint        = "1709.07321",
    archivePrefix = "arXiv",
    primaryClass  = "astro-ph.HE",
    url           = "https://arxiv.org"
}

@article{PierreAuger:2024Anisotropy19Y,
    author        = "{Abdul Halim}, Adila and others",
    collaboration = "Pierre Auger Collaboration",
    title         = "{Large-scale cosmic ray anisotropies with 19 years of data}",
    journal       = "The Astrophysical Journal",
    volume        = "976",
    number        = "1",
    pages         = "48",
    year          = "2024",
    doi           = "10.3847/1538-4357/ad843b"
}

@article{PierreAuger:2013AnisotropyApJL,
    author        = "{Abreu}, P. and others",
    collaboration = "Pierre Auger Collaboration",
    title         = "{Constraints on the Origin of Cosmic Rays above $10^{18}$ eV from Large-scale Anisotropy Searches}",
    journal       = "The Astrophysical Journal Letters",
    volume        = "762",
    number        = "1",
    pages         = "L13",
    year          = "2013",
    doi           = "10.1088/2041-8205/762/1/L13",
    url           = "https://iopscience.iop.org/article/10.1088/2041-8205/762/1/L13"
}

@article{Auger:2012_anisotropy,
   title={LARGE-SCALE DISTRIBUTION OF ARRIVAL DIRECTIONS OF COSMIC RAYS
                    DETECTED ABOVE 10<sup>18</sup> eV AT THE PIERRE AUGER
                    OBSERVATORY},
   volume={203},
   ISSN={1538-4365},
   url={http://dx.doi.org/10.1088/0067-0049/203/2/34},
   DOI={10.1088/0067-0049/203/2/34},
   number={2},
   journal={The Astrophysical Journal Supplement Series},
   author = "{Abreu}, P. and others",
   collaboration = "Pierre Auger Collaboration",
   year=2012}

@article{Abbasi:2017_Galprotons,
   title={Search for EeV protons of galactic origin},
   volume={86},
   ISSN={0927-6505},
   url={http://dx.doi.org/10.1016/j.astropartphys.2016.11.001},
   DOI={10.1016/j.astropartphys.2016.11.001},
   journal={Astroparticle Physics},
   publisher={Elsevier BV},
   author={Abbasi, R.U. and others},
   collaboration = "Teelscope Array Collaboration",
   year={2017},
   pages={21–26} }

@article{Valade:2026_CF4,
    author = {Valade, Aur{\'e}lien and Libeskind, Noam and Pomar{\`e}de, Daniel and Stiskalek, Richard and Hoffman, Yehuda and Gottl{\"o}ber, Stefan and Tully, R. Brent},
    title = "{Constraining cosmological simulations with peculiar velocities: a forward-modeling approach}",
    eprint = "2602.03699",
    archivePrefix = "arXiv",
    year = "2026"
}

@article{Hoffman:2018_CF2,
	doi = {10.1038/s41550-018-0502-4},
	url = {https://doi.org/10.1038%2Fs41550-018-0502-4},
	year = 2018,
	month = {jul},
	volume = {2},
	number = {8},
	pages = {680--687},
	author = {Yehuda Hoffman and Edoardo Carlesi and Daniel Pomar{\`{e}}de and R. Brent Tully and H{\'{e}}l{\`{e}}ne M. Courtois and Stefan Gottlöber and Noam I. Libeskind and Jenny G. Sorce and Gustavo Yepes},
	title = {The quasi-linear nearby Universe},
	journal = {Nature Astronomy}
}

@article{unger:2015_ufa,
	title = {Origin of the ankle in the ultrahigh energy cosmic ray spectrum, and of the extragalactic protons below it},
	volume = {92},
	url = {https://link.aps.org/doi/10.1103/PhysRevD.92.123001},
	doi = {10.1103/PhysRevD.92.123001},
	number = {12},
	urldate = {2022-01-11},
	journal = {PRD},
	author = {Unger, Michael and Farrar, Glennys R. and Anchordoqui, Luis A.},
	year = {2015},
	note = {Publisher: American Physical Society},
	pages = {123001},
}

@article{pierog:2025_eposlhcr,
      title={EPOS.LHC-R : a global approach to solve the muon puzzle}, 
      author={Tanguy Pierog and Klaus Werner},
      year={2025},
      eprint={2508.07105},
      archivePrefix={arXiv},
      primaryClass={astro-ph.HE},
      url={https://arxiv.org/abs/2508.07105}, 
}

@article{Riehn:2020_Sibyll,
  title = {Hadronic interaction model sibyll 2.3d and extensive air showers},
  author = {Riehn, Felix and Engel, Ralph and Fedynitch, Anatoli and Gaisser, Thomas K. and Stanev, Todor},
  journal = {Phys. Rev. D},
  volume = {102},
  issue = {6},
  pages = {063002},
  numpages = {28},
  year = {2020},
  month = {Sep},
  publisher = {American Physical Society},
  doi = {10.1103/PhysRevD.102.063002},
  url = {https://link.aps.org/doi/10.1103/PhysRevD.102.063002}
}

@article{Ding:2021,
    title = {The {Imprint} of {Large}-scale {Structure} on the {Ultrahigh}-energy {Cosmic}-{Ray} {Sky}},
    volume = {913},
    issn = {2041-8205},
    url = {https://doi.org/10.3847/2041-8213/abf11e},
    doi = {10.3847/2041-8213/abf11e},
    language = {en},
    number = {1},
    urldate = {2022-02-08},
    journal = {ApJL},
    author = {Ding, Chen and Globus, Noémie and Farrar, Glennys R.},
    
    year = {2021},
    pages = {L13},
}

@article{Allard:2022,
   title={What can be learnt from UHECR anisotropies observations: I. Large-scale anisotropies and composition features},
   volume={664},
   ISSN={1432-0746},
   url={http://dx.doi.org/10.1051/0004-6361/202142491},
   DOI={10.1051/0004-6361/202142491},
   journal={Astronomy \& Astrophysics},
   publisher={EDP Sciences},
   author={Allard, D. and Aublin, J. and Baret, B. and Parizot, E.},
   year={2022},
   month=Aug, pages={A120} }

@article{Li:2026kxd,
    author = "Li, Ben and Blasi, Pasquale and Amato, Elena",
    title = "{Suppressed diffusion and gamma-ray emission from the Cygnus Bubble}",
    eprint = "2606.03881",
    archivePrefix = "arXiv",
    primaryClass = "astro-ph.HE",
    month = "6",
    year = "2026"
}

@article{Blasi:2023quw,
    author = "Blasi, Pasquale and Morlino, Giovanni",
    title = "{High-energy cosmic rays and gamma-rays from star clusters: the case of Cygnus OB2}",
    eprint = "2306.03762",
    archivePrefix = "arXiv",
    primaryClass = "astro-ph.HE",
    doi = "10.1093/mnras/stad1662",
    journal = "Mon. Not. Roy. Astron. Soc.",
    volume = "523",
    number = "3",
    pages = "4015--4028",
    year = "2023"
}

@article{Menchiari:2024tqw,
    author = "Menchiari, Stefano and Morlino, Giovanni and Amato, Elena and Bucciantini, Niccol{\`o} and Beltr{\'a}n, Maria Teresa",
    title = "{Cygnus OB2 as a test case for particle acceleration in young massive star clusters}",
    eprint = "2402.07784",
    archivePrefix = "arXiv",
    primaryClass = "astro-ph.HE",
    month = "2",
    year = "2024"
}

@article{Vieu:2024qjx,
    author = {Vieu, Thibault and Larkin, Cormac J. K. and H{\"a}rer, Lucia and Reville, Brian and Sander, Andreas A. C. and Ramachandran, Varsha},
    title = "{Hydrodynamic simulation of Cygnus OB2: the absence of a cluster wind termination shock}",
    eprint = "2406.13589",
    archivePrefix = "arXiv",
    primaryClass = "astro-ph.GA",
    doi = "10.1093/mnras/stae1627",
    journal = "Mon. Not. Roy. Astron. Soc.",
    volume = "532",
    number = "2",
    pages = "2174--2188",
    year = "2024"
}

@article{Haerer:2025ull,
    author = "Haerer, L. and Vieu, T. and Schulze, F. and Larkin, C. J. K. and Reville, B.",
    title = "{Deciphering the gamma-ray emission in the Cygnus region}",
    eprint = "2508.21644",
    archivePrefix = "arXiv",
    primaryClass = "astro-ph.HE",
    doi = "10.1051/0004-6361/202555531",
    journal = "Astron. Astrophys.",
    volume = "703",
    pages = "A111",
    year = "2025"
}

@article{Deligny:2019,
   title={Measurements and implications of cosmic ray anisotropies from TeV to trans-EeV energies},
   volume={104},
   ISSN={0927-6505},
   url={http://dx.doi.org/10.1016/j.astropartphys.2018.08.005},
   DOI={10.1016/j.astropartphys.2018.08.005},
   journal={Astroparticle Physics},
   publisher={Elsevier BV},
   author={Deligny, O.},
   year={2019},
   
   month=Jan, pages={13–41} }

@article{Jansson:2012pc,
    author        = "Jansson, Ronnie and Farrar, Glennys R.",
    title         = "{A New Model of the Galactic Magnetic Field}",
    journal       = "The Astrophysical Journal",
    volume        = "757",
    number        = "1",
    pages         = "14",
    year          = "2012",
    doi           = "10.1088/0004-637X/757/1/14",
    eprint        = "1204.3662",
    archivePrefix = "arXiv",
    primaryClass  = "astro-ph.GA",
    url           = "https://arxiv.org/abs/1204.3662"
}

@article{Jansson:2012b_Random,
    author        = "Jansson, Ronnie and Farrar, Glennys R.",
    title         = "{The Galactic Magnetic Field}",
    journal       = "The Astrophysical Journal Letters",
    volume        = "761",
    number        = "1",
    pages         = "L11",
    year          = "2012",
    doi           = "10.1088/2041-8205/761/1/L11",
    eprint        = "1210.7820",
    archivePrefix = "arXiv",
    primaryClass  = "astro-ph.GA",
    url           = "https://arxiv.org/abs/1210.7820"
}

@article{Unger:2024uf23,
    author        = "Unger, Michael and Farrar, Glennys R.",
    title         = "{The Coherent Magnetic Field of the Milky Way}",
    journal       = "The Astrophysical Journal",
    volume        = "970",
    number        = "1",
    pages         = "95",
    year          = "2024",
    doi           = "10.3847/1538-4357/ad4a54"
}

@article{Korochkin:2025kst,
    author        = "Korochkin, Alexander and Semikoz, Dmitri and Tinyakov, Peter",
    title         = "{A new model of the coherent Galactic magnetic field}",
    journal       = "Astronomy \& Astrophysics",
    volume        = "693",
    pages         = "A284",
    year          = "2025",
    doi           = "10.1051/0004-6361/202451310",
    url           = "https://aanda.org"
}

@article{Planck:2016GMF,
   title={<i>Planck</i>intermediate results: XLII. Large-scale Galactic magnetic fields},
   volume={596},
   ISSN={1432-0746},
   url={http://dx.doi.org/10.1051/0004-6361/201528033},
   DOI={10.1051/0004-6361/201528033},
   journal={Astronomy \& Astrophysics},
   publisher={EDP Sciences},
   author={Adam, R. and Ade, P. A. R. and Alves, M. I. R. and Ashdown, M. and Aumont, J. and Baccigalupi, C. and Banday, A. J. and Barreiro, R. B. and Bartolo, N. and Battaner, E. and Benabed, K. and Benoit-Lévy, A. and Bernard, J.-P. and Bersanelli, M. and Bielewicz, P. and Bonavera, L. and Bond, J. R. and Borrill, J. and Bouchet, F. R. and Boulanger, F. and Bucher, M. and Burigana, C. and Butler, R. C. and Calabrese, E. and Cardoso, J.-F. and Catalano, A. and Chiang, H. C. and Christensen, P. R. and Colombo, L. P. L. and Combet, C. and Couchot, F. and Crill, B. P. and Curto, A. and Cuttaia, F. and Danese, L. and Davis, R. J. and de Bernardis, P. and de Rosa, A. and de Zotti, G. and Delabrouille, J. and Dickinson, C. and Diego, J. M. and Dolag, K. and Doré, O. and Ducout, A. and Dupac, X. and Elsner, F. and Enßlin, T. A. and Eriksen, H. K. and Ferrière, K. and Finelli, F. and Forni, O. and Frailis, M. and Fraisse, A. A. and Franceschi, E. and Galeotta, S. and Ganga, K. and Ghosh, T. and Giard, M. and Gjerløw, E. and González-Nuevo, J. and Górski, K. M. and Gregorio, A. and Gruppuso, A. and Gudmundsson, J. E. and Hansen, F. K. and Harrison, D. L. and Hernández-Monteagudo, C. and Herranz, D. and Hildebrandt, S. R. and Hobson, M. and Hornstrup, A. and Hurier, G. and Jaffe, A. H. and Jaffe, T. R. and Jones, W. C. and Juvela, M. and Keihänen, E. and Keskitalo, R. and Kisner, T. S. and Knoche, J. and Kunz, M. and Kurki-Suonio, H. and Lamarre, J.-M. and Lasenby, A. and Lattanzi, M. and Lawrence, C. R. and Leahy, J. P. and Leonardi, R. and Levrier, F. and Liguori, M. and Lilje, P. B. and Linden-Vørnle, M. and López-Caniego, M. and Lubin, P. M. and Macías-Pérez, J. F. and Maggio, G. and Maino, D. and Mandolesi, N. and Mangilli, A. and Maris, M. and Martin, P. G. and Martínez-González, E. and Masi, S. and Matarrese, S. and Melchiorri, A. and Mennella, A. and Migliaccio, M. and Miville-Deschênes, M.-A. and Moneti, A. and Montier, L. and Morgante, G. and Munshi, D. and Murphy, J. A. and Naselsky, P. and Nati, F. and Natoli, P. and Nørgaard-Nielsen, H. U. and Oppermann, N. and Orlando, E. and Pagano, L. and Pajot, F. and Paladini, R. and Paoletti, D. and Pasian, F. and Perotto, L. and Pettorino, V. and Piacentini, F. and Piat, M. and Pierpaoli, E. and Plaszczynski, S. and Pointecouteau, E. and Polenta, G. and Ponthieu, N. and Pratt, G. W. and Prunet, S. and Puget, J.-L. and Rachen, J. P. and Reinecke, M. and Remazeilles, M. and Renault, C. and Renzi, A. and Ristorcelli, I. and Rocha, G. and Rossetti, M. and Roudier, G. and Rubiño-Martín, J. A. and Rusholme, B. and Sandri, M. and Santos, D. and Savelainen, M. and Scott, D. and Spencer, L. D. and Stolyarov, V. and Stompor, R. and Strong, A. W. and Sudiwala, R. and Sunyaev, R. and Suur-Uski, A.-S. and Sygnet, J.-F. and Tauber, J. A. and Terenzi, L. and Toffolatti, L. and Tomasi, M. and Tristram, M. and Tucci, M. and Valenziano, L. and Valiviita, J. and Van Tent, F. and Vielva, P. and Villa, F. and Wade, L. A. and Wandelt, B. D. and Wehus, I. K. and Yvon, D. and Zacchei, A. and Zonca, A.},
   year={2016},
   month=Dec, pages={A103} }

@article{Pshirkov:2011,
   title={DERIVING THE GLOBAL STRUCTURE OF THE GALACTIC MAGNETIC FIELD FROM FARADAY ROTATION MEASURES OF EXTRAGALACTIC SOURCES},
   volume={738},
   ISSN={1538-4357},
   url={http://dx.doi.org/10.1088/0004-637X/738/2/192},
   DOI={10.1088/0004-637x/738/2/192},
   number={2},
   journal={The Astrophysical Journal},
   publisher={American Astronomical Society},
   author={Pshirkov, M. S. and Tinyakov, P. G. and Kronberg, P. P. and Newton-McGee, K. J.},
   year={2011},
   month=Aug, pages={192} }

@article{Farrar:ICRC2021,
  author = "Farrar, Glennys R.  and  Ding, Chen",
  title = "{Transient Source for the Highest Energy Galactic Cosmic Rays}",
  doi = "10.22323/1.395.0488",
  journal = "PoS",
  year = 2021,
  volume = "ICRC2021",
  pages = "488"
}

@article{Compton_Getting:1935,
  title = {An Apparent Effect of Galactic Rotation on the Intensity of Cosmic Rays},
  author = {Compton, Arthur H. and Getting, Ivan A.},
  journal = {Phys. Rev.},
  volume = {47},
  issue = {11},
  pages = {817--821},
  numpages = {0},
  year = {1935},
  month = {Jun},
  publisher = {American Physical Society},
  doi = {10.1103/PhysRev.47.817},
  url = {https://link.aps.org/doi/10.1103/PhysRev.47.817}
}

@article{Mollerach:2022,
   title={Large-scale anisotropies of extragalactic cosmic rays below the ankle},
   volume={2022},
   ISSN={1475-7516},
   url={http://dx.doi.org/10.1088/1475-7516/2022/12/021},
   DOI={10.1088/1475-7516/2022/12/021},
   number={12},
   journal={Journal of Cosmology and Astroparticle Physics},
   publisher={IOP Publishing},
   author={Mollerach, S. and Roulet, E. and Taborda, O.},
   year={2022},
   month=Dec, pages={021} }

@ARTICLE{2009ARA&A..47..481A,
       author = {{Asplund}, Martin and {Grevesse}, Nicolas and {Sauval}, A. Jacques and {Scott}, Pat},
        title = "{The Chemical Composition of the Sun}",
      journal = {\araa},
         year = 2009,
        month = sep,
       volume = {47},
       number = {1},
        pages = {481-522},
          doi = {10.1146/annurev.astro.46.060407.145222},
archivePrefix = {arXiv},
       eprint = {0909.0948},
 primaryClass = {astro-ph.SR},
       adsurl = {https://ui.adsabs.harvard.edu/abs/2009ARA&A..47..481A}
}
\bibliographystyle{aasjournalv7}

\end{document}